\documentclass[a4paper,12pt]{article}
\usepackage{amsmath,amsthm,amsfonts,amssymb,bm,mathrsfs}
\usepackage{enumerate}
\usepackage{graphicx}
\usepackage{natbib}
\usepackage{xcolor}
\usepackage{sgame}
\usepackage{indentfirst}
\usepackage{autobreak}
\usepackage{comment}
\usepackage{multirow,array}
\usepackage{subfigure}
\usepackage[scale=0.76]{geometry}

\usepackage{float}

\usepackage{tikz}
\usetikzlibrary{intersections}
\usetikzlibrary{decorations.markings}

\usepackage{titlesec}

\allowdisplaybreaks

\usepackage{setspace}
\renewcommand{\baselinestretch}{1.5}
\allowdisplaybreaks

\newtheorem{defn}{Definition}
\newtheorem{thm}{Theorem}
\newtheorem{lem}{Lemma}
\newtheorem{assum}{Assumption}
\newtheorem{prop}{Proposition}
\newtheorem{rmk}{Remark}
\newtheorem{exam}{Example}
\newtheorem{coro}{Corollary}

\newtheorem{claim}{Claim}
\newtheorem{Fact}{Fact}

\newcommand{\bE}{\mathbb{E}}

\newcommand{\bR}{\mathbb{R}}

\newcommand{\cA}{\mathcal{A}}

\newcommand{\cF}{\mathcal{F}}

\newcommand{\cL}{\mathcal{L}}
\newcommand{\cM}{\mathcal{M}}

\newcommand{\cT}{\mathcal{T}}

\DeclareMathOperator{\rmd}{d\!}

\usepackage{hyperref}

\hypersetup{colorlinks=true,
	citecolor=blue,
	linkcolor=blue,
	urlcolor=blue,
	bookmarksopen=true,
	pdfstartview={XYZ null null 1.00},
	pdfpagelayout={SinglePage}
}

\everydisplay{\setstretch{1}} 

\begin{document}

\title{Monotone Equilibrium, Admissibility and Perfection\footnote{This paper has been presented in the 7th World Congress of the Game Theory Society, Beijing, China, August 2024; 2025 NUS-WHU Economics Symposium, Wuhan University, March 2025; 24th SAET Conference, Ischia Italy, June 2025; 13th World Congress of the Econometric Society, Seoul, Korea, August 2025. We thank the 
participants at these conferences for their valuable comments. The authors are also grateful to Oriol Carbonell-Nicolau, Jinwoo Kim, David McAdams, Pavlo Prokopovych, Satoru Takahashi and Nicholas C. Yannelis for helpful suggestions and encouragements. An earlier version of this paper was circulated under the title ``Monotone Perfection.’’
}}

	
\author{Wei~He\thanks{Department of Economics, The Chinese University of Hong Kong, Hong Kong. E-mail: hewei@cuhk.edu.hk.}
\and
Yeneng~Sun\thanks{Departments of Economics and Mathematics, National University of Singapore, Singapore. Email: ynsun@nus.edu.sg.}
\and
Hanping~Xu\thanks{Institute for Advanced Study in Mathematics, Harbin Institute of Technology, Harbin. Email: xuhanping@u.nus.edu.}
}

\date{July 9, 2026}
	
\pagenumbering{gobble} 

\maketitle

\abstract{


This paper studies monotone equilibrium, admissibility, and perfection in Bayesian games. \cite{A2001}, \cite{M2003}, and \cite{R2011} prove existence of pure-strategy monotone equilibria under single-crossing and quasi-supermodularity. We construct a counterexample satisfying these assumptions but with no admissible monotone equilibrium, showing these results cannot generally be strengthened without further conditions. 
To address this existence problem, we introduce perfect monotone equilibrium, combining monotonicity with trembling-hand robustness. We show that stronger assumptions such as increasing differences and supermodularity ensure existence of perfect (and hence admissible) monotone equilibria. We also demonstrate the practical relevance of our results through auctions and Bertrand competitions.

\bigskip

}

\newpage
\tableofcontents
\newpage

\pagenumbering{arabic} 
\setcounter{page}{1}


\section{Introduction}
As a foundational element of game theory, the model of Bayesian games has provided a standard analytical tool and has been applied in many areas. Auctions, for instance, are among the most successful applications of this model. In auction theory and many other applications, attention often focuses on pure strategy monotone equilibria. Such equilibria capture a natural economic prediction: bidders with higher valuations tend to submit higher bids. The existence of pure strategy monotone equilibrium has been established in Bayesian games with great success. The seminal work of \cite{A2001} first demonstrated the existence of a pure strategy monotone equilibrium based on the single crossing condition. \cite{M2003} generalized this result to settings with multidimensional and partially ordered type and action spaces. \cite{R2011} showed that contractibility is automatically satisfied given any nonempty monotone best responses, and employed a powerful fixed-point theorem to establish the general existence of pure strategy monotone equilibria. This line of research has proven to be useful across a wide range of applications.\footnote{For more applications and developments of monotone equilibria, see, for example, \cite{Reny1999}, \cite{JSSZ2002}, \cite{RZ2004}, \cite{JS2005}, \cite{M2006}, and \cite{PY2017, PY2019}.}

A monotone equilibrium, however, does not by itself always deliver a strategically convincing prediction. 
Specifically, it may permit some types to choose weakly dominated actions, thereby violating the basic criterion of admissibility. While monotonicity is valuable for yielding ordered and economically interpretable predictions, admissibility remains a fundamental requirement for rational decision-making.
As pointed out by \citet[p.429]{Arrow1951}, ``The rule that one should restrict oneself to admissible actions is extremely reasonable.''\footnote{\cite{LR1957} adopted  admissibility as a decision-making axiom, and \cite{KM1986} considered it a core requirement for strategic stability.
Furthermore,  \cite{BFK2008} characterized admissibility, emphasizing that ``Weak-dominance concepts give sharp predictions in many games of applied interest.''} In the following simple example, we illustrate the issue that monotonicity and admissibility may diverge.

Consider a first-price auction with two bidders. Bidder~$1$'s private value $v_1$ is uniformly drawn from $[0,5]$, and bidder~$2$'s value $v_2$ is uniformly drawn from $[7,8]$. Each bidder $i$ submits a bid $b_i \in \{0,1,2,\cdots,8\}$ after observing her value $v_i$. The bidder who submits a higher bid wins the good, and ties are broken with equal probability. Here, $\hat{b}_1 \equiv  5$ and $\hat{b}_2 \equiv 6$ is a monotone equilibrium. To see it, given that bidder~$2$ bids $6$, bidder~$1$'s payoff is at most $0$, achievable by bidding any number lower than $6$. When bidder~$1$ bids $5$,  the best response of bidder~$2$ is to bid $6$ for any $v_2$. Thus, $(\hat{b}_1, \hat{b}_2)$ is an equilibrium, which is trivially monotone. However, it is obvious that bidding $5$ is weakly dominated by bidding $0$ for bidder~$1$, as her payoff is at most $0$ with a bid of $5$ and at least $0$ with a bid of $0$. When bidder~$2$ bids $0$, bidder~$1$ is strictly better off by bidding $0$.

Earlier literature cautions against trivial monotone equilibria involving weakly dominated strategies.\footnote{\cite{M2007} provided two auction examples in which all weakly undominated equilibria are non-monotone. It can be verified that these two examples fall outside the conditions under which \cite{A2001}, \cite{M2003}, and \cite{R2011} establish the existence of monotone equilibria. \citet[p.~518]{R2011} noted that ``Multi-unit uniform-price auctions always have trivial equilibria in weakly dominated strategies in which some player always bids very high on all units and all others always bid zero.'' }
Since equilibrium existence results can often be strengthened by imposing admissibility, it is natural to ask whether the monotone-equilibrium existence theorems of \cite{A2001}, \cite{M2003}, and \cite{R2011} can likewise be strengthened to guarantee admissible monotone equilibria. 
Example~\ref{exam-2} shows that this strengthening is not generally possible: although the example satisfies the conditions imposed in those papers, it admits no admissible monotone equilibrium. Thus, admissibility is a substantive constraint on monotone equilibrium rather than a routine strengthening of existing equilibrium results.
This leaves open a fundamental question in the theory of monotone equilibrium: under what alternative conditions, if any, can the existence of admissible monotone equilibria be guaranteed?

To study this question, we adopt a widely used behavioral criterion that is stronger than admissibility and allows for the possibility of small mistakes, commonly known as a ``trembling hand.'' As emphasized by \cite{S1975}, ``a satisfactory interpretation of equilibrium ... seems to require that the possibility of mistakes is not completely excluded.'' 
This consideration motivates a refinement of monotone equilibrium that accommodates small deviations while preserving the useful monotone structure.
In this paper, we introduce and study a new equilibrium concept, which we call  ``perfect monotone equilibrium.'' The concept captures the idea that players may occasionally make mistakes, but only with negligible probability. Crucially, players must continue to play best responses even when others' strategies are subject to small trembles.\footnote{In the above first-price auction example, suppose that bidder~$2$ might choose unintended strategies due to a trembling hand, and adopt a completely mixed strategy that places strictly positive probability to every bid with probability $1 - \epsilon$ on bid $6$. In this case, bidder~$1$ strictly prefers bidding $0$ to $5$, as it yields a strictly higher payoff when bidder~$2$ bids $0$. Thus,  bidding $5$ is never a best response for bidder~$1$ for any $0< \epsilon < 1$. Bidder $1$ should therefore prefer bidding $0$ to $5$, anticipating the possibility that bidder 2 may bid $0$ with positive probability. } By incorporating trembling-hand robustness, perfect monotone equilibrium selects monotone equilibria that are not only admissible, but also robust to small perturbations.

We begin with Bayesian games with finite actions in the multidimensional setting. \cite{A2001} established the existence of monotone equilibria in the one-dimensional setting under the assumption that each player's interim payoff satisfies the single crossing condition when opponents adopt monotone strategies. \cite{M2003} showed that a monotone equilibrium exists in settings with multidimensional and partially ordered type and action spaces under the additional quasi-supermodularity condition. In Section~\ref{subsec-one dimensional}, we present a counterexample in Example~\ref{exam-2} demonstrating that these conditions are not sufficient to guarantee the existence of perfect monotone equilibria. In this example, the game admits a monotone equilibrium and a (non-monotone) perfect equilibrium, but no perfect monotone equilibrium. Indeed we show something stronger: even when interim payoffs satisfy the single crossing condition and quasi-supermodularity, there is no admissible monotone equilibrium. Thus, under the conventional conditions, monotonicity and admissibility may not always be compatible. The failure in this example is due to the absence of complementarity across the coordinates of one player's action, which highlights the need to search for stronger conditions. We prove the existence of a perfect monotone equilibrium in Theorem~\ref{thm-main existence} under two assumptions:  when opponents adopt monotone strategies, each player's interim payoff has increasing differences in own actions and types, and is supermodular in own actions. These two conditions imply that each player's interim payoff satisfies monotone incremental returns in own types and complementarity in own actions.\footnote{For more discussions and developments of those conditions, see, for example, \cite{T1978}, \cite{MS1994}, \cite{Shannon1995}, \cite{QS2009, QS2012}, and \cite{AR2025}. }  Importantly, it yields an admissible monotone equilibrium in finite-action Bayesian games.

Notably, Example~\ref{exam-2} in fact satisfies the increasing differences condition, but its interim payoffs are only quasi-supermodular, not supermodular. In the one-dimensional setting, supermodularity is automatic. Nevertheless, the single crossing condition still does not suffice for monotone perfection. In Example~\ref{exam-SCC}, we provide a simple two-player Bayesian game with independent types and binary actions. In this example, each player's interim payoff satisfies the single crossing condition when opponents play monotone strategies. The game admits both a monotone equilibrium and a (non-monotone) perfect equilibrium, but no perfect monotone equilibrium. 
 The reason is that the single crossing condition does not guarantee that monotone best responses remain monotone under fully supported trembles. This identifies increasing differences, rather than single crossing, as the right condition for perfect monotone equilibrium. 
Together, these two examples establish that both the increasing differences condition and supermodularity are tight.

Theorem~\ref{thm-main existence} applies to supermodular games, in which the existence of perfect monotone equilibria is established in this commonly used class of games.\footnote{Example~\ref{exam-super} presents a simple supermodular game with two equilibria: a monotone but imperfect equilibrium where players choose the maximal action, and a perfect monotone equilibrium where both players choose the minimal action.} We also revisit the multi-unit uniform price auction models studied in \cite{M2003, M2007} and \cite{R2011}. Their results are strengthened by showing that a perfect monotone equilibrium exists.\footnote{We also discuss examples showing that the conclusion may fail when the relevant increasing-differences structure breaks down.}

In Section~\ref{subsec-Reny}, we extend the existence result to general Bayesian games as in \cite{R2011}.\footnote{Weaker versions of the single crossing condition and quasi-supermodularity were imposed in \cite{R2011}.} We show in Theorem~\ref{thm-Reny} that the existence of perfect monotone equilibrium continues to hold in the general setting with  action spaces being compact and complete metric lattices and type spaces being partially ordered probability spaces, under the same assumptions of increasing differences and supermodularity. A perfect monotone equilibrium satisfies the property of limit admissibility in Bayesian games with infinitely many actions, which extends the notion of admissibility to the setting with general action spaces. The equilibrium existence result is applied to a multi-unit auction with risk-averse bidders and uniform pricing, which has been studied in \cite{M2007} and \cite{R2011}. It was pointed out that a monotone equilibrium does not exist in this model under the standard partial order on Euclidean spaces. \cite{R2011} established a non-trivial monotone equilibrium in the sense that bidders play admissible monotone strategies.\footnote{\cite{R2011} constructed a novel endogenous partial order tailored to the setting, and proved the existence of a non-trivial monotone equilibrium under this new partial order.} Our Theorem~\ref{thm-Reny} is applicable in this model and guarantees the existence of a perfect monotone equilibrium, which rules out trivial, weakly dominated actions.

In many important applications of Bayesian games, players' payoffs are often discontinuous. For example, in first-price auctions for a single object, a bidder experiences a discrete change in payoffs when her bid shifts from being below opponents' highest bid to above it. Similarly, in price competitions, a firm's market share can jump significantly if its price slightly undercuts the current market price. Due to the payoff discontinuities, Theorem~\ref{thm-Reny} does not apply directly to those environments. In Section~\ref{sec-application}, we study three economic applications with affiliated types and a continuum of actions; namely, first-price auctions, all-pay auctions, and Bertrand competitions. We introduce a class of ``generalized auctions'' that encompasses these examples, and prove in Theorem~\ref{thm-discontinuous} the existence of perfect monotone equilibria. 

Our paper is related to the literature that provides equilibrium refinement aiming to eliminate undesirable equilibria. The classic work of \cite{S1975} proposed the notion of perfect equilibrium by introducing completely mixed strategies. \cite{SS1995} studied the possible issues with the notion of perfect equilibrium in strategic form games with compact action spaces. 
\cite{Bajoori2016} proved the existence of perfect equilibrium in Bayesian games with finite action spaces and continuous payoffs over countable type spaces. \cite{Carbonell-Nicolau2021} showed that a perfect equilibrium exists in discontinuous Bayesian games under the conditions of strong uniform payoff security and absolute continuity.\footnote{The notion of perfect equilibrium has also been studied in other general environments; see, for example, \cite{CM2014} for potential games, and \cite{Rath1994, Rath1998} and \cite{SZ2020} for large games.} All those papers do not invoke the monotone method, while our paper focuses on monotone strategies in Bayesian games with ordered structures and studies the notion of perfect monotone equilibrium.

There is another stream of literature on games with strategic complementarity, which focuses on the existence of pure strategy equilibria and monotone comparative statics in supermodular games; see, for example, \cite{T1978}, \cite{MR1990}, \cite{Vives1990} and  \cite{MS1994}.\footnote{\cite{Quah2007}, \cite{DQ2024}, and \cite{CKK2026} studied various novel concepts of set orders and demonstrated their applications across a wide range of contexts.} \cite{CM2015} further studied pure strategy perfect equilibria in supermodular games with discontinuous payoffs. In these works, the strategies themselves need not be monotone in types. \cite{ZV2007} studied the class of monotone supermodular games and provided a constructive proof for the existence of a greatest and a least Bayesian Nash equilibrium, each one in strategies that are monotone in type.
In the current paper, we develop monotone perfection methods applying to Bayesian games that may not exhibit complementarity across action profiles, but satisfy the monotone incremental returns in own type when others adopt monotone strategies.

This paper is also related to the literature that aims to provide sufficient conditions for the existence of equilibria in Bayesian games. \cite{RR1982} worked with the conditions of independent atomless types and private values. \cite{MW1985} considered 
general Bayesian games with the absolute continuity condition on type spaces. 
\cite{CM2018} studied Bayesian games with discontinuous payoffs. \cite{FY2018} provided sufficient conditions that ensure the existence of Pareto-undominated and socially-maximal pure strategy equilibria. \cite{HS2019} introduced a general condition called  ``coarser inter-player information''  and showed that it is necessary and sufficient for the existence of pure strategy equilibria. All those papers do not consider monotone strategies and perfect equilibria.

The remainder of the paper is organized as follows. Section~\ref{sec-bayesian} introduces the model of Bayesian games. Section~\ref{sec-results} defines perfect monotone equilibrium, presents two counterexamples showing that the conventional single-crossing and quasi-supermodularity conditions do not ensure its existence, and establishes existence under increasing differences and supermodularity. Section~\ref{sec-application} provides applications to auctions and Bertrand competitions. All proofs are collected in the appendices.


\section{Bayesian Games}\label{sec-bayesian}


Before describing the model, we first introduce the concept of lattices. Let $(L , \ge)$ be a partially ordered set.\footnote{It means that  $\geq$ is transitive  $\left( x \geq y \text{ and } y \geq z \implies x \geq z \right), \text{ reflexive } \left( x \geq x \right), \text{ and antisymmetric }$ $\left( x \geq y \text{ and } y \geq x \implies x = y \right).
$} For any $S \subseteq L$, let $\vee S$ denote the least upper bound of $S$ in $L$ (if it exists), satisfying 
\begin{itemize}
\item $\vee S \ge s$  for any $s \in S$, and 
\item for $c \in L$, if $c \ge s$ for any $s \in S$, then $c \ge \vee S$. 
\end{itemize}
Similarly, let $\wedge S$ denote the greatest lower bound of $S$ in $L$ (if it exists), satisfying  
\begin{itemize}
\item $\wedge S \le s$ for any $s \in S$, and

\item for $c \in L$, if $c \le s$ for any $s \in S$, then $c \le \wedge S$. 
\end{itemize}
For example, if $S = \{a,b\}$, then $a \vee b = \vee S$ and $a \wedge b = \wedge S$.

\begin{defn}
\begin{enumerate}
\item A lattice is a partially ordered set $(L , \ge)$ such that for all $a, b \in L$,  $a \vee b \in L$ and $a \wedge b \in L$. A subset $E$ of $L$ is a sublattice  if  and only if $a \vee b \in E$,  $a \wedge b \in E$  for all $a, b \in E$. 

\item A metric lattice  $L$ is a lattice endowed with a metric under which the operators $\vee $ and $\wedge$ are continuous.\footnote{A lattice in $\mathbb{R}^m$ is called an Euclidean metric lattice if it is endowed with the coordinate-wise partial order and the Euclidean metric.}

\item A lattice $L$ is complete if for each nonempty $S \subseteq L$, $\vee S \in L$ and $\wedge S \in L$. Every finite lattice is complete.

\end{enumerate}
 \label{defn -  lattices}
\end{defn}

Now we are ready to present the model of Bayesian games. 

\begin{itemize}

\item The set of players is denoted by $I = \{1, 2, \ldots, n \}$, $n \ge 2$.
		
\item For each $i \in I$, the action space of player~$i$ is $A_i$, which is a compact metric space and a 
lattice.
Denote $A= \prod_{i=1}^n A_i$. Let $\cA_i$ be the induced Borel $\sigma$-algebra on $A_i$, and $\cA = \otimes_{i = 1}^n \cA_i.$

\item Player $i$'s type $t_i$ is drawn from the type space $T_i$, which is endowed with the $\sigma$-algebra $\cT_i$. The set $T_i$ is partially ordered.
    Denote $T = \prod_{i=1}^n T_i$. 

\item The common prior over the players' type spaces is a countably additive probability measure $\lambda$ on $T$. Let $(T_i, \cT_i, \lambda_i)$ be an atomless probability space, where $\lambda_i$ is the marginal of $\lambda$ on $T_i$ for each $i \in I$.\footnote{We adopt the convention that a probability space is a complete countably additive measure space with the full set having measure one.} 
		
\item Given action profile $a \in A$ and type profile $t \in T$, player~$i$'s payoff is $u_i(a, t)$, which is bounded, jointly measurable, and continuous in $a$ for each $t \in T$.

\item  For each player~$i \in I$, a pure strategy $g_{i}^*$  is a measurable mapping from $(T_i, \cT_i, \lambda_i)$ to $A_i$. A mixed strategy $g_{i}$  is a mapping from $T_i$ to $\cM(A_i)$ such that for each $B \in \cA_i$, $g_{i}(t_i) (B )$ (also denoted by $g_{i}(t_i, B )$) is measurable in $t_i \in T_i$, where $\cM(A_i)$ is the set of Borel probability measures on $A_i$.\footnote{It is well-known that a pure strategy $g_{i}^*$ can be viewed as a mixed strategy $g_{i}$ by letting $g_{i}(t_i) = \delta_{g_{i}^*(t_i)}$, where $\delta_{g_{i}^*(t_i)}$ is the Dirac measure concentrated at the point $g_{i}^*(t_i)$. }

\item A monotone (or increasing) strategy $\alpha_i$ is a pure strategy profile that is increasing in player~$i$'s types; that is, $\alpha_i(t_i^H) \ge \alpha_i(t_i^L)$ for $t_i^H \ge t_i^L$.   For each $i \in I$, let $\cF_i$ be the set of monotone strategies of player $i$. As usual, $\cF = \prod_{i = 1}^n \cF_i $ and $\cF_{-i} = \prod_{j \ne i,  j \in I} \cF_j $.

\item Given a mixed strategy profile $g=(g_1,\ldots,g_n)$, the interim payoff of player~$i$ depends on her own action  $a_i$, own type $t_i$, and  other players' strategies $g_{-i} = \left(g_1, \cdots, g_{i-1}, g_{i+1}, \cdots, g_n \right)$, 
$$ V_i (a_i, t_i; g_{-i}) = \int_{T_{-i}}  \int_{A_{-i}}u_i(a_i, a_{-i}, t_i, t_{-i})  \mathop{\otimes}\limits_{j  \in I, j  \ne i} g_j(t_j, \rmd a_j)  \rmd  \lambda_{-i}( t_{-i}|t_i),
$$
where $ \lambda_{-i}( t_{-i} | t_i)$ is a version of the conditional probability measure on $T_{-i}$ given $t_i$. 
\end{itemize}
	
Let $\mathrm{BR}_i(t_i, g_{-i})= \left\{a_i \in A_i: V_i (a_i, t_i; g_{-i}) \ge V_i (a_i', t_i; g_{-i}) \text{ for all } a_i' \in A_i \right\}$ be the set of player $i$'s best responses in $A_i$ to other players' strategies $g_{-i}$ at type $t_i$. 

\begin{defn}
\begin{enumerate}
\item A Bayesian Nash equilibrium is a pure strategy profile $g^*=(g^*_1,g_2^*,\ldots,g^*_n)$ such that for each player $i \in I$, $g^*_i(t_i) \in \text{BR}_i(t_i, g^*_{-i})$ for $\lambda_i$-almost all $t_i \in T_i$.

\item A monotone equilibrium is a Bayesian Nash equilibrium $g^*=(g^*_1,g_2^*,\ldots,g^*_n)$ with $g^*_i$ being an increasing strategy for each player $i \in I$. 
\end{enumerate}
\label{def-BNE}
\end{defn}


\vspace{-0.5cm}

\section{Main Results}\label{sec-results}

In this section, we introduce the new solution concept ``perfect monotone equilibrium'' and study its existence in Bayesian games. Section~\ref{subsec-one dimensional} focuses on finite-action Bayesian games with multidimensional types and actions. In this environment, a perfect monotone equilibrium is automatically admissible. We present a counterexample to demonstrate that the conventional single crossing condition and quasi-supermodularity are insufficient to guarantee the existence of perfect monotone equilibrium. We then prove an existence theorem under the conditions of increasing differences and supermodularity. In Section~\ref{subsec-Reny}, we extend the analysis to Bayesian games with general type and action spaces, demonstrating their applicability beyond the standard multidimensional setting.


\subsection{Bayesian games with finite action spaces}\label{subsec-one dimensional}

Throughout this subsection, we work with Bayesian games with finite action spaces.


\subsubsection{Admissibility and perfect monotone equilibrium}\label{subsec-perfect}

The notion of monotone equilibrium has been widely studied in Bayesian games; see \cite{A2001}, \cite{M2003}, and \cite{R2011}. Yet, as demonstrated by the simple example in the introduction, a monotone equilibrium may use weakly dominated actions. Below, we propose the notion of ``perfect monotone equilibrium'' in the finite-action setting, which strengthens the notion of monotone equilibrium by requiring that such equilibria be perfect. Specifically, we consider a refinement of monotone equilibrium that remains trembling-hand robust, while retaining the useful monotone structure. Importantly, a perfect monotone equilibrium rules out weakly dominated actions and is robust to small deviations. 

We first recall several relevant dominance notions for Bayesian games.

\begin{defn}[Dominance]\label{defn-dominance}
\begin{enumerate}
\item For player $i \in I$, an action $a_i \in A_i$ is said to be {strictly dominated} at $t_i \in T_i$ if there exists a mixed action $\sigma_i \in \cM(A_i)$ such that
\[
V_i(a_i,t_i;g_{-i})<\sum_{a_i' \in A_i}V_i(a_i',t_i;g_{-i})\sigma_i(a_i')
\]
for every strategy profile $g_{-i}$.

\item For player $i \in I$, an action $a_i \in A_i$ is said to be {weakly dominated} at $t_i \in T_i$ if there exists a mixed action $\sigma_i \in \cM(A_i)$ such that
\begin{itemize}
\item[(i)] 
\[
V_i(a_i,t_i;g_{-i})\le \sum_{a_i' \in A_i}V_i(a_i',t_i;g_{-i})\sigma_i(a_i')
\]
for every strategy profile $g_{-i}$; and

\item[(ii)] there exists a strategy profile $\hat g_{-i}$ such that
\[
V_i(a_i,t_i;\hat g_{-i})< \sum_{a_i' \in A_i}V_i(a_i',t_i;\hat g_{-i})\sigma_i(a_i').
\]
\end{itemize}

\item An action $a_i \in A_i$ is said to be {undominated} at type $t_i$ if there is no mixed action in $\cM(A_i)$ that weakly dominates $a_i$.
\end{enumerate}
\end{defn}

\begin{defn}
A pure strategy profile $g$ is said to be {admissible} if for each player~$i$, the action $g_i(t_i)$ is undominated at type $t_i$ for $\lambda_i$-almost all $t_i \in T_i$.
\end{defn}

Next, we define the notion of ``perfect monotone equilibrium.'' A mixed strategy profile $g=(g_1,\ldots,g_n)$ is called a completely mixed strategy profile if for each player $i$, each type $t_i \in T_i$, and each action $a_i \in A_i$, one has $g_i(t_i, a_i)>0$.

\begin{defn}[Monotone perfection with finite actions]
\

\begin{enumerate}[a.]
\item A pure strategy profile $g=(g_1,\ldots,g_n)$ is said to be perfect if there exists a sequence of completely mixed strategy profiles $\{g^k=(g_1^k,\ldots,g_n^k)\}_{k=1}^{\infty}$ such that for every player~$i$ and $\lambda_i$-almost all $t_i \in T_i$, the following properties hold:
\begin{itemize}
\item[(1)] $\lim\limits_{k\to\infty}g_i^k(t_i)=\delta_{g_i(t_i)}$ in $\cM(A_i)$; and
\item[(2)] $g_i(t_i)\in \mathrm{BR}_i(t_i,g_{-i}^k)$ for sufficiently large $k$.
\end{itemize}
A Bayesian Nash equilibrium $g$ is said to be a perfect equilibrium if it is perfect.

\item A pure strategy profile $g$ is said to be a perfect monotone equilibrium if it is both a perfect equilibrium and a monotone equilibrium.
\end{enumerate}
\label{def-perfect}
\end{defn}

The definition above states that the pure action prescribed at each type must remain optimal against arbitrarily small fully supported perturbations of the opponents' strategies. This is consistent with the the notion of perfect equilibrium introduced in \cite{S1975}. 
It is well known that any strategy profile that is perfect in a complete-information game must be an equilibrium. In addition, if the action set  is finite for each player, then a perfect equilibrium satisfies the property of admissibility; that is, players do not play weakly dominated strategies. These desirable properties continue to hold in our setting. The following proposition summarizes the results for Bayesian games with finite actions.

\begin{prop}
\begin{enumerate}
\item Any pure strategy profile that is perfect is a Bayesian Nash equilibrium.

\item Any perfect monotone equilibrium is admissible in Bayesian games with finite actions.
\end{enumerate}
\label{prop-admissibility}
\end{prop}


\subsubsection{Multidimensional types and actions}\label{subsec-multi}

We focus on the multidimensional setting with finite actions. As most ingredients have already been specified in Section~\ref{sec-bayesian}, here we only describe the necessary additional structures.

\begin{itemize}
\item For each $i \in I$, player~$i$'s action space $A_i$ is a finite sublattice of $\bR^{s_i}$ with respect to the product order.
\item Player~$i$'s type space is $T_i=[0,1]^{l_i}$ for some $l_i \in \mathbb{Z}_{+}$.\footnote{Here, $[0,1]^{l_i}$ can be replaced by $[\underline{t}_i,\overline{t}_i]^{l_i}$, where $\overline{t}_i>\underline{t}_i$ and $\underline{t}_i,\overline{t}_i\in \bR$.} The joint distribution over types has a density $f$ on $T=\prod_{i\in I}T_i$ with respect to the Lebesgue measure. The type space is endowed with the product order and the usual Euclidean topology.\footnote{In \cite{M2003}, it is assumed that $A_i$ is a finite sublattice of $\bR^{s_i}$ or $[0,1]^{s_i}$, and the joint density over the types is bounded with a positive lower bound. The more general set-up considered in Section~\ref{subsec-Reny} is covered by Reny's framework; see \citet[p.~509]{R2011}.}
\end{itemize}

In the literature, the single crossing condition and quasi-supermodularity are widely adopted to guarantee the existence of a monotone equilibrium. 

\begin{defn}[Single crossing]\label{def-SCC}
Let $(X,\ge,\vee,\wedge)$ be a lattice and $(Y,\ge)$ a partially ordered set. A function $h\colon X\times Y \to \bR$ is said to satisfy the  {single crossing condition} in $(x,y)\in X\times Y$ if for any $x'>x$ and $y'>y$,
\[
h(x',y)-h(x,y)\ge (>)0
\quad \Longrightarrow \quad
h(x',y')-h(x,y')\ge (>)0.
\]
\end{defn}

\begin{defn}[Quasi-supermodularity]\label{def-quasi supermodular}
A function $h\colon X \to \bR$ is said to be {quasi-supermodular} in $x$ if for any $x,x' \in X$,
\[
h(x') \ge (>) h(x\wedge x')
\quad \Longrightarrow \quad
h(x\vee x') \ge (>) h(x).
\]
\end{defn}

\cite{A2001} showed that a monotone equilibrium exists in the one-dimensional setting if each player's interim payoff satisfies the single crossing condition when opponents adopt monotone strategies. The results in the setting with multidimensional and partially ordered type and action spaces are obtained in \cite{M2003} under the condition of quasi-supermodularity in own actions. 
Below, we provide a simple two-player Bayesian game with independent types. In this game, each player's interim payoff satisfies the single crossing condition and is quasi-supermodular in actions. The game admits a monotone equilibrium and a (non-monotone) perfect equilibrium, but no admissible monotone equilibrium. By Proposition~\ref{prop-admissibility}, it does not possess any perfect monotone equilibrium. 
 

\begin{exam}\label{exam-2}
There are two players: $I = \{1,2\}$. Their types are independently and uniformly drawn from the unit interval $[0,1]$. Each player's action space is
\[
A_i=\{0,1\}^2=\{(0,0),(0,1),(1,0),(1,1)\}
\]
endowed with the standard product order, thereby forming a complete lattice. Both players' payoffs do not depend on types, which are given in the payoff tables below. Note that player~2's payoff depends only on her own action and the first component of player~1's action.
\begin{table}[h]
			\centering
			\renewcommand{\arraystretch}{1.5}
			
			\begin{minipage}{0.48\textwidth}
				\centering
				\small
				\begin{tabular}{cc|c|c|c|c|}
					\multicolumn{2}{c}{} & \multicolumn{4}{c}{Player 2} \\
					\multicolumn{2}{c}{} & \multicolumn{1}{c}{(0,0)} & \multicolumn{1}{c}{(0,1)} & \multicolumn{1}{c}{(1,0)} & \multicolumn{1}{c}{(1,1)} \\ \cline{3-6}
					& (0,0) & $1$ & $0$ & $0$ & $0$ \\ \cline{3-6}
					Player 1 & (0,1) & $2$ & $-7$ & $1$ & $0$ \\ \cline{3-6}
					& (1,0) & $0$ & $7$ & $-\frac{1}{2}$ & $0$ \\ \cline{3-6}
					& (1,1) & $1$ & $0$ & $0$ & $0$ \\ \cline{3-6}
				\end{tabular}
				\caption{Player 1's Payoff $u_1(a_1, a_2)$}
			\end{minipage}
			\hfill
			\begin{minipage}{0.48\textwidth}
				\centering
				\small
				\begin{tabular}{cc|c|c|c|c|}
					\multicolumn{2}{c}{} & \multicolumn{4}{c}{Player 2} \\
					\multicolumn{2}{c}{} & \multicolumn{1}{c}{(0,0)} & \multicolumn{1}{c}{(0,1)} & \multicolumn{1}{c}{(1,0)} & \multicolumn{1}{c}{(1,1)} \\ \cline{3-6}
					& (0,$\ast$) & $-1$ & $1$ & $-100$ & $-50$ \\ \cline{3-6}
					Player 1 & (1,$\ast$) & $7$ & $-1$ & $-100$ & $-50$ \\ \cline{3-6}
					\multicolumn{6}{c}{} \\ 
					\multicolumn{6}{c}{} \\
				\end{tabular}
				\caption{Player 2's Payoff $u_2(a_1, a_2)$}
			\end{minipage}
\end{table}

The single crossing condition holds as both players' payoff functions do not depend on types. In Appendix~\ref{proof - no MPE}, we show that each player's interim payoff is quasi-supermodular in own action when the other player adopts a monotone strategy. It is easy to see that the strategy profile $(s_1,s_2)$ is a monotone equilibrium:
\[
s_1(t_1)=
\begin{cases}
(0,0), & t_1\in \left[0,\frac45\right),\\
(1,1), & t_1\in \left[\frac45,1\right],
\end{cases}
\qquad
s_2(t_2)=
\begin{cases}
(0,0), & t_2\in \left[0,\frac78\right),\\
(0,1), & t_2\in \left[\frac78,1\right].
\end{cases}
\]
Given $s_1$, player~2 is indifferent between $(0,0)$ and $(0,1)$. Given $s_2$, all four actions in $A_1$ are player~1's best responses. However, $(s_1,s_2)$ is not an admissible monotone equilibrium, as player~1's strategy involves weakly dominated actions.

For player~1, the only undominated actions are the two incomparable middle points $(0,1)$ and $(1,0)$. The smallest and largest actions, $(0,0)$ and $(1,1)$, are weakly dominated by the mixed action $\frac{1}{2}(0,1)+\frac{1}{2}(1,0)$ at any type. Hence, once we impose admissibility, player~1 can use only the two middle actions $(0,1)$ and $(1,0)$. Because $(0,1)$ and $(1,0)$ are incomparable under the product order, a monotone strategy can use at most one of them, so any admissible monotone strategy of player~1 must be constant. For player~2, the only undominated actions are $(0,0)$ and $(0,1)$, so any admissible monotone strategy is a cutoff rule between these two actions. 

Once we focus on these admissible monotone strategies, no monotone fixed point remains. Indeed, if player~1 chooses $(0,1)$, then player~2 prefers $(0,1)$; if player~2 chooses $(0,1)$, then player~1 prefers $(1,0)$; if player~1 chooses $(1,0)$, then player~2 prefers $(0,0)$; and if player~2 chooses $(0,0)$, then player~1 prefers $(0,1)$. Thus, these  monotone best responses cycle, and there is no admissible monotone equilibrium. Since every perfect monotone equilibrium must be admissible, a perfect monotone equilibrium does not exist.
\end{exam}

The following claim summarizes the findings, and its proof is deferred to the appendix.

\begin{claim}\label{claim-quasi}
\begin{enumerate}
\item For each player~$i$ and type~$t_i$, the interim payoff $V_i(a_i, t_i;\alpha_{-i})$ is quasi-supermodular in $a_i$ whenever the opponent adopts a monotone strategy.
\item The game admits a monotone equilibrium.
\item The game admits a perfect equilibrium.
\item The game admits no admissible monotone equilibrium.
\item The game admits no perfect monotone equilibrium.
\end{enumerate}
\end{claim}

The counterexample above demonstrates that the standard conditions are insufficient to guarantee the existence of a perfect monotone equilibrium. Furthermore, under these conditions, monotonicity itself need not be compatible with admissibility. Therefore, stronger conditions are required to address this issue. We show that the increasing differences condition and supermodularity are well-suited for ensuring the existence of a perfect monotone equilibrium.

\begin{defn}[Increasing differences]\label{def-ID}
Let $(X,\ge,\vee,\wedge)$ be a lattice and $(Y,\ge)$ a partially ordered set. A function $h\colon X\times Y \to \bR$ is said to satisfy the {increasing differences condition} in $(x,y)\in X\times Y$ if for any $x'>x$ and $y'>y$,
\[
h(x',y')-h(x,y') \ge h(x',y)-h(x,y).
\]
\end{defn}

\begin{defn}[Supermodularity]\label{def-supermodular}
Let $(X,\ge,\vee,\wedge)$ be a lattice. A function $h\colon X \to \bR$ is said to be  {supermodular} in $x$ if for any $x,x'\in X$,
\[
h(x\vee x')+h(x\wedge x') \ge h(x)+h(x').
\]
\end{defn}

\begin{assum}
For each $i \in I$ and monotone strategies $g_{-i} \in \cF_{-i}$, player $i$'s interim payoff $V_i(a_i,t_i;g_{-i})$ has increasing differences in $(a_i,t_i)$.
\label{assum-1}
\end{assum}

\begin{assum}
For each $i \in I$, $t_i \in T_i$, and monotone strategies $g_{-i} \in \cF_{-i}$, player $i$'s interim payoff $V_i(a_i,t_i;g_{-i})$ is supermodular in $a_i$.
\label{assum-2}
\end{assum}

Assumption~\ref{assum-1} implies that each player's interim payoff satisfies monotone incremental returns in own types given any increasing strategies by others.\footnote{Given $g_{-i} \in \mathcal{F}_{-i}$, if $V_i(a_i, t_i; g_{-i})$ is supermodular in $(a_i, t_i)$, then it is clear that $V_i$ has increasing differences in $(a_i, t_i)$. However, the converse may not hold when the action and type spaces are not one-dimensional.}  
Assumption~\ref{assum-2} imposes complementarity across the coordinates of a player's own action. In Example~\ref{exam-2}, both players' payoffs do not depend on types. As a result, the increasing differences condition is satisfied. The failure in this example arises from the absence of complementarity across the coordinates of player~1's action. Indeed, player~1's two action coordinates work more like substitutes than complements, and moving both coordinates up to $(1, 1)$ does not make the action more attractive in the way supermodularity requires.  Furthermore, the lattice structure fails when restricted to weakly undominated best-response actions.

\begin{thm}
Under Assumptions~\ref{assum-1} and \ref{assum-2}, a perfect monotone equilibrium exists, which is admissible.
\label{thm-main existence}
\end{thm}

If $A_i$ is one-dimensional, then supermodularity is automatic, implying that Assumption~\ref{assum-2} is trivially satisfied. As the increasing differences condition is satisfied in Example~\ref{exam-2}, it indicates that Theorem~\ref{thm-main existence} fails if Assumption~\ref{assum-2} is weakened to quasi-supermodularity. Below, we present another example to show that Assumption~\ref{assum-1} cannot be weakened to the single crossing condition in the one-dimensional setting, even when the supermodularity condition holds. In this counterexample, there are two players with independent one-dimensional types, and each player has binary actions. We shall verify that each player's interim payoff satisfies the single crossing condition when opponents adopt monotone strategies. This game has a monotone equilibrium and a (non-monotone) perfect equilibrium, but has no perfect monotone equilibrium. 

\begin{exam}\label{exam-SCC}
There are two players $\{1,2\}$. Their types are independently and uniformly drawn from the unit interval $[0,1]$. Let $t_i$ denote player $i$'s type, and let $a_i$ denote player $i$'s action. Both players have binary actions $\{1,2\}$. The payoff tables are given below.
\begin{table}[htbp]
\begin{minipage}[t]{0.5\textwidth}
\centering
\begin{tabular}{cc|c|c|}
& \multicolumn{1}{c}{} & \multicolumn{2}{c}{Player $2$}\\
& \multicolumn{1}{c}{} & \multicolumn{1}{c}{$1$} & \multicolumn{1}{c}{$2$} \\\cline{3-4}
\multirow{2}*{Player $1$}
& $1$ & $\frac{5}{2}-\frac{4}{3}t_1$ & $\frac{1}{2}+\frac{1}{2}t_2$ \\\cline{3-4}
& $2$ & $\frac{55}{24}-\frac{t_2}{6}-\frac{7}{6}t_1$ & $2+t_2-\frac{7}{6}t_1$ \\\cline{3-4}
\end{tabular}
\caption{Player 1's payoff}
\label{tab1}
\end{minipage}
\begin{minipage}[t]{0.5\textwidth}
\centering
\begin{tabular}{cc|c|c|}
& \multicolumn{1}{c}{} & \multicolumn{2}{c}{Player $2$}\\
& \multicolumn{1}{c}{} & \multicolumn{1}{c}{$1$} & \multicolumn{1}{c}{$2$} \\\cline{3-4}
\multirow{2}*{Player $1$}
& $1$ & $-1$ & $1$ \\\cline{3-4}
& $2$ & $7$ & $-1$ \\\cline{3-4}
\end{tabular}
\caption{Player 2's payoff}
\label{tab2}
\end{minipage}
\end{table}

Since the action space is one-dimensional, the obstruction cannot come from the lattice geometry: with only two ordered actions, supermodularity is automatic. The issue is instead that the monotone equilibrium depends on a knife-edge cancellation. To see this, suppose that player~2 adopts an increasing cutoff strategy $s_2$ with cutoff $x_2$, so that player~2 chooses action $1$ on $[0,x_2)$ and action $2$ on $[x_2,1]$. Then player~1's interim payoff difference between actions $2$ and $1$ is
\[
V_1(2,t_1;s_2)-V_1(1,t_1;s_2)
=
\Bigl(\frac{7}{4}-2x_2\Bigr)\Bigl(1+\frac{x_2}{6}-\frac{2}{3}t_1\Bigr).
\]
As the second factor is strictly positive, the sign of this difference is determined by the first factor, namely by player~2's cutoff $x_2$. Thus, if $x_2<\frac78$, all types of player~1 prefer action $2$; if $x_2>\frac78$, all types prefer action $1$; and if $x_2=\frac78$, all types are indifferent. This logic supports the existence of a monotone equilibrium. In this game, there is a unique monotone equilibrium with cutoff pair $(x_1^*,x_2^*)=(\frac45,\frac78)$. However, it also suggests that the equilibrium is fragile: the monotone pattern is sustained by exact indifference at the critical cutoff $x_2=\frac78$.

If player~2 chooses action $1$, then player~1's gain from switching from action $1$ to $2$ is
\[
u_1(2,1,t_1,t_2)-u_1(1,1,t_1,t_2)
=
-\frac{5}{24}-\frac{t_2}{6}+\frac{t_1}{6},
\]
which is increasing in $t_1$. By contrast, if player~2 chooses action $2$, then the same gain is
\[
u_1(2,2,t_1,t_2)-u_1(1,2,t_1,t_2)
=
\frac{3}{2}+\frac{t_2}{2}-\frac{7}{6}t_1,
\]
which is decreasing in $t_1$. That is, these two effects drive player~1's incentives in opposite directions. In the monotone equilibrium, these two effects cancel exactly, and it is why a monotone cutoff equilibrium exists. 

This exact cancellation disappears once player~2's strategy is perturbed to be completely mixed. Below the cutoff $\frac78$, player~2 still places most probability on action $1$, but now assigns a small positive probability to action $2$; above the cutoff, player~2 still chooses action $2$, but trembles a bit to action $1$ with a small probability. These trembles matter because they upset the balance between the two opposing effects described above. 
The perturbed payoff difference between actions $2$ and $1$ is now strictly decreasing difference in $t_1$. Thus, low types strictly prefer action $2$, whereas high types strictly prefer action $1$. A perfect monotone equilibrium fails to exist, because any equilibrium surviving arbitrarily small trembles would have to reverse the monotone cutoff structure.
\end{exam}

The following claim summarizes the results.

\begin{claim}
\begin{enumerate}
\item Each player~$i$'s interim payoff satisfies the single crossing condition in $(a_i,t_i)$ when the opponent adopts a monotone strategy.
\item This game has a monotone equilibrium.
\item This game has a perfect equilibrium.
\item This game does not possess a perfect monotone equilibrium.
\end{enumerate}
\label{claim- SCC but not super}
\end{claim}

\begin{rmk}
The two counterexamples identify distinct sources of failure. In Example~\ref{exam-2}, quasi-supermodularity may leave monotone fixed points supported by dominated actions. Once dominated actions are removed, admissible monotone best responses need not admit a fixed point. In Example~\ref{exam-SCC}, by contrast, supermodularity is automatic. However, the single crossing condition does not ensure that monotone best responses remain monotone under fully supported trembles, as the gain from the higher action need not move monotonically with type. Together, these two examples explain why Assumptions~\ref{assum-1} and \ref{assum-2} are the natural conditions for the existence of perfect monotone equilibria.
\end{rmk}

\begin{rmk}
The notion of perfect monotone equilibrium refines that of monotone equilibrium. To study the existence of monotone equilibria, previous works (\textit{e.g.}, \cite{A2001}, \cite{M2003}, and \cite{R2011}) typically rely on the (weak) single crossing condition and quasi-supermodularity, which are weaker than the increasing differences and supermodularity conditions considered here. \cite{T1978} indicated that the increasing differences condition and supermodularity are easy to work with. \cite{MS1994} provided characterizations for these conditions.

Many previous works seek to either apply the single crossing and quasi-supermodularity conditions to new economic environments, or to weaken them in order to accommodate broader classes of Bayesian games. Examples~\ref{exam-2} and \ref{exam-SCC} demonstrate that strengthening these conditions is necessary to ensure the existence of perfect monotone equilibria. Importantly, both the increasing differences condition and supermodularity are tight. 
\end{rmk}


\subsubsection{Applications}\label{subsec-finite-app}

We now discuss two classes of applications of Theorem~\ref{thm-main existence}. The first concerns supermodular games. The second concerns multi-unit uniform price auctions.

\paragraph{Supermodular games.}
\

Supermodular games have been extensively studied in the literature; see, for example, \cite{T1978}, \cite{MR1990}, \cite{Vives1990}, \cite{MS1994}, and \cite{ZV2007}. The applications include coordination games, public good contribution games, and games with network externalities. \cite{A2001} showed the existence of monotone equilibrium in supermodular games in the one-dimensional setting. We apply Theorem~\ref{thm-main existence} to this class of games and show that a perfect monotone equilibrium exists when players have affiliated types. The result strengthens the standard monotone-equilibrium existence statement by incorporating admissibility and trembling-hand robustness.

\begin{defn}\label{assum-d2}
Types are said to be  {affiliated} if for any $t$ and $t'$,
\[
f(t\vee t')\cdot f(t\wedge t') \ge f(t)\cdot f(t').
\]
\end{defn}

\begin{prop}
Suppose that for each player $i$, the payoff function $u_i(a,t)$ is supermodular in $a$ and in $(a_i,t_j)$ for $1\le j\le n$, and the types are affiliated. Then the game possesses a perfect monotone equilibrium.
\label{prop-supermodular}
\end{prop}

The proposition follows directly from Theorem~\ref{thm-main existence}. Following Facts~(i)--(v) in \citet[p.~871--872]{A2001}, if the players $-i$ adopt monotone strategies $\phi_{-i}$, then the interim payoff $V_i(a_i,t_i;\phi_{-i})$ is supermodular in $(a_i,t_i)$, and hence it has increasing differences in $(a_i,t_i)$. Therefore, Assumptions~\ref{assum-1} and \ref{assum-2} hold, and Theorem~\ref{thm-main existence} yields the conclusion.

Below we present a simple supermodular game with two equilibria. The first is a monotone but imperfect equilibrium, as both players play the maximum action that is weakly dominated. Applying the notion of  perfect monotone equilibrium, one selects an admissible extremal equilibrium in this coordination game, in which both players play the minimal action.

\begin{exam}\label{exam-super}
There are two players $\{1,2\}$. Their types are independently and uniformly drawn from the unit interval $[0,1]$. Both players have binary actions $\{0,1\}$. The payoff table is given below.

\begin{table}[htbp]
\centering
\begin{tabular}{cc|c|c|}
& \multicolumn{1}{c}{} & \multicolumn{2}{c}{Player $2$}\\
& \multicolumn{1}{c}{} & \multicolumn{1}{c}{$0$} & \multicolumn{1}{c}{$1$} \\\cline{3-4}
Player $1$ & $0$ & $1,1$ & $0,0$ \\\cline{3-4}
& $1$ & $0,0$ & $0,0$ \\\cline{3-4}
\end{tabular}
\end{table}
\end{exam}

This is a symmetric supermodular game in which payoffs do not depend on types. By Proposition~\ref{prop-supermodular}, perfect monotone equilibria exist. There are indeed two equilibria: $(g_1,g_2)$ and $(f_1,f_2)$.
\begin{itemize}
\item In the first equilibrium, $g_1=g_2\equiv 1$. Both players always choose the maximal action. This equilibrium is monotone, but action $1$ is weakly dominated by action $0$ at every type for each player. Hence $(g_1,g_2)$ is not admissible.
\item In the second equilibrium, $f_1=f_2\equiv 0$. Both players always choose the minimal action. This is a perfect monotone equilibrium.
\end{itemize}

\begin{Fact}
Example~\ref{exam-super} is a supermodular game. In addition, $(g_1,g_2)$ is not a perfect monotone equilibrium, while $(f_1,f_2)$ is a perfect monotone equilibrium.
\label{claim-super}
\end{Fact}

\paragraph{Multi-unit uniform price auctions.}
\

We further illustrate the application by revisiting the multi-unit uniform price auction in \cite{M2003}.\footnote{ Here we allow bidders to have heterogeneous demands and type spaces.}  He proved that a monotone equilibrium exists in this auction setting.
 We shall apply Theorem~\ref{thm-main existence} and show that a perfect monotone equilibrium exists in McAdams's auction model, thereby strengthening the original existence result.

\begin{exam}[Multi-unit uniform price auction]
Consider a setting with $n$ bidders competing for $S$ identical units of a good. Each bidder $i$ receives value $V_i(q, t)$ from an allocation $q = (q_1, \ldots, q_n)$, where $q_i$ is the quantity allocated to bidder~$i$ and $t = (t_1, \ldots, t_n)$ denotes the profile of private types. Each $t_i$ is drawn independently from a uniform distribution with support $[\underline{t}_i,\overline{t}_i]$, where $\overline{t}_i > \underline{t}_i$ and  $\underline{t}_i, \overline{t}_i \in \mathbb{R}$.\footnote{The distribution can be replaced by any atomless distribution over $[\underline{t}_i,\overline{t}_i]$ with $\overline{t}_i  >  \underline{t}_i$ and  $\underline{t}_i, \overline{t}_i \in \mathbb{R}$.} The value function $V_i$ is piecewise continuous in the types and satisfies the following monotonicity condition: for any two allocations $q$ and $q'$, the difference $V_i(q', t) - V_i(q, t)$ is increasing in $t_i$ whenever $q_i' > q_i$ and $q_j' \leq q_j$ for all $j \neq i$.

Bidders aim to maximize their expected surplus, defined as the difference between value and payment.  Each bidder $i$ has a demand $Q_i \leq S$, where the aggregate demand satisfies $\sum_{i=1}^n Q_i > S$.
 A bid from bidder $i$ consists of a non-increasing vector $b_i = (b_i(1), \ldots, b_i(Q_i))$, where each component $b_i(q)$ lies in a discrete price grid $\{p^{\min}, p^{\min} + 1, \ldots, p^{\max}\}$.
 Let $b^k$ denote the $k$th-highest unit bid across all submitted bids. For each bidder, define $\underline{q}_i = \max\{q : b_i(q) > b^S\}$ and $\bar{q}_i = \max\{q : b_i(q) \geq b^S\}$, which represent the minimum and maximum quantity that bidder~$i$ can receive in any market-clearing allocation. Provided that total supply satisfies $\sum_i \underline{q}_i \leq S \leq \sum_i \bar{q}_i$, bidder~$i$ is first assigned $\underline{q}_i$ units. The remaining quantity $r = S - \sum_i \underline{q}_i$ is allocated by randomly ordering bidders and sequentially assigning up to $\bar{q}_i - \underline{q}_i$ additional units, until the supply is exhausted. A bidder's total allocation is thus $q_i^* = \underline{q}_i + \min\{\bar{q}_i - \underline{q}_i, r\}$, depending on her position in the random order. For payments, consider two commonly used uniform pricing rules: under the $S$th-price rule, all winning bidders pay the lowest winning bid $b^S$ per unit; under the $(S+1)$st-price rule, they pay the highest losing bid $b^{S+1}$. In either case, a bidder's total payment equals her quantity times the applicable price, denoted by $Z_i = q_i^* b^S$ (or $Z_i = q_i^* b^{S+1}$ ).
 \label{exam-mcadams}
\end{exam}

\cite{M2003} showed that given the actions of other bidders $b_{-i}$ and a random ordering $r$ of bidders,  the following holds for any bidder $i$. 
\begin{itemize}
	\item[(1)] The allocation vector $q(b_i,b_{-i};r)$ satisfies that $q_i$ is increasing in $b_i$, while $q_j$ is decreasing in $b_i$ for every $j \ne i$. It implies that $V_i$ has increasing differences in $(b_i,t_i)$,
	and thus the interim payoff function of bidder $i$ also satisfies increasing differences conditions in $(b_i,t_i)$ regardless of the strategies of the other bidders.  
	\item[(2)] For bidder~$i$'s bid $b_i^1$, $b_i^2$, $b_i^{1 \wedge 2}$ or $b_i^{1 \vee 2}$,\footnote{Here, $b_i^{1 \wedge 2}$ and $b_i^{1 \vee 2}$ denote the coordinate-wise minimum and maximum of $b_i^1$ and $b_i^2$.} let $p^1$, $p^2$, $p^{1 \wedge 2}$, and $p^{1 \vee 2}$ denote the $S$-th (or $(S+1)$-th) price, and $q^1$, $q^2$, $q^{1 \wedge 2}$, and $q^{1 \vee 2}$ be the corresponding allocation vectors. Then 
	\[
	\{ (q^1,p^1), (q^2,p^2) \}
	=
	\{ (q^{1\wedge 2},p^{1 \wedge 2}), (q^{1 \vee 2}, p^{1 \vee 2}) \},
	\]
	and hence $V_i - Z_i$ is modular in $b_i$.\footnote{\cite{M2003} showed that modularity holds for any function $u_i(V_i,Z_i)$.} It follows that the interim payoff function of bidder $i$ is modular in $b_i$, regardless of the strategies of the other bidders.
\end{itemize}
The argument above also applies when bidders have heterogeneous demands and  type spaces.  As a consequence of our Theorem~\ref{thm-main existence}, we obtain the following corollary. 

\begin{coro} 
A perfect monotone equilibrium exists in the multi-unit uniform price auction.
\label{cor-PME1}
\end{coro}

In the model above, bidder $i$'s payoff is given by the ex post surplus $V_i - Z_i$, implying that the bidder is risk neutral over the ex post surplus.
Consider the setting where bidder~$i$'s payoff is allowed to be $u_i(V_i - Z_i)$ with $u_i$ being concave (\textit{i.e.}, bidder $i$ is risk-averse), Corollary \ref{cor-PME1} may fail to hold. Below, we modify Example 2 in \cite{M2007} to illustrate the issue.

\begin{exam}
The following is an example of a multi-unit uniform price auction with both risk-neutral and risk-averse bidders. 

Two bidders compete for two identical units of a good. Bidder~1 wants both units, while bidder~2 demands only one unit. Bidder $i$ receives value $V_i(q, t) = q_i \, t_i$ from an allocation $q = (q_1, q_2)$, where $q_i$ is the quantity allocated to bidder~$i$ and $t = (t_1, t_2)$ denotes the profile of private types. Types $t_1$ and $t_2$ are drawn independently from uniform distributions with supports $[3,5]$ and $[0,2]$, respectively.

Bidders $1$ and $2$ submit the bids $b_1 = (b_{11},b_{12})$ and $b_2$ simultaneously. Each of $b_{11}$, $b_{12}$, and $b_2$ is drawn from a finite set $B = \{0, 1, 2, 3, 4, 5\}$. In addition, the bids of bidder $1$ need to be ordered as $b_{11} \geq b_{12}$.

Under the uniform pricing rule, the price $p$ is set equal to the third-highest bid among all three submitted unit bids. Every unit bid strictly exceeding $p$ wins a unit at price $p$. Any remaining units are allocated among the bids exactly equal to $p$ via a random tie-breaking rule, which selects winners in a uniformly random bidder order.  

Given a profile of bids $b = (b_1, b_2)$, let $q_i(b)$ denote the quantity that bidder $i$ wins, and $Z_i = q_i \, p$ be his payment. Bidder $i$'s ex post surplus is $V_i - Z_i$, and the ex post utility is $u_i(V_i - Z_i)$. Bidder $2$ is risk-neutral, while bidder $1$ is satiated at $4.5$ units of the surplus. In particular, $u_1(x) = \min \{ 4.5, x \}$, where $x$ is bidder $1$'s ex post surplus. Thus, bidder $1$ exhibits risk aversion. 
\label{exam-multi-unit fail PME}
\end{exam}

\begin{Fact}
	\phantomsection
	\begin{itemize}
		\item[1.] The strategies 
		\[
		f_1(t_1) = 
		\begin{cases}
			(1,0) & \text{if } t_1 \in [3, \tfrac{7}{2}), \\ 
			(3,3) & \text{if } t_1 \in [\tfrac{7}{2}, 5],
		\end{cases}
		\quad\text{and}\quad
		f_2(t_2) = 
		\begin{cases}
			1 & \text{if } t_2 \in [0, \tfrac{2}{3}), \\ 
			2 & \text{if } t_2 \in [\tfrac{2}{3}, 2],
		\end{cases}
		\]
		form a monotone equilibrium. 
		\item[2.] 	This game admits no perfect monotone equilibrium.
	\end{itemize}\label{fact:non constant ME}
\end{Fact}
Part~(1) above shows that a monotone equilibrium exists, in which both bidders adopt non-constant monotone strategies and win with non-trivial probabilities. 

It is clear that $V_i$ is continuous in types and satisfies the monotonicity condition as in Example~\ref{exam-mcadams}. However, since bidder~1 is not risk-neutral, $u_1(V_1 - Z_1)$ fails to satisfy the increasing differences condition in $(b_1, t_1)$. This leads to the non-existence of a perfect monotone equilibrium in (2). Moreover, even when bidders are risk-neutral, as long as their types are not independent, Corollary \ref{cor-PME1} may fail to hold. The following example demonstrates this point by modifying \citet[Example~1]{M2007} and Example~\ref{exam-multi-unit fail PME} above. In particular, the increasing differences condition fails in both Example~\ref{exam-multi-unit fail PME} and the following modified example.

\begin{Fact}
Suppose that bidders are risk-neutral. The types $t_1$ and $t_2$ are affiliated. With probability $50\%$, $t_1$ and $t_2$ are independent with $t_1 \sim U[2,3]$ and $t_2 \sim U[0,1]$. With probability $50\%$, $t_1$ and $t_2$ are independent with $t_1 \sim U[3,4]$ and $t_2 \sim U[4,5]$. Under this modification in Example~\ref{exam-multi-unit fail PME}, there does not exist a perfect monotone equilibrium.
\label{fact-no PME type}
\end{Fact}


\subsection{Bayesian games with general type and action spaces}\label{subsec-Reny}

We now extend the analysis to Bayesian games with general type and action spaces, following the general framework in \cite{R2011}.


\subsubsection{Perfect monotone equilibrium in the general setting}\label{subsec-general-perfect}

We first extend the notion of perfect monotone equilibrium to the general setting. A mixed strategy profile $g=(g_1,\ldots,g_n)$ is called a completely mixed strategy profile if for each player $i$, each type $t_i \in T_i$, and any nonempty open subset $O_i$ of $A_i$, $g_i(t_i,O_i)>0$.\footnote{The definition of a completely mixed strategy has been adopted in previous works; see, for example, \cite{SS1995}, \cite{Bajoori2016}, and \cite{MR2020}.}

\begin{defn}[Monotone perfection in general spaces]
\

\begin{enumerate}[a.]
\item A pure strategy profile $g=(g_1,\ldots,g_n)$ is said to be  {perfect} if there exists a sequence of completely mixed strategy profiles $\{g^k=(g_1^k,\ldots,g_n^k)\}_{k=1}^{\infty}$ such that for every player $i$ and $\lambda_i$-almost all $t_i \in T_i$, the following properties hold:
\begin{itemize}
\item[(1)] $\lim\limits_{k \to \infty} \rho\bigl(g_i^k(t_i),\delta_{g_i(t_i)}\bigr)=0$;\footnote{Let $(X,d)$ be a compact metric space, where $d$ denotes the metric on $X$. 
 The space $\mathcal{M}(X)$ of Borel probability measures on $X$ is endowed with the topology of weak convergence. Such a topology can be induced by the Prohorov metric $\rho$ on $\mathcal{M}(X)$, which is defined by
\[
\rho(\nu,\mu)=\inf\Bigl\{\epsilon:\nu(B)\le \mu(B^\epsilon)+\epsilon \text{ and } \mu(B)\le \nu(B^\epsilon)+\epsilon \text{ for any Borel set } B\subseteq X\Bigr\},
\]
where $B^\epsilon=\{b\in X:d(b,B)<\epsilon\}$ and $d(b,B)=\inf_{b'\in B}d(b,b')$. Throughout the paper, $d$ denotes the metric on any compact metric space and $\rho$ denotes the corresponding Prohorov metric on the space of Borel probability measures.}
\item[(2)] $\lim\limits_{k \to \infty} d\bigl(g_i(t_i),\mathrm{BR}_i(t_i,g_{-i}^k)\bigr)=0$.
\end{itemize}
A Bayesian Nash equilibrium $g$ is said to be a  {perfect equilibrium} if it is perfect.

\item A pure strategy profile $g$ is said to be a  {perfect monotone equilibrium} if it is both a perfect equilibrium and a monotone equilibrium.
\end{enumerate}
\label{def-perfect-general}
\end{defn}

In general action spaces, the finite-action notion of perfection is no longer the appropriate benchmark.  The definition above extends Definition~\ref{def-perfect} to the general setting. Intuitively, exact best-response conditions under perturbations are too strong to serve as a robust refinement. The natural requirement is instead that the limiting pure action be arbitrarily close to the perturbed best-response correspondence. \citet{SS1995} presented strong and weak notions of perfection in infinite normal-form games. The strong notion requires perturbed play to put high probability on the set of pure best responses, whereas the weak notion only requires concentration near that set.\footnote{Technically, the difference is that the strong perfect equilibrium uses the strong topology (the one induced by the norm) in place of the weak topology (induced by Prohorov metric).} These two notions coincide in finite–action games.
Our definition corresponds to the weak perfect equilibrium in \citet{SS1995}. It requires vanishing fully supported trembles and asymptotic proximity of the limiting pure action to the perturbed best-response correspondence, rather than concentration of the perturbing mixed strategy itself on exact best responses. 

In the setting with infinite actions, an admissible perfect equilibrium may not exist. This issue has been demonstrated in \citet[Example~2.1]{SS1995}, and still appears even if one considers monotone equilibria. 
To address this issue, we adopt the notion of limit admissibility, which has been studied by \citet{SS1995} in infinite normal-form games.\footnote{A closely related notion called limit undominatedness has been studied in \cite{Bajoori2016} for Bayesian games.}

\begin{defn}
An action $a_i \in A_i$ is said to be  {limit undominated} at type $t_i$ if there exists a sequence of undominated actions $\{a_i^k\}_{k=1}^{\infty}$ such that\footnote{The relevant  dominance notions in the infinite-action setting can be defined analogously as those in Definition~\ref{defn-dominance}. We avoid the repetition for simplicity.}
\[
\lim\limits_{k\to\infty} d(a_i^k,a_i)=0.
\]
A pure strategy profile $g$ is said to be  {limit admissible} if for each player $i$, the action $g_i(t_i)$ is limit undominated for $\lambda_i$-almost all $t_i \in T_i$.
\end{defn}

The next proposition is the infinite-action analogue of Proposition~\ref{prop-admissibility}.\footnote{The proof is lengthy but rather standard; see Appendix C in the arXiv version of \cite{HSX2025}.} 

\begin{prop}
\begin{enumerate}
\item Any pure strategy profile that is perfect in the sense of Definition~\ref{def-perfect-general} must be a Bayesian Nash equilibrium.

\item If the action spaces are general compact spaces, then any perfect monotone equilibrium is limit admissible.
\end{enumerate}
\label{prop-limit-admissibility}
\end{prop}


\subsubsection{General existence theorem and applications}\label{subsec-general-model}

In this subsection, we extend our results to settings with general action and type spaces. The following framework is from \cite{R2011} and is imposed here.

Let $(L,\ge)$ be a partially ordered set endowed with a $\sigma$-algebra $\cL$. The partial order $\ge$ on $L$ is said to be  {measurable} if
\[
\{(a,b)\in L\times L:a\ge b\}\in \cL \otimes \cL.
\]
If $L$ is endowed with a topology, the partial order is said to be  {closed} if
\[
\{(a,b)\in L\times L:a\ge b\}
\]
is closed in the product topology on $L\times L$. If $L$ is a subset of a real vector space, the partial order is said to be  {convex} if the set above is convex. The additional assumptions below are needed on Bayesian games as presented in Section~\ref{sec-bayesian}.\footnote{\cite{R2011} provided a detailed discussion of these assumptions, and Proposition~3.1 therein presents several sufficient conditions that are satisfied in common settings.}
\begin{itemize}
\item The partial order on each action space $A_i$ is closed.
\item The partial order on each type space $T_i$ is measurable.
\item Either (i) $A_i$ is a convex subset of a locally convex topological vector space and the partial order on $A_i$ is convex, or (ii) $A_i$ is a complete metric lattice. It is possible for (i) to hold for some players and (ii) to hold for others.
\item There exists a countable subset $T_i^0$ of $T_i$ such that every set in $\cT_i$ with positive probability under $\lambda_i$ contains two points between which lies a point in $T_i^0$.\footnote{For $a,b,c\in L$, we say that $b$ lies between $a$ and $c$ if $a\ge b\ge c$.}
\end{itemize}

The substantive economic assumptions remain Assumptions~\ref{assum-1} and \ref{assum-2}. Under those conditions, the finite-action result extends to the general setting.

\begin{thm}
Under Assumptions~\ref{assum-1} and \ref{assum-2}, there exists a perfect monotone equilibrium.\footnote{As Theorem~\ref{thm-Reny} covers Theorem~\ref{thm-main existence}, we only prove Theorem~\ref{thm-Reny} in Appendix~A for simplicity.}
\label{thm-Reny}
\end{thm}


\begin{rmk}
\begin{enumerate}
\item While we explicitly work with the weak perfect monotone equilibrium, the proof of Theorem~\ref{thm-Reny} in fact establishes the existence of a strong perfect monotone equilibrium as well. 

\item A Bayesian game is symmetric if (1) the action and type spaces and the associated partial orders are the same among all the players, and (2) for any permutation $\pi$ of $\{1, 2, \ldots, n\}$, $\lambda(D) = \lambda(\{t_{\pi} \colon t \in D\} )$ for any $D \in \cT$ and $u(a_{\pi}, t_{\pi}) = u_{\pi}(a, t)$ for any $(a, t) \in A \times T$. It is pointed out in \citet[Theorem~4.5]{R2011} that a symmetric monotone equilibrium exists in a symmetric Bayesian game therein. Following the same argument in the proof of Theorem~\ref{thm-Reny}, a symmetric perfect monotone equilibrium exists in a symmetric Bayesian game under Assumptions~\ref{assum-1} and \ref{assum-2}.\footnote{To prove this claim, one can focus on symmetric monotone equilibrium in the first step of Appendix~\ref{proof -thm finite existence multidimension}. }

\item One may impose additional structure on the perturbations; for example, restricting them to be mixtures over monotone strategies. Under the standard single crossing and quasi‑supermodularity conditions, a perfect monotone equilibrium may fail to exist even with arbitrary perturbations. As a result, it cannot exist under restricted perturbations either. In contrast, under increasing differences and supermodularity, the proof of Theorem~\ref{thm-Reny} directly implies that a perfect monotone equilibrium exists, even when the perturbations must satisfy such a monotonicity condition.

\item \citet[Theorem 4.3]{R2011} showed that in certain auctions one can restrict attention to strategies that never bid above one's value. This is a very useful observation in those  settings. Our Example~\ref{exam-2} shows that this result does not extend to the general setting. Even if one only cares about eliminating weakly dominated strategies, the standard single crossing and quasi‑supermodularity conditions are still insufficient to guarantee the existence of an admissible monotone equilibrium.
\end{enumerate}
\end{rmk}

In the following, we apply Theorem~\ref{thm-Reny}  to a multi-unit auction with risk-averse bidders and uniform pricing, which has been studied in \cite{M2007} and \cite{R2011}.

\begin{exam}[Multi-unit auction with risk-averse bidders and uniform pricing]
Consider a uniform-price auction with $n$ bidders and $m$ identical units of a single good for sale. Each bidder $i$ simultaneously submits a bid vector $b_i = (b_{i1}, \ldots, b_{im})$, where bids are ordered such that $b_{i1} \geq \cdots \geq b_{im}$, and each $b_{ik}$ is drawn from a finite set $B \subseteq [0,1]$ containing both 0 and 1. Let $b_{ik}$ denote bidder $i$'s $k$th unit bid. Under the uniform pricing rule, the price $p$ is set equal to the $(m+1)$st highest bid among all $nm$ submitted unit bids. Every unit bid strictly exceeding $p$ wins a unit at price $p$. Any remaining units are allocated among the bids exactly equal to $p$ via a random tie-breaking rule that selects winners in a uniformly random bidder order.

Bidder $i$'s private type is a nonincreasing vector $t_i = (t_{i1}, \ldots, t_{im}) \in [0,1]^m$, and the type space is $T_i = \{ t_i \in [0,1]^m : t_{i1} \geq \cdots \geq t_{im} \}$. Bidder $i$ is risk averse, with a utility function $u_i : [-m, m] \rightarrow \mathbb{R}$ that is increasing and concave; i.e., $u_i' > 0$ and $u_i'' \leq 0$. Let $t = (t_1, \ldots, t_n)$ denote the full type profile, and $t_{-i}$ the types of all bidders other than $i$. Following \cite{R2011}, we focus on the special case with private values. When bidder $i$ is allocated $k$ units, her utility depends on her marginal values $t_{ik}$.  Therefore, bidder $i$'s expected utility when winning $k$ units at price $p$ is given by 
\[ u_i\Bigl( \sum_{j=1}^k t_{ij} - kp \Bigr). \] 
Types are drawn independently across bidders, and bidder $i$'s type vector is distributed according to a density $f_i$, which is positive on all of $T_i$.
\end{exam}

Both \cite{M2007} and \cite{R2011} pointed out the issue in establishing monotone equilibrium existence under the standard partial order on Euclidean spaces. In particular, the model fails to satisfy the single crossing condition under this conventional order, implying that the existence result on monotone equilibrium does not directly apply. To address this issue, Reny introduced the following novel endogenous partial order tailored to the economic environment.

For each bidder $i$, let 
\[
\alpha_i = \frac{u_i'(-m)}{u_i'(m)} - 1 \geq 0
\]
and consider the partial order $\succeq_i$ on $T_i$ defined as follows: $t_i' \succeq_i t_i$ if
\begin{equation}
\label{eq:partialorder}
\begin{aligned}
t'_{i1} &\geq t_{i1} \quad \text{and} \\
t'_{ik} - \alpha_i(t'_{i1} + \cdots + t'_{i,k-1}) &\geq t_{ik} - \alpha_i(t_{i1} + \cdots + t_{i,k-1}) \quad \text{for all } k \in \{2, \ldots, m\}. \nonumber
\end{aligned}
\end{equation}

\cite{R2011} verified that the increasing differences condition and supermodularity are satisfied under this partial order, and his general existence theorem then guarantees the existence of a monotone equilibrium. As pointed out in \cite{R2011}, a trivial equilibrium always exists, where bidders may employ weakly dominated actions. His focus is on non-trivial monotone equilibrium. A perfect monotone equilibrium inherently eliminates such trivial, weakly dominated actions. Our Theorem~\ref{thm-Reny} is applicable in this setting, and implies the existence of a perfect monotone equilibrium.

\begin{coro}
A perfect monotone equilibrium exists under the type-space partial order $\succeq_i$ and the usual coordinatewise partial order on bids.
\end{coro}

\vspace{-0.5cm}

\section{Applications to Auctions and Bertrand Competitions}\label{sec-application}

In many applications of Bayesian games, such as auctions and price competitions, it is natural to focus on monotone equilibria, since players with higher valuations or costs tend to submit higher bids or charge higher prices. However, the equilibrium existence results in Section~\ref{sec-results} are not directly applicable to these settings, as payoffs are typically discontinuous due to ties. In this section, we introduce a model of ``generalized auctions'' that encompasses first-price auctions, all-pay auctions, and Bertrand competitions. The existence of perfect monotone equilibria in generalized auctions is established in Theorem~\ref{thm-discontinuous}, which in turn implies the corresponding results for first-price auctions, all-pay auctions, and Bertrand competitions.


\subsection{A class of generalized auctions}\label{sec-discontinuous model}

There are $n$ players. Each player~$i$ receives a private type $v_i \in [0,1]$. The joint density is $f:[0,1]^n \to \bR_{+}$. After knowing the type, player~$i$ takes an action from $A_i = \{Q\} \cup [0 ,\overline{b}_i]$, where $\overline{b}_i > 0$, $Q < 0$ and taking the action $Q$ means that the player chooses to quit the game.

If player~$i$ chooses quitting the game, then his payoff is $0$. Suppose that player~$i$ chooses to stay and takes an action $b_i$. He needs to buy a ``ticket'' for staying, denoted by $\Phi_i(b_i)$. That is, $\Phi_i(b_i)$ is the induced entry fee that player~$i$ needs to pay. Besides, if he gets an object, then player~$i$ receives an interdependent payoff $W_i(b_i, v)$ and pays the per-unit cost $C_i(b_i)$. Let $D(b)$ denote the total quantity/demand of the goods, and $P_i(b)$ the winning probability of player~$i$. The player who plays the highest action above \( Q \) wins the game, with tie-breaking determined randomly and uniformly.
Player~$i$'s payoff is
$$u_i(b,v) = 
\begin{cases}
	0,  &    \text{ if } b_i = Q; \\
	\left( W_i(b_i,v) - C_i(b_i) \right) D(b)P_i(b) - \Phi_i(b_i), &  \text{ if }    b_i = \max_{j \in I} b_j > Q; \\
	-    \Phi_i(b_i),  & \text{ otherwise}.
\end{cases}
$$ 
Let $P_i(Q, b_{-i}) = 0$ for any $b_{-i} \in B_{-i}$ and $W_i(Q, v) = 0$ for any $v \in [0,1]^n$, with $\Phi_i(Q) = 0$ and $C_i(Q) = 0$.
 We can simplify the payoff as
$$u_i(b,v) = ( W_i(b_i,v) - C_i(b_i) )D(b)P_i(b)  -    \Phi_i(b_i).
$$

\begin{rmk}\label{rmk-discontinuous}
The model above covers various economic applications. For example, when $D(b) \equiv 1, C_i(b) \equiv 0$, $P_i$ is the winning probability and $\Phi_i$ captures the forfeited cost,  it includes the all-pay auction with interdependent values as an example. When $\Phi_i \equiv 0$ and $D(b) \equiv 1$, it is the standard first-price auction. In general, it can be viewed as an auction with entry fees. In addition, one can interpret \( D(b) \) as the total demand, which is not necessarily a constant function.\footnote{We can further extend the model to allow each player to have individual quantity/demand $D_i$. All the results continue to hold.} Then it covers the model of price competitions as well. We shall explain these examples in more details in the next subsection. 
\end{rmk}

We make the following assumptions. 
\begin{assum}
\begin{enumerate}[(1)]
\item The interdependent payoff $W_i(b_i,v)$ is  continuous  in $ [0, \overline{b}_i] \times [0,1]^n$, strictly increasing in $v_i$ and increasing in $v_{-i}$ for each $b_i \in   [0, \overline{b}_i]$. Additionally, for any $v_i > 0$, we have $W_i(\cdot, v_i, \cdot) > 0$ for all $(b_i, v_{-i}) \in [0, \overline{b}_i] \times [0, 1]^n$.

\item The cost functions $C_i$ and $\Phi_i$ are increasing and continuous in $b_i$ with $C_i(0) = \Phi_i(0) = 0$.

\item The function $D$ is  positive, continuous and increasing in $b$.

\item The difference $W_i(b_i^H, v) - W_i(b_i^L, v)$ is increasing in $v$ for $b_i^L < b_i^H$, where $b_i^L, b_i^H \in \{Q\} \cup [0, \overline{b}_i]$.

\item  The sign of $W_i(b_i,v) - C_i(b_i)$ is independent of $v_{-i}$. That is, for $b_i \in [0, \overline{b}_i]$ and $v_i \in [0,1]$, it falls into one of the following three cases:
		(i) $W_i(b_i,v_i,v_{-i}) - C_i(b_i)  >  0$ for all $v_{-i} \in [0,1]^{n-1}$,  (ii) $W_i(b_i,v_i, v_{-i}) - C_i(b_i)  <  0$ for all $v_{-i} \in [0,1]^{n-1}$, or (iii) $W_i(b_i,v_i,v_{-i}) - C_i(b_i) \equiv 0$.

\item The winning payoff that player $i$ receives is strictly decreasing in her action. That is, for $v \in [0,1]^n$, and $b_j \in \{Q\} \cup [0 , \overline{b}_j]$ for $j \ne i$, $(W_i(b_i,v) - C_i(b_i))D(b_i,b_{-i}) - \Phi_i(b_i)$ is strictly decreasing in $b_i$. 
\end{enumerate}
\label{assum-d1}
\end{assum}

\begin{rmk}
In the assumption above, (1)--(3) are the common continuity and monotonicity conditions. Condition~(4) is the increasing differences condition in $(b_i, v)$. In the private-value setting, 
 Condition~(5) is trivially satisfied. In general, $W_i(b_i,v) - C_i(b_i)$ represents player~$i$'s variable payoff, which depends on whether she wins. This condition captures how interdependence in values affects the payoff structure. Condition~(6) implies that playing a higher action is costly when the player wins.
 \label{rmk: assum-d1}
\end{rmk}

\begin{thm}
Under Assumption~\ref{assum-d1}, perfect monotone equilibria exist in the class of  generalized auctions with affiliated types, which are limit admissible.
\label{thm-discontinuous}
\end{thm}

\begin{rmk}
To prove Theorem~\ref{thm-discontinuous}, given a Bayesian game $G$ with discontinuous payoffs, we first construct a sequence of games $\{ G^m\}_{m = 2}^{\infty}$ converging to $G$. Each $G^m$ is a slight perturbation that differs from the limit game $G$ only in its payoff functions: when players play the action profile~$b$ at type profile $v$, player~$i$'s payoff $u_i^m(b, v)$ in $G^m$ equals $u_i(\tilde b^m, v)$ -- player~$i$'s payoff based on the mixed action profile $\tilde b^m$ in the original game. Here $\tilde b_j^m$ is a completely mixed probability measure on $A_j$ putting probability $1 - \frac{1}{m}$ at the action $b_j$ for each $j$. Next, we further discretize $G^m$ to $\{G^{mk}\}_{k \ge 1}$ for each $m \ge 2$. 	
\begin{enumerate}
\item In each $G^{mk}$, a monotone equilibrium $g^{mk}$ exists.

\item Taking \( k \to \infty \), we obtain a monotone strategy \( g^m \) in \( G^m \), which can be interpreted as a completely mixed strategy in \( G \).

\item Taking $m \to \infty$, we obtain a monotone strategy $g$ in $G$. The aim is to show that $g$ is a monotone equilibrium in the original game $G$.
\end{enumerate}

One difficulty in this double-limit argument is to show that $g^m$ is a monotone equilibrium in $G^m$. To understand the difficulty, one can regard $G^m$ as an auction game. The standard asymptotic approach in the literature is to take steps~(1) and (2) described above, and show that $g^m$ is indeed a monotone equilibrium in $G^m$. When invoking such a limit argument,  the common technique is to show that the probability under ${g}^m$, that two or more players simultaneously play the same highest action above $Q$, is zero.\footnote{See, for example, \cite{RZ2004} for a careful treatment of the first-price auctions. } Towards this aim, one needs to approximate player~$i$'s interim payoff at type $v_i$ with action $b_i$ by its right limit whenever this interim payoff is nonnegative. In our game $G^m$, as $g^m$ is viewed as a completely mixed strategy in $G$, player~$i$'s payoff induced by $g^m$ is written as the summation of finite many payoff terms. The summation itself can be still shown to be nonnegative. However, to adopt the asymptotic argument, we need every term in this summation to be nonnegative, which is unclear. In the proof, we show that this issue can be addressed under Assumption~{\ref{assum-d1}}.
\label{rmk-dis}
\end{rmk}


\vspace{-0.5cm}

\subsection{Applications}\label{sec-discontinous applications}

In this subsection, we demonstrate the usefulness of Theorem~\ref{thm-discontinuous} through three economic examples: first-price auctions, all-pay auctions, and Bertrand competitions.


\paragraph{First-price auctions}

\

Consider the following first-price auction with affiliated types. There are $n \ge 2$ bidders. The value of bidder $i$ is $v_i \in [0, 1]$.  The joint density for bidders' values is $f \colon [0,1]^{n} \to \bR_+$.  Players have affiliated types.

After receiving their values, each bidder $i$ submits a bid $b_i$ from  $\{Q\} \cup [0, \overline{b}_i]$, where  $Q <  0$ and $Q$ corresponds to not participating in the auction. The bidder  submitting the highest bid above $Q$ wins the object,  with tie-breaking done randomly and uniformly. If bidder $i$ wins the object, then her payoff is $v_i - b_i$. Otherwise, bidder $i$ receives payoff $0$. To be precise, the payoff of bidder $i$ is
$$u_i(b,v) = 
\begin{cases}
\frac{1}{ |I^w|}   (v_i - b_i), &  \text{ if }    b_i = \max_{j \in I} b_j > Q,  \\
0,  &  \text{ otherwise}, 
\end{cases}
$$ 
where \( I^w = \{ i \in I \colon b_i = \mathop{\max}\limits_{j \in I} b_j > Q \} \) is the set of winning bidders, and \( |I^w| \) denotes the number of elements in the set \( I^w \).

This example is a special case of the model presented in Section~\ref{sec-discontinuous model}. For each bidder~$i$, let $W_i(b_i,v) = v_i$, $C_i(b_i) = b_i$, $D(b) \equiv 1$ and $\Phi_i(b_i) \equiv 0$. Then bidder $i$'s payoff $u_i(b,v) = (v_i - b_i)P_i(b)$.

Note that $W_i(b_i^H, v) - W_i(b_i^L,v) =  0$ for any $b_i^H  > b_i^L$, which is trivially monotone in $v$.  Besides, $(W_i(b_i,v) - C_i(b_i))D(b) = v_i - b_i$, which is strictly decreasing in $b_i$ and does not depend on $v_{-i}$. Thus, Assumption~\ref{assum-d1} holds in this example.

\begin{coro}
Perfect monotone equilibria exist in first-price auctions with affiliated types.\label{coro-first}
\end{coro}

In first-price auctions with symmetric bidders, affiliated types and interdependent values, \cite{M2007a} showed that there is a unique monotone equilibrium in symmetric strategies. In the example above, if $f$ is symmetric and continuously differentiable, then by \citet[Theorem~1]{M2007a} and our Corollary~\ref{coro-first}, there is a unique monotone equilibrium that is perfect and in symmetric strategies.


\paragraph{All-pay auctions}

\

Next, we consider an all-pay auction game with interdependent payoffs and affiliated types. Each bidder~$i$ receives  a private signal $v_i \in [0, 1]$. The joint density is $f \colon [0,1]^n  \to \bR_+$. After receiving their signals, each bidder~$i$ submits a bid from  $B_i = \{Q\} \cup [0, \overline{b}_i]$, where the quit option $Q < 0$.

The bidder submitting the highest bid above $Q$ wins the object,  with tie-breaking done randomly and uniformly. All the bidders who bid above $Q$ need to pay their bids. If bidder~$i$ wins the object at bid $b_i$, then her payoff is $w_i(b_i,v) - b_i$; otherwise, bidder~$i$ receives payoff $- b_i$. If bidder~$i$ quits the game, then she receives payoff $0$. Specifically, bidder~$i$'s payoff is 
$$u_i(b, v) =
\begin{cases}
0,  &  \text{ if }   b_i =  Q;  \\
\frac{1}{ |I^w|} w_i(b_i, v)  -  b_i, &  \text{ if }   b_i = \max_{j \in I} b_j > Q;  \\
- b_i , &  \text{ otherwise}. 
\end{cases}
$$

We make the following assumption on the payoff functions.

\begin{assum}
\begin{enumerate}
\item The payoff function \( w_i(b_i, v) \) is positive for any \( v_i > 0 \) and \( w_i(b_i, 0, v_{-i}) \equiv 0 \), continuous on \( [0, \overline{b}_i] \times [0,1]^n \), increasing in \( v_{-i} \), and strictly increasing in \( v_i \).

\item The difference $w_i(b_i^H, v) - w_i(b_i^L, v)$ is increasing in $v$ for $b_i^H > b_i^L \ge  0$.

\item   The function $w_i(b_i, v) - b_i$ is strictly decreasing in $b_i$.
\end{enumerate}\label{assum-auction payoff}
\end{assum}

This example is also covered by the model in Section~\ref{sec-discontinuous model}. For each bidder $i$, let $W_i(b_i,v) = \omega_i(b_i,v)$, $C_i(b_i) \equiv 0$, $D(b) \equiv 1$, $\Phi_i(b_i) = b_i$ for $b_i \in [0, \overline{b}_i]$ and $\Phi_i(Q) = 0$. Then, bidder $i$'s payoff function $u_i(b,v) = w_i(b_i,v)P_i(b) -  b_i$ if $b_i \ne Q$, and  $0$ otherwise. 

We have that \( W_i(b_i^H, v) - W_i(b_i^L, v) = w_i(b_i^H, v) - \omega_i(b_i^L, v) \), which is monotone in \( v \) by Assumption~\ref{assum-auction payoff}. Furthermore, \( W_i(b_i, v) - C_i(b_i) = w_i(b_i, v) \) is positive for any \( v_i > 0 \), and \( W_i(b_i, 0, v_{-i}) - C_i(b_i) = - C_i(b_i) \), hence Assumption~\ref{assum-d1} (5) holds. 
Since \( (W_i(b_i, v) - C_i(b_i)) D(b_i, b_{-i}) - \Phi_i(b_i) = w_i(b_i, v) - b_i \) is strictly decreasing in \( b_i \) by Assumption~\ref{assum-auction payoff}, it follows that Assumption~\ref{assum-d1} holds in this example.

\begin{coro}
Under Assumption~\ref{assum-auction payoff}, a perfect monotone equilibrium exists in all-pay auctions with affiliated types. \label{coro-discontinuous}
\end{coro}


\paragraph{Bertrand competitions with unknown costs}

\

The third application is a price competition game in which firms have private costs. There are $n$ firms that compete by setting prices  $p_i \in    [0,\overline p_i]$, where $\overline{p}_i \ge 1$ for each $i$. Each firm~$i$ knows its marginal cost $c_i \in [0,1]$. The joint density is $f \colon [0,1]^n \to \bR_+$.  The market demand is given by a continuous, positive and componentwise decreasing function $\hat{D} (p_1, p_2, \cdots, p_n)$ such that $p_i \hat{D} (p_i, p_{-i})$ is strictly increasing in $p_i$, which implies that the price of the demand is relatively inelastic. If $p_i = \mathop{\text{min}}\limits_{1 \le j \le n} p_j  $, then firm~$i$'s demand $\hat D_i (p)   = \frac{\hat{D}(p)}{|\{ j : p_j = p_i  \}|  }$; otherwise, $\hat D_i (p) = 0$. Firm $i$'s profit is 
$$w_i(c_i,p_i, p_{-i})  =  (p_i - c_i)  \hat D_i (p).
$$

This example is also covered by the model presented in Section~\ref{sec-discontinuous model}. Denote $b_i = \overline p_i -p_i$ and $v_i = 1 - c_i$.
For each firm~$i$, let $W_i(b_i,v) =  \overline{p}_i  + v_i - 1$, $C_i(b_i) = b_i$, $D(b) = \hat{D}(\overline{p} - b)$, and $\Phi_i(b_i) \equiv 0$. Then
$$( W_i(b_i,v) - C_i(b_i) )D(b)P_i(b)  -    \Phi_i(b_i) =  (p_i - c_i)  \hat D_i (p) = w_i(c_i,p_i, p_{-i}).
$$

Note that $W_i(b_i^H, v) - W_i(b_i^L,v) = 0$, which is a constant function in $v$.  Besides, $W_i(b_i, v) - C_i(b_i)$ does not depend on $v_{-i}$, and 
$$( W_i(b_i,v) - C_i(b_i) )D(b) - \Phi_i(b_i)=  (\overline{p}_i  - b_i) \hat{D}(\overline{p} - b)  - ( 1 - v_i  )\hat{D}(\overline{p} - b),
$$
which is strictly decreasing in $b_i$ by the relative inelasticity of the demand. Thus, Assumption~\ref{assum-d1} holds in this example.

\begin{coro}
A perfect monotone equilibrium exists in Bertrand competitions with affiliated unknown costs.\label{coro-bertrand} 
\end{coro}

\vspace{-0.5cm}

\section{Conclusion}\label{sec-conclusion}

%
%

This paper studies the role of admissibility in monotone equilibria of Bayesian games. While monotone equilibria provide ordered and economically interpretable predictions, they may rely on weakly dominated actions and therefore fail to satisfy a basic requirement of rational decision-making. We show that this concern cannot be resolved simply by strengthening existing monotone-equilibrium existence results: the standard single-crossing and quasi-supermodularity conditions used in the literature do not generally guarantee the existence of admissible monotone equilibria.

To address this issue, we introduce perfect monotone equilibrium, which incorporates trembling-hand robustness into the study of monotone equilibrium. This refinement requires players to remain optimal against small, fully supported perturbations of opponents' strategies, thereby ruling out monotone equilibria sustained by weakly dominated actions. We establish the existence of perfect monotone equilibria in finite-action Bayesian games under increasing differences in own actions and types and supermodularity in own actions. These conditions ensure that best responses preserve the relevant monotone structure even under small trembles. We also show, through counterexamples, that these assumptions are tight. 
The analysis is further extended to Bayesian games with general type and action spaces. Under the same structural conditions, we prove the existence of perfect monotone equilibria when action spaces are compact complete metric lattices and type spaces are partially ordered probability spaces. In this setting, perfect monotone equilibrium satisfies limit admissibility, extending the admissibility interpretation beyond finite action spaces.

Finally, we demonstrate the applicability of the results in several economic environments. Our findings strengthen existing monotone-equilibrium results for multi-unit auctions and supermodular games, and also cover generalized auction environments with discontinuous payoffs, including first-price auctions, all-pay auctions, and Bertrand competitions. Overall, the paper provides a framework for selecting monotone equilibria that are not only economically interpretable, but also admissible and robust to small mistakes.


\vspace{-0.2cm}

\renewcommand{\baselinestretch}{1.25}






\vspace{-0.2cm}

\appendix
\renewcommand{\thesection}{\Alph{section}}  
\setcounter{section}{0}  

\section*{Appendix A}
\addcontentsline{toc}{section}{Appendix A}  
\setcounter{section}{1}  
\label{appendixA}  


The proof of Proposition~\ref{prop-admissibility}   is presented in Appendix~\ref{proof -prop1}. The proof of Theorem~\ref{thm-Reny}  is  provided in Appendix~\ref{proof -thm finite existence multidimension}. Appendix~\ref{proof - no MPE} includes the proofs of Claims~\ref{claim-quasi} and \ref{claim- SCC but not super}.

\vspace{-0.2cm}
\subsection{Proof of Proposition~\ref{prop-admissibility}  }\label{proof -prop1}

	\begin{proof}[(1)]
	Suppose that a pure strategy profile $g = (g_1, \dots, g_n)$ is perfect. 
	Then there exists a sequence of completely mixed strategy profiles 
	$\{g^k = (g_1^k, \dots, g_n^k)\}_{k=1}^{\infty}$ satisfying 
	Conditions (1) and (2) in Definition~\ref{def-perfect}.
	
	By Condition (2), for $\lambda_i$-almost every $t_i \in T_i$, 
	there exists $K(t_i) \in \mathbb{Z}_{+}$ such that, for all $k \ge K(t_i)$,
	\[
	g_i(t_i) \in \operatorname{BR}_i\bigl(t_i, g_{-i}^k\bigr).
	\]
	Hence, for every such $k$ and any alternative action $a_i' \in A_i$,
	\[
	V_i\bigl(g_i(t_i), t_i; g_{-i}^k\bigr) 
	\ge 
	V_i\bigl(a_i', t_i; g_{-i}^k\bigr).
	\]
	By the boundedness and continuity of $u_i$ on the action space, letting $k \to \infty$ yields
	\[
	V_i\bigl(g_i(t_i), t_i; g_{-i}\bigr) 
	\ge 
	V_i\bigl(a_i', t_i; g_{-i}\bigr), 
	\quad \text{for all } a_i' \in A_i.
	\]
	Since this holds for $\lambda_i$-almost every $t_i$, $g$ is a Bayesian Nash equilibrium.
	\end{proof}
	\medskip

The following lemma is the key for showing that a perfect monotone equilibrium is admissible.	
	\begin{lem}\label{lem-undominated against completely mixed strategies}
		If an action $a_i \in A_i$ is a best response against a completely mixed strategy profile $\hat{\alpha}_{-i}$ at some type $t_i \in T_i$, \textit{i.e.},
		$$
		V_i(a_i, t_i; \hat{\alpha}_{-i}) \geq V_i(a_i', t_i; \hat{\alpha}_{-i}) 
		$$
		for all $a_i' \in A_i$, then $a_i$ is undominated at type $t_i$.
	\end{lem}

	\begin{proof}
		We prove the contrapositive. Suppose $a_i \in A_i$ is weakly dominated at type $t_i$ by some mixed strategy $\sigma_i \in \cM(A_i)$.
		Then, by definition of weak dominance, for every strategy profile $\alpha_{-i} $,
		\[
		V_i(\sigma_i, t_i; \alpha_{-i}) \geq V_i(a_i, t_i; \alpha_{-i}),
		\]
		with strict inequality for at least one strategy profile $\alpha_{-i}^* $:
		\[
		V_i(\sigma_i, t_i; \alpha_{-i}^*) > V_i(a_i, t_i; \alpha_{-i}^*).
		\]

Assume, for contradiction, that $a_i$ is a best response against a completely mixed strategy profile $\hat{\alpha}_{-i}$ at type $t_i$. That is,
		\[
		V_i(a_i, t_i; \hat{\alpha}_{-i}) \geq V_i(a_i', t_i; \hat{\alpha}_{-i}) \quad \text{for all } a_i' \in A_i.
		\]
		Since $A_i$ is finite, $V_i$ is linear in $\sigma_i$:
		\[
		V_i(\sigma_i, t_i; \hat{\alpha}_{-i}) 
		= \sum_{a_i' \in A_i} \sigma_i(a_i') \, V_i(a_i', t_i; \hat{\alpha}_{-i})
		\leq V_i(a_i, t_i; \hat{\alpha}_{-i}).
		\]
		Weak dominance gives the reverse inequality, so equality holds:
		\[
		V_i(\sigma_i, t_i; \hat{\alpha}_{-i}) = V_i(a_i, t_i; \hat{\alpha}_{-i}).
		\]
		
		Now, for each player $j \neq i$, we want to construct a strategy $g_j^m$ that is close to $\alpha_j^*$ and can be embedded into $\hat{\alpha}_j$ with small probability $1/m$. 
		
		Since $\hat{\alpha}_j$ is completely mixed, for every $t_j \in T_j$ and every $a_j \in A_j$, we have $\hat{\alpha}_j(t_j, a_j) > 0$. For each $m \in \mathbb{Z}_+$, define the measurable set
		\[
		B_j(m) := \left\{ t_j \in T_j : \min_{a_j \in A_j} \hat{\alpha}_j(t_j, a_j) \leq \frac{1}{m} \right\}.
		\]
		Since $\hat{\alpha}_j(t_j,a_j) > 0$ for all $t_j$ and all $a_j$, we have $\bigcap_{m=1}^{\infty} B_j(m) = \emptyset$. Moreover, $B_j(m)$ is a decreasing sequence of measurable sets. Therefore, by continuity from above,
		\[
		\lambda_j(B_j(m)) \to 0 \quad \text{as } m \to \infty.
		\]
		For $t_j \notin B_j(m)$, we have
		\[
		\hat{\alpha}_j(t_j,a_j) \geq \frac{1}{m} \quad \text{for all } a_j \in A_j.
		\]
		
		Define the strategy $g_j^m$ as follows:
		\[
		g_j^m(t_j) := 
		\begin{cases}
			\alpha_j^*(t_j), & \text{if } t_j \notin B_j(m), \\
			\hat{\alpha}_j(t_j), & \text{if } t_j \in B_j(m).
		\end{cases}
		\]
		That is, $g_j^m$ equals $\alpha_j^*$ on the set where $\hat{\alpha}_j$ is sufficiently mixed (i.e., where the weight $1/m$ can be embedded), and equals $\hat{\alpha}_j$ on the exceptional set $B_j(m)$ where $\hat{\alpha}_j$ is too close to zero.
		
		For $t_j \notin B_j(m)$, we have $g_j^m(t_j) = \alpha_j^*(t_j)$. Since $\hat{\alpha}_j(t_j, a_j) \geq 1/m$ for all $a_j \in A_j$, we can define
		\[
		\overline{\alpha}_j(t_j,a_j) := \frac{\hat{\alpha}_j(t_j,a_j) - \frac{1}{m} \alpha_j^*(t_j,a_j)}{1 - \frac{1}{m}}.
		\]
		This is a well-defined probability distribution because:
		\begin{itemize}
			\item $\hat{\alpha}_j(t_j,a_j) - \frac{1}{m} \alpha_j^*(t_j,a_j) \geq \frac{1}{m} - \frac{1}{m} \cdot 1 = 0$;
			\item $\sum_{a_j \in A_j} \overline{\alpha}_j(t_j,a_j) =1$.
		\end{itemize}
		Thus, for $t_j \notin B_j(m)$,
		\[
		\hat{\alpha}_j(t_j) = \frac{1}{m} g_j^m(t_j) + \left(1 - \frac{1}{m}\right) \overline{\alpha}_j(t_j).
		\]
		For $t_j \in B_j(m)$, since $g_j^m(t_j) = \hat{\alpha}_j(t_j)$, we have trivially
		\[
		\hat{\alpha}_j(t_j) = \frac{1}{m} g_j^m(t_j) + \left(1 - \frac{1}{m}\right) \hat{\alpha}_j(t_j).
		\]
		
		Hence, for each $j \neq i$, we have the convex combination representation:
		\[
		\hat{\alpha}_j = \frac{1}{m} g_j^m + \left(1 - \frac{1}{m}\right) \widetilde{\alpha}_j,
		\]
		where $\widetilde{\alpha}_j$ is defined as  $\overline{\alpha}_j$ above on $T_j \setminus B_j(m)$ and as $\hat{\alpha}_j$ on $B_j(m)$.
		
		Therefore, the product strategy profile can be expanded as:
		\begin{align}\label{equa-expan}
			\hat{\alpha}_{-i} = \frac{1}{m^{n-1}} g_{-i}^m + \sum_{\alpha_{-i} \in \prod_{j \neq i} \{g_j^m, \widetilde{\alpha}_j\} \setminus \{g_{-i}^m\}} \prod_{j \neq i} \left( \frac{1}{m} \mathbf{1}_{g_j^m}(\alpha_j) + \left(1 - \frac{1}{m}\right) \mathbf{1}_{\widetilde{\alpha}_j}(\alpha_j) \right) \alpha_{-i}.
		\end{align}

		Now observe that $g_{-i}^m \to \alpha_{-i}^*$ as $m \to \infty$ in the sense that $g_j^m(t_j) = \alpha_j^*(t_j)$ for all $t_j \notin B_j(m)$, and $\lambda_j(B_j(m)) \to 0$. Therefore, by the dominated convergence theorem (since $u_i$ is bounded),
		\[
		V_i(a_i', t_i; g_{-i}^m) \to V_i(a_i', t_i; \alpha_{-i}^*) \quad \text{for every } a_i' \in A_i.
		\]
		By the strict inequality $	V_i(\sigma_i, t_i; \alpha_{-i}^*) > V_i(a_i, t_i; \alpha_{-i}^*)$, for sufficiently large $m$, we have 
		\[
		V_i(\sigma_i, t_i; g_{-i}^m) > V_i(a_i, t_i; g_{-i}^m).
		\]
		
		Now apply weak dominance to each term in the expansion (\ref{equa-expan}) . For every $\alpha_{-i} \in \prod_{j \neq i} \{g_j^m, \widetilde{\alpha}_j\}$, weak dominance gives
		\[
		V_i(\sigma_i, t_i; \alpha_{-i}) \geq V_i(a_i, t_i; \alpha_{-i}).
		\]
		For the term $\alpha_{-i} = g_{-i}^m$, the above inequality is strict. Therefore, taking the convex combination in (\ref{equa-expan}),
		\[
		V_i(\sigma_i, t_i; \hat{\alpha}_{-i}) > V_i(a_i, t_i; \hat{\alpha}_{-i}),
		\]
		because the term $g_{-i}^m$ appears with strictly positive weight $1/m^{n-1} > 0$, and all other terms satisfy weak inequality.
		
		This contradicts to that $a_i$ is a best response to $\hat{\alpha}_{-i}$. 
		Therefore, $a_i$ cannot be weakly dominated; that is, $a_i$ is undominated at type $t_i$.
	\end{proof}

\begin{proof}[(2)]
	Let $\hat g$ be a perfect monotone equilibrium. By Definition~\ref{def-perfect}, there exists a sequence of completely mixed strategy profiles
	$\{\hat g^k\}_{k\in\mathbb{Z}_+}$ satisfying Conditions~(1) and~(2).
	
	Fix a type $t_i$ satisfying Conditions~(1) and~(2). By the definition of perfection,
	$\hat g_i(t_i)$ is a best response to $\hat g_{-i}^k$ for all sufficiently large $k$.
	Since each $\hat g^k$ is completely mixed, Lemma~\ref{lem-undominated against completely mixed strategies} implies that
	$\hat g_i(t_i)$ is undominated at type $t_i$.
	Therefore, $\hat g$ is admissible.
\end{proof}

\subsection{Proof of Theorem~\ref{thm-Reny}}\label{proof -thm finite existence multidimension}

We first provide a sketch of the proof. 

\begin{itemize}
\item To prove the existence of a perfect monotone equilibrium in the game $G$, we construct in Step~1 a sequence of games $\{ G^m\}_{m = 2}^{\infty}$ converging to $G$. Each $G^m$ is a slight perturbation of $G$ in the following sense. The game $G^m$ differs from the limit game $G$ only in its payoff functions: when players play the action profile~$a$ at type profile $t$, player~$i$'s payoff $u_i^m(a, t)$ in $G^m$ equals $u_i(\tilde a^m, t)$ -- player~$i$'s payoff based on the action profile $\tilde a^m$ in the original game. Here $\tilde a_j^m$ is a completely mixed probability measure putting probability $1 - \frac{1}{m}$ at the action $a_j$ for each player~$j$.

\item In Step~2, it will be shown that each game $G^m$ satisfies Assumptions~\ref{assum-1} and \ref{assum-2}, and thus has a monotone equilibrium $g^m$.
Importantly, $g^m$ can be interpreted as a completely mixed strategy $\tilde g^m$ in the game $G$ such that $\tilde{g}^m_i$ is an approximate best response of $\tilde{g}^m_{-i}$ for each player~$i$. The sequence $\{g^m\}$ has a convergent subsequence  $\{g^{m_r}\}_{r = 1}^{\infty}$ with the limit $g$.

\item In Step~3, we will show that $g$ is a perfect equilibrium in the original game $G$. Given that $g$ is an increasing strategy, it follows that  $g$ is a perfect monotone equilibrium.
\end{itemize}



\noindent	\textbf{Step 1.}  Since $A_i$ is  a compact metric space, there exists a countable dense subset $S_i= \{s_i^k\}_{k = 1}^{\infty} \subseteq A_i$ for each $i \in I$. For each $a_i \in A_i$ and each positive integer $m \ge 2$, let $\tilde a_i^m$ denote a completely mixed probability measure on $A_i$ putting probability $1 - \frac{1}{m}$ on action $a_i$,
\begin{equation}\label{eq-tilde a_i^m}
\tilde a_i ^ m = \left(1 - \frac{1}{m}\right)\delta_{a_i} + \frac{1}{m}\sum_{k = 1}^{ \infty}  \frac{1}{2^k}\delta_{s_i^k},
\end{equation}
where $F_i = \{\delta_{s_i^k}\}_{k = 1}^{\infty}$ is the set of Dirac measures that concentrate at the points in $S_i$. 
It is easy to see that $\tilde a_i ^ m = \sum_{s_i \in  S_i \cup \{a_i\}} \tilde a_i^m(s_i) \delta_{s_i}$. 
In game $G^m$,  player $i$'s payoff function is
\begin{equation}\label{eq-u_i^m}
u_i^m(a,t)= u_i(\tilde a^m,t) = \sum_{s_1 \in  S_1 \cup \{a_1\}} \ldots \sum_{s_n \in  S_n \cup \{a_n\}} \tilde a_1^m(s_1) \cdots \tilde a_n^m(s_n)  u_i(s_1, \ldots, s_n,t),
\end{equation}
which is bounded, jointly measurable, and continuous in $a$. 
It is easy to see that the Prohorov metric $\rho \left(\tilde a_i^m, \delta_{a_i} \right)$ is less than or equal to $\frac {1} {m}$. 
Clearly, $u_i^m$  converges pointwise to $u_i$.

\noindent	\textbf{Step 2.} Fix an increasing strategy profile $\phi$. For each $m \ge 2$ and $i \in I$, 
we consider a completely mixed strategy profile $\tilde \phi_i^m$,  where $\tilde \phi_i^m(t_i)$ takes the completely mixed probability measure $\widetilde{\phi_i(t_i)}^m$ on $A_i$. By Equation (\ref{eq-tilde a_i^m}), $\tilde \phi_i^m(t_i)$ places probability $1 - \frac{1}{m}$ at the action $\phi_i(t_i)$, and probability $\frac{1}{m} \cdot \frac{1}{2^k}$ at the action $s_i^k$ for each $k \ge 1$; those probability weights are independent of \( t_i \) for each player \( i \). 
Let $\Gamma_i^m : S_i \cup \{\phi_i\} \to [0,1]$ be such that $\Gamma_i^m(\phi_i) = 1- \frac{1}{m}$ and  $\Gamma_i^m(s_i^k) = \frac{1}{m} \cdot \frac{1}{2^k}$ for all $k$, where $s_i^k$ is viewed as a strategy in which player~$i$ chooses action  $s_i^k$ at any type $t_i \in T_i$.
It is clear that 
\begin{equation}\label{eq-Gamma_i^m}
\Gamma_i^m(\phi_i) = 1 - \frac{1}{m} = \tilde \phi_i^m(t_i) (\phi_i(t_i)) 
\quad \text{and} \quad
\Gamma_i^m(s_i^k) = \frac{1}{m} \cdot \frac{1}{2^k} = \tilde \phi_i^m(t_i) (s_i^k). 
\end{equation}
We can obtain the following identities about player $i$'s interim payoff in game $G^m$:
\begin{align}\label{equa-m}
&V_i^m(a_i,t_i; \phi_{-i}) \notag   \\
&  =  \int_{T_{-i}}   u_i^m(a_i,\phi_{-i}(t_{-i}),t_i, t_{-i})  \rmd \lambda_{-i}(t_{-i} |t_i)  \notag  \\
&  = \int_{T_{-i}}    \sum_{s_i \in S_i \cup \{a_i\} } 
   \sum_{s_{-i} \in \prod_{j \ne i}  \left(S_j \cup \{\phi_j(t_j)\} \right)} 
\tilde a_i^m(s_i) 
\prod_{j \ne i} \tilde \phi_j^m(t_j)(s_j) 
  u_i(s_1, \ldots, s_n,t)   \rmd \lambda_{-i}(t_{-i} |t_i)   \notag  \\
&    = \int_{T_{-i}}  \sum_{s_i \in  S_i \cup \{a_i\} }  
 \sum_{\alpha_{-i} \in \prod_{j \ne i}  \left(S_j \cup \{\phi_j\}\right)} 
\tilde a_i^m(s_i) 
  \prod_{j \ne i}\Gamma^m_{j}  (\alpha_{j})
  u_i(s_i, \alpha_{-i}(t_{-i}),t_i,t_{-i})   \rmd \lambda_{-i}(t_{-i} |t_i)   \notag  \\           
&    =  \sum_{s_i \in  S_i \cup \{a_i\}}  
\sum_{\alpha_{-i} \in \prod_{j \ne i} \left( S_j \cup \{\phi_j\}\right)} \tilde a_i^m(s_i)  \prod_{j \ne i}\Gamma^m_{j}  (\alpha_{j}) \int_{T_{-i}}  u_i(s_i, \alpha_{-i}(t_{-i}),t_i,t_{-i})   \rmd \lambda_{-i}(t_{-i} |t_i)   \notag  \\                                                                                        
&  =  \sum_{s_i \in  S_i \cup \{a_i\}}  
\sum_{\alpha_{-i} \in \prod_{j \ne i} \left(S_j \cup \{\phi_j\}\right)}   \tilde a_i^m(s_i)  \prod_{j \ne i}\Gamma^m_{j} (\alpha_{j}) V_i(s_i, t_i; \alpha_{-i}).
\end{align}
The second equality above is a direct consequence of Equation~(\ref{eq-u_i^m}). The third equality uses Equation~(\ref{eq-Gamma_i^m}) and relies on the fact that the probability weights of $\tilde{\phi}^m_i(t_i)$ are independent of $t_i$ for each player $i$. Finally, the fourth equality is obtained by interchanging the order of integration and summation.
It follows from Equation (\ref{equa-m}) that 
\begin{align*}
V_i^m(a_i,t_i; \phi_{-i})  =  &
\left(1 - \tfrac{1}{m}\right) \sum_{\alpha_{-i} \in \left(\prod_{j \ne i} S_j \cup \{\phi_j\}\right)}     \prod_{j \ne i}\Gamma^m_{j} (\alpha_{j}) V_i(a_i, t_i; \alpha_{-i}) \\
& + 
\sum_{s_i \in  S_i}  
\sum_{\alpha_{-i} \in \left(\prod_{j \ne i} S_j \cup \{\phi_j\}\right)}   \tilde a_i^m(s_i)  \prod_{j \ne i}\Gamma^m_{j} (\alpha_{j}) V_i(s_i, t_i; \alpha_{-i}),
\end{align*}
where the formula after the plus sign does not depend on $a_i$.

Since the strategies $\alpha_{j} \in S_j \cup \{\phi_j\}$ are monotone, Assumptions \ref{assum-1} and \ref{assum-2} imply that $V_i(a_i,t_i;\alpha_{-i})$  satisfies IDC in $(a_i,t_i)$ and $V_i(a_i,t_i;\alpha_{-i})$ is supermodular in $a_i$. 
Given that these two conditions are preserved under positive linear combinations, 
we obtain that 
\begin{itemize}
\item $V_i^m(a_i,t_i;\phi_{-i})$  satisfies IDC in $(a_i,t_i)$;
\item $V_i^m(a_i,t_i;\phi_{-i})$ is supermodular in $a_i$.
\end{itemize}
By \citet[Theorem 4.1 and Propositon 4.4]{R2011}, $G^m$ possesses a monotone equilibrium $g^m$.  Applying the generalized Helly's selection theorem in \citet[Lemma A.10]{R2011}, there exists a subsequence   $\{g^{m_r}\}_{r = 1}^{\infty}$   of $\{g^m\}_{m = 2}^{\infty}$ such that for each player $i$, $\{g_i^{m_r}\}_{r = 1}^{\infty}$ converges to a measurable monotone strategy $g_i$ almost everywhere. That is, $\lim\limits_{r \to \infty} d (g_i^{m_r}(t_i), g_i(t_i)) = 0$ for $\lambda_i$-almost all $t_i \in T_i$.

\noindent	\textbf{Step 3.} For each $m_r \in \mathbb{Z}_{+}$, let  $\tilde {g}_i^{m_r}$ be the completely mixed strategy corresponding to $g_i^{m_r}$ as defined in the beginning of Step 2.
By the facts that 
$\rho(\tilde{g}_i^{m_r}(t_i), \delta_{{g}^{m_r}_i(t_i)}) \le \tfrac {1} {m_r}$ and
  $\lim\limits_{r \to \infty} d(g_i^{m_r}(t_i), g_i(t_i)) = 0$,\footnote{By the definition of Prohorov metric, it is easy to see that  $\lim\limits_{r \to \infty} d\left(g_i^{m_r}(t_i), g_i(t_i)\right) = 0$ iff  $\lim\limits_{r \to \infty} \rho\left(\delta_{g_i^{m_r}(t_i)}, \delta_{g_i(t_i)}\right) = 0$.
} we  obtain that for each player $i$, $\lim\limits_{r \to \infty} \rho(\tilde{g}_i^{m_r}(t_i), \delta_{g_i(t_i)}) = 0$ for $\lambda_i$-almost all $t_i \in T_i$. 
We can obtain the following identities about player $i$'s interim payoff in game $G$ in terms of the relevant completely mixed actions/strategies:
\begin{align}\label{equa-v_im}
& 	V_i(\tilde{a}_i^{m_r},t_i; \tilde{\phi}_{-i}^{m_r}) \notag  \\
	&= \sum_{s_i \in  S_i \cup \{a_i\} } \tilde{a}_i^{m_r}(s_i) V_i(s_i, t_i; \tilde{\phi}_{-i}^{m_r}) \notag \\
	&= \sum_{s_i \in  S_i \cup \{a_i\} } \tilde{a}_i^{m_r}(s_i) 	\int_{T_{-i}} \int_{A_{-i}} u_i(s_i, s_{-i}, t) 	\mathop{\otimes}_{j \ne i} \tilde \phi^{m_r}_j(t_j, \rmd s_j) \rmd \lambda_{-i}(t_{-i}|t_i)  \notag \\
	& = \sum_{s_i \in  S_i \cup \{a_i\} } \tilde{a}_i^{m_r}(s_i) 	\int_{T_{-i}}  \sum_{s_{-i} \in  \prod_{j \ne i} \left(S_j \cup \{\phi_j(t_j)\} \right)}  \prod_{j \ne i} \tilde \phi_j^{m_r}(t_j)(s_j) 
	u_i(s_i,s_{-i},t)   \rmd \lambda_{-i}(t_{-i} |t_i)  \notag \\
&  = \int_{T_{-i}}    \sum_{s_i \in S_i \cup \{a_i\} } 
\sum_{s_{-i} \in  \prod_{j \ne i} \left( S_j \cup \{\phi_j(t_j)\} \right)} 
\tilde a_i^{m_r}(s_i) 
\prod_{j \ne i} \tilde \phi_j^m(t_j)(s_j) 
u_i(s_1, \ldots, s_n,t)   \rmd \lambda_{-i}(t_{-i} |t_i)   \notag  \\
& = V_i^{m_r}(a_i,t_i; \phi_{-i}),
\end{align}
where the last equality follow from Equation~\eqref{equa-m}.
It follows from Equation~(\ref{equa-v_im}) that 
\begin{equation}\label{eq-V_i_tilde{a}_i^{m_r}}
V_i(\tilde{a}_i^{m_r},t_i; \tilde{\phi}_{-i}^{m_r}) 
= \left(1 - \tfrac{1}{m_r}\right) V_i(a_i, t_i; \tilde{\phi}_{-i}^{m_r})
+ 
\frac{1}{m_r}\sum_{k = 1}^{ \infty}  \frac{1}{2^k} V_i(s_i^k, t_i; \tilde{\phi}_{-i}^{m_r}).
\end{equation}

Since $g^{m_r}$ is a Bayesian Nash equilibrium of the game $G^{m_r}$ for each $r \in \mathbb{Z}_{+}$,  we know that for $\lambda_i$-almost  all $t_i \in T_i$, 
\begin{equation}\label{eq-Vg}
 V_i^{m_r}(g^{m_r}(t_i),t_i;g^{m_r}_{-i}) \ge V_i^{m_r}(a_i,t_i;g^{m_r}_{-i}) \text{ for all } a_i \in A_i.
 \end{equation}
By combining countably many $\lambda_i$-null sets, there exists a set $C_i \in \mathcal{T}_i$ with $\lambda_i (C_i) =1$ such that for all $t_i \in C_i$, 
$\lim\limits_{r \to \infty} d (g_i^{m_r}(t_i), g_i(t_i)) = 0$ and 
Equation (\ref{eq-Vg}) holds for all $m_r$. 
Fix a $t_i \in C_i$. 
By Equation~(\ref{equa-v_im}), 
Equation (\ref{eq-Vg}) is equivalent to $V_i(\tilde{g}_i^{m_r}(t_i),t_i,\tilde{g}_{-i}^{m_r}) \ge V_i(\tilde a_i^{m_r},t_i,\tilde{g}^{m_r}_{-i})$ for all $a_i \in A_i$.  
It then follows from Equation (\ref{eq-V_i_tilde{a}_i^{m_r}}) that
\begin{align*}
	V_i(\tilde{g}_i^{m_r}(t_i),t_i,\tilde{g}_{-i}^{m_r})  -  V_i(\tilde a_i^{m_r},t_i,\tilde{g}^{m_r}_{-i}) 
 = \left(1 - \tfrac{1}{m_r}\right)    \left(V_i(g_i^{m_r}(t_i),t_i,\tilde {g}_{-i}^{m_r}) - V_i(a_i,t_i,\tilde{g}^{m_r}_{-i})\right) \ge 0
\end{align*}
for all $a_i \in A_i$.  
Thus,  we have $g_i^{m_r}(t_i) \in \text{BR}_i(t_i,\tilde{g}_{-i}^{m_r})$ in the game $G$, which means that $ d \left({g}_i^{m_r}(t_i), \mathrm{BR}_i(t_i, \tilde{g}_{-i}^{m_r})\right) =  0$ for all $m_r$. 
By the triangle inequality, we have
$$
0 \le \lim_{r \to \infty} d \big(g_i(t_i), \mathrm{BR}_i(t_i, \tilde{g}_{-i}^{m_r}) \big) 
\le
 \lim_{r \to \infty} d \big( {g_i(t_i)}, {g}_i^{m_r}(t_i) \big)  + 
 \lim_{r \to \infty} d \big( {g}_i^{m_r}(t_i), \mathrm{BR}_i(t_i, \tilde{g}_{-i}^{m_r}) \big) 
= 0.
$$
By taking $r$ to infinity in Equation (\ref{eq-Vg}), 
we obtain that 
$ V_i(g(t_i),t_i;g_{-i}) \ge V_i(a_i,t_i;g_{-i})$ for all $a_i \in A_i$.\footnote{By \citet[Theorem 4]{MS1994}, the best response correspondence  $\text{BR}_i(t_i,g_{-i})$  is monotone in $t_i$.  When  $g_i(t_i) \in \text{BR}_i(t_i,g_{-i})$ for $\lambda_i$-almost all $t_i \in T_i$,  there exists a monotone selection of the best response correspondence  $ \text{BR}_i(t_i,g_{-i})$, denoted by $g_i^*$, such that it equals  $g_i$  for $\lambda_i$-almost all $t_i \in T_i$. Moreover, if $g$ is a Bayesian Nash equilibrium, then $g^*$ is a Bayesian Nash equilibrium with $g^*_i(t_i) \in \text{BR}_i(t_i,g^*_{-i}) $  for every type $t_i \in T_i$ and each player $i \in I$.} 
Hence, $g$ is a perfect monotone equilibrium. This completes the proof.

%
%


\vspace{-0.2cm}

\subsection{Proofs of Claims~\ref{claim-quasi} and \ref{claim- SCC but not super}}\label{proof - no MPE}

\begin{proof}[Proof of Claim~\ref{claim-quasi}]
Because the payoff functions are type-independent, the increasing differences condition holds trivially for both players. We prove each part of the claim in turn.

\medskip

\begin{enumerate}
\item[\textbf{1.}] Let
\[
D_i(a_i,a_i',t_i;\alpha_j)=V_i(a_i,t_i;\alpha_j)-V_i(a_i',t_i;\alpha_j)
\]
denote player~$i$'s interim payoff difference between actions $a_i$ and $a_i'$ when the opponent adopts a monotone strategy $\alpha_j$.

In each player's action space, the only incomparable pair is $(0,1)$ and $(1,0)$. Hence, to verify quasi-supermodularity, it suffices to check the two implications
\[
D_i((0,1),(0,0),t_i;\alpha_j)\ge (>)0
\;\Longrightarrow\;
D_i((1,1),(1,0),t_i;\alpha_j)\ge (>)0,
\]
and
\[
D_i((1,0),(0,0),t_i;\alpha_j)\ge (>)0
\;\Longrightarrow\;
D_i((1,1),(0,1),t_i;\alpha_j)\ge (>)0.
\]

Consider player~1. Any monotone strategy $\alpha_2$ must take one of the following two forms, for some $p_{00},p_{01},p_{10}\in [0,1]$:
\[
\alpha_2(t_2)=
\begin{cases}
(0,0), & t_2\in [0,p_{00}],\\
(0,1), & t_2\in (p_{00},p_{00}+p_{01}],\\
(1,1), & t_2\in (p_{00}+p_{01},1],
\end{cases}
\qquad\text{or}\qquad
\alpha_2(t_2)=
\begin{cases}
(0,0), & t_2\in [0,p_{00}],\\
(1,0), & t_2\in (p_{00},p_{00}+p_{10}],\\
(1,1), & t_2\in (p_{00}+p_{10},1].
\end{cases}
\]

If $\alpha_2$ takes the first form, then
\begin{align*}
D_1((0,1),(0,0),t_1;\alpha_2) &= p_{00}-7p_{01},\\
D_1((1,1),(1,0),t_1;\alpha_2) &= p_{00}-7p_{01},\\
D_1((1,0),(0,0),t_1;\alpha_2) &= -p_{00}+7p_{01},\\
D_1((1,1),(0,1),t_1;\alpha_2) &= -p_{00}+7p_{01}.
\end{align*}
If $\alpha_2$ takes the second form, then
\begin{align*}
D_1((0,1),(0,0),t_1;\alpha_2) &= p_{00}+p_{10},\\
D_1((1,1),(1,0),t_1;\alpha_2) &= p_{00}+\frac12 p_{10},\\
D_1((1,0),(0,0),t_1;\alpha_2) &= -p_{00}-\frac12 p_{10},\\
D_1((1,1),(0,1),t_1;\alpha_2) &= -p_{00}-p_{10}.
\end{align*}
In both cases, the two quasi-supermodularity implications follow immediately. Hence $V_1(a_1,t_1;\alpha_2)$ is quasi-supermodular in $a_1$ for every monotone strategy $\alpha_2$.

Next consider player~2. Action $(1,0)$ is strictly dominated by $(1,1)$, and action $(1,1)$ is strictly dominated by both $(0,0)$ and $(0,1)$. Thus
\[
D_2((1,1),(1,0),t_2;\alpha_1)>0
\qquad\text{and}\qquad
D_2((1,0),(0,0),t_2;\alpha_1)<0
\]
for every monotone strategy $\alpha_1$ and every $t_2$. Hence the antecedent in the second quasi-supermodularity implication is never satisfied, and the relevant inequalities hold vacuously. Therefore $V_2(a_2,t_2;\alpha_1)$ is quasi-supermodular in $a_2$.

\medskip

\item[\textbf{2.}] Consider the pair of monotone strategies $(s_1,s_2)$ defined by
\[
s_1(t_1)=
\begin{cases}
(0,0), & t_1\in \left[0,\frac45\right),\\
(1,1), & t_1\in \left[\frac45,1\right],
\end{cases}
\qquad
s_2(t_2)=
\begin{cases}
(0,0), & t_2\in \left[0,\frac78\right),\\
(0,1), & t_2\in \left[\frac78,1\right].
\end{cases}
\]
Given $s_1$, player~2 is indifferent between $(0,0)$ and $(0,1)$ at every type. Given $s_2$, all four actions in $A_1$ are best responses for player~1 at every type. Thus $(s_1,s_2)$ is a monotone equilibrium.

\medskip

\item[\textbf{3.}] Consider the profile
\[
s_1'(t_1)=
\begin{cases}
(0,1), & t_1\in \left[0,\frac45\right),\\
(1,0), & t_1\in \left[\frac45,1\right],
\end{cases}
\qquad
s_2'(t_2)=
\begin{cases}
(0,0), & t_2\in \left[0,\frac78\right),\\
(0,1), & t_2\in \left[\frac78,1\right].
\end{cases}
\]
This profile is a Bayesian Nash equilibrium. The detailed perturbation argument verifying that $(s_1',s_2')$ is a perfect equilibrium is relegated to the appendix. There, a sequence of completely mixed strategy profiles converging to $(s_1',s_2')$ is constructed such that
\[
\mathrm{BR}_2(t_2;s_1^k)=\{(0,0),(0,1)\}
\qquad\text{and}\qquad
\mathrm{BR}_1(t_1;s_2^k)=\{(0,1),(1,0)\}
\]
for all $k$ and all types.

\medskip

\item[\textbf{4.}] For player~1, the actions $(0,0)$ and $(1,1)$ are weakly dominated by the mixed action $\frac12(0,1)+\frac12(1,0)$. Hence any undominated strategy of player~1 can use only $(0,1)$ and $(1,0)$. Since these two actions are incomparable, any \emph{monotone} strategy of player~1 must be constant: it must be either the constant strategy $(0,1)$ or the constant strategy $(1,0)$.

For player~2, the actions $(1,0)$ and $(1,1)$ are strictly dominated. Hence any admissible monotone strategy of player~2 must be of the form
\[
\beta_2(t_2)=
\begin{cases}
(0,0), & t_2\in [0,p),\\
(0,1), & t_2\in [p,1]
\end{cases}
\]
for some $p\in [0,1]$.

Now consider admissible monotone best responses. If player~1 always plays $(0,1)$, then player~2's best response is to play $(0,1)$ always, that is, $p=0$. If player~2 always plays $(0,1)$, then player~1's best response is to play $(1,0)$ always. If player~1 always plays $(1,0)$, then player~2's best response is to play $(0,0)$ always, that is, $p=1$. Finally, if player~2 always plays $(0,0)$, then player~1's best response is to play $(0,1)$ always. Hence admissible monotone best responses cycle, and there is no admissible monotone equilibrium.

\medskip

\item[\textbf{5.}] By Proposition~\ref{prop-admissibility}, any perfect monotone equilibrium in a finite Bayesian game must be admissible. Part~4 shows that there is no admissible monotone equilibrium. Therefore the game has no perfect monotone equilibrium.
\end{enumerate}
\end{proof}


\begin{proof}[Proof of Claim~\ref{claim- SCC but not super}]

\begin{enumerate}
\item[\textbf{1.}] Since player $2$'s payoff function doesn't depend on types, his interim payoff function $V_2(a_2,t_2;s_1)$ satisfies the increasing differences condition in $(a_2,t_2)$ for any $s_1 \in \cF_1$, which implies that   $V_2(a_2,t_2; s_1)$ also satisfies the single crossing condition in $(a_2,t_2)$ for any $s_1 \in \cF_1$.  Below,  we shall focus on player $1$'s interim payoff function.

Given that player $2$ plays an increasing strategy
		\[
		s_2(t_2) = 
		\begin{cases}
			1 & \text{if } t_2 \in [0, x_2], \\
			2 & \text{if } t_2 \in (x_2, 1].
		\end{cases}
		\]
		
we compute the difference in player $1$'s expected payoff from choosing action $2$ versus action $1$:
			\begin{align}\label{exam-c1-1}
				& V_1(2,t_1;s_2) - V_1(1,t_1;s_2) \notag \\
				&= \int_{0}^{x_2} \left[u_1(2,1,t_1,t_2) - u_1(1,1,t_1,t_2)\right] \mathrm{d}t_2 
			+ \int_{x_2}^1 \left[u_1(2,2,t_1,t_2) - u_1(1,2,t_1,t_2)\right] \mathrm{d}t_2 \notag \\
				&= \int_{0}^{x_2} \left[\left(\tfrac{55}{24} - \tfrac{1}{6}t_2 - \tfrac{7}{6}t_1\right) - \left(\tfrac{5}{2} - \tfrac{4}{3}t_1\right)\right] \mathrm{d}t_2
			+ \int_{x_2}^1 \left[\left(2 + t_2 - \tfrac{7}{6}t_1\right) - \left(\tfrac{1}{2} + \tfrac{1}{2}t_2\right)\right] \mathrm{d}t_2  \notag \\
				&= \left(\tfrac{7}{4} - 2x_2\right)\left(1 + \tfrac{1}{6}x_2 - \tfrac{2}{3}t_1\right).
			\end{align}
Since the inequality 
$ 1 + \tfrac{1}{6}x_2 - \tfrac{2}{3}t_1 > 0 $
holds for all $t_1 \in [0,1]$, the sign of 
$V_1(2,t_1;s_2) - V_1(1,t_1;s_2) $
is independent of $t_1$, which implies that the interim payoff function $V_1(a_1,t_1;s_2)$ satisfies the single crossing condition in $(a_1,t_1)$ for any $s_2$ in the set of increasing strategies $\mathcal{F}_2$. 
However, for $x_2 < \tfrac{7}{8}$, the difference $V_1(2,t_1;s_2) - V_1(1,t_1;s_2) $  is decreasing in $t_1$ rather than increasing. 
Therefore,  $V_1(a_1,t_1;s_2)$ does not satisfy the increasing differences condition in $(a_1,t_1)$ for any $s_2 \in \mathcal{F}_2$.

			
			\item[\textbf{2.}] Suppose that player $1$ adopts the following increasing strategy: 
		\[
		s_1(t_1) = 
		\begin{cases}
			1 & \text{if } t_1 \in [0, x_1], \\
			2 & \text{if } t_1 \in (x_1, 1].
		\end{cases}
		\]
A straightforward calculation shows that	
			\begin{align*}
			&	V_2(2,t_2;s_1) - V_2(1,t_2;s_1) \\
			&  =  \int_{0}^{x_1} \left[  u_2(1,2,t_1,t_2) -     u_2(1,1,t_1,t_2)   \right]  \rmd t_1   +   	\int_{x_1}^1 \left[  u_2(2,2,t_1,t_2) -     u_2(2,1,t_1,t_2)  \right]     \rmd t_1  
			\\
			&= 10x_1 - 8.
			\end{align*}
Given that player~1 employs an increasing strategy $s_1$ with cutoff point $x_1$, we slightly abuse notation by using $x_1$ to represent the strategy $s_1$ itself. For each type $t_2$ of player~2, we have 
\begin{itemize}
			\item    $\text{BR}_2(t_2 ,x_1) \equiv   \{ 1 \}$, if $x_1 < \frac{4}{5}$;
			\item    $\text{BR}_2(t_2 ,x_1) \equiv   \{ 1,2 \}$, if $x_1 =  \frac{4}{5}$;
			\item    $\text{BR}_2(t_2 ,x_1) \equiv   \{ 2 \}$, if $x_1 > \frac{4}{5}$.
		\end{itemize}
Let $\mathrm{MBR}_2(x_1)$ denote the collection of all monotone functions $s_2$ with $s_2(t_2) \in \mathrm{BR}_2( t_2, x_1)$ for any $t_2 \in T_2$. 
Each $ s_2 \in \text{MBR}_2(x_1) $ can be identified by a (unique) cutoff type $x_2 \in [0,1]$,
\[
s_2(t_2) = 
\begin{cases}
	1 & \text{if } t_2 \in [0, x_2), \\
	2 & \text{if } t_2 \in [x_2, 1],
\end{cases}
\quad \text{or} \quad
s_2(t_2) = 
\begin{cases}
	1 & \text{if } t_2 \in [0, x_2], \\
	2 & \text{if } t_2 \in (x_2, 1].
\end{cases}
\]
These two monotone functions are equivalent as they only differ at a measure zero point.  The optimal cutoff point $x_2$ is
\[
x_2 
\begin{cases}
	= 1 & \text{if } x_1 < \frac{4}{5}, \\
	\in [0,1] & \text{if } x_1 = \frac{4}{5}, \\
	 = 0 & \text{if } x_1 > \frac{4}{5}.
\end{cases}
\]

Given the expression for $V_1(2,t_1;s_2) -V_1(1,t_1;s_2)$ computed in Equation~(\ref{exam-c1-1}), player~$1$'s best response correspondence and 
the optimal cutoff are given by
$$\text{BR}_1(t_1,x_2) = 
\begin{cases}
 \{2\} &  \text{if } x_2  < \frac{7}{8},\\
\{1,2\} &  \text{if } x_2 = \frac{7}{8}, \\
\{1\}&  \text{if }x_2  > \frac{7}{8},
\end{cases}
\quad 
\mbox{and}
\quad 
x_1 
\begin{cases}
= 0 &  \text{if } x_2  < \frac{7}{8},\\
\in [0,1]  &   \text{if }x_2 = \frac{7}{8}, \\
= 1   &   \text{if }x_2  > \frac{7}{8}.
\end{cases}				
$$
By solving the fixed point conditions for the cutoff strategies,
we obtain a unique pair of optimal cutoff types $(\frac{4}{5}, \frac{7}{8}).$ This  implies that the game has a unique monotone equilibrium $(s_1^*,s_2^*)$,\footnote{Note that $s_1^*$ and $s_2^*$ may take different actions at types $\frac{4}{5}$ and $\frac{7}{8}$. 
	A monotone equilibrium is considered unique if it coincides with any other such monotone equilibrium almost everywhere.
	
}  where 
			\begin{flalign*}
				& &s_1^*(t_1)=\left\{
				\begin{aligned}
					1 & \quad &   \text{if }t_1 \in [0,\tfrac{4}{5}], \\
					2 & \quad  &   \text{if }t_1 \in (\tfrac{4}{5}, 1], 
				\end{aligned}
				\right.     &   \text{ and }
				s_2^*(t_2)=\left\{
				\begin{aligned}
					1 & \quad &  \text{if } t_2 \in [0,\tfrac{7}{8}], \\
					2 & \quad  &   \text{if }t_2 \in (\tfrac{7}{8}, 1].
				\end{aligned}
				\right.  &  
			\end{flalign*}

\item[\textbf{3.}]	Let	$r =  \frac{-173+\sqrt{34889}}{80}$. It is clear that $\tfrac{1}{6} < r < \tfrac{1}{5}$.	Define a strategy profile $(g_1^*,g_2^*)$ as follows:     
\begin{flalign*}
	& &g_1^*(t_1)=\left\{
	\begin{aligned}
		2 & \quad &  \text{if } t_1 \in [0,\tfrac{1}{5}], \\
		1 & \quad  & \text{if }  t_1 \in (\tfrac{1}{5}, 1],
	\end{aligned}
	\right.     &   \quad  \text{ and }  \quad
	g_2^*(t_2)=\left\{
	\begin{aligned}
		2 & \quad &  \text{if } t_2 \in [0,r], \\
		1 & \quad  &   \text{if }t_2 \in (r, 1].
	\end{aligned}
	\right.  &    
\end{flalign*}
We shall show that $(g_1^*,g_2^*)$ is a perfect equilibrium in this game.

Firstly, we show that  $(g_1^*,g_2^*)$ is a Bayesian Nash equilibrium. Given that player $1$ plays strategy $g_1^*$, we know that 
\begin{align*}
 V_2(2,t_2;g_1^*) - V_2(1,t_2;g_1^*)  
 = \left[ (-1)\times \tfrac{4}{5} + 7 \times \tfrac{1}{5}  \right] -     \left[ 1 \times \tfrac{4}{5} + (-1) \times \tfrac{1}{5}  \right]  
 =  0.
\end{align*}
The first equality holds because player~$2$'s payoff is type-independent and player~$1$ follows strategy $g_1^*$, randomizing between actions with $\mathbb{P}( \{t_1 : g_1^*(t_1)=1\})=\frac{4}{5}$ and $\mathbb{P}(\{t_1 : g_1^*(t_1)=2\})=\frac{1}{5}$.
Hence, player $2$ has no incentive to deviate from $g_2^*.$

Given that player~2 plays strategy $g_2^*$, a direct computation yields
\begin{align}\label{exam-c1-2}
	& V_1(2,t_1;g_2^*) - V_1(1,t_1; g_2^*)  \notag \\
&	=  \int_{0}^{r} \left[u_1(2,2,t_1,t_2) - u_1(1,2,t_1,t_2)\right] \mathrm{d}t_2 
	 + \int_{r}^1 \left[u_1(2,1,t_1,t_2) - u_1(1,1,t_1,t_2)\right] \mathrm{d}t_2  \notag \\
&	=  \int_{0}^r \left[(2 + t_2 - \tfrac{7}{6}t_1) - (\tfrac{1}{2} + \tfrac{t_2}{2})\right] \mathrm{d}t_2  + \int_{r}^{1} \left[(\tfrac{55}{24} - \tfrac{t_2}{6} - \tfrac{7}{6}t_1) - (\tfrac{5}{2} - \tfrac{4}{3}t_1)\right] \mathrm{d}t_2  \notag \\
&	=  -\tfrac{7}{24} + \tfrac{41}{24}r + \tfrac{r^2}{3} +\tfrac{4}{3}t_1 (\tfrac{1}{8} - r)  \notag \\
&	=  \tfrac{4}{3}\left(t_1 - \tfrac{1}{5}\right)\left(\tfrac{1}{8} - r\right)   -\tfrac{31}{120} + \tfrac{173}{120}r  + \tfrac{r^2}{3}\notag \\
& =  \tfrac{4}{3}\left(t_1 - \tfrac{1}{5}\right)\left(\tfrac{1}{8} - r\right).
\end{align}
The final equality follows from the fact that $r$ satisfies the quadratic equation:
$$
-\tfrac{31}{120} + \tfrac{173}{120}r + \tfrac{1}{3}r^2 = 0.
$$
Since $r > \tfrac{1}{8}$, the difference
$
V_1(2,t_1;g_2^*) - V_1(1,t_1;g_2^*)
$
is strictly decreasing in $t_1$, as seen from the coefficient $\left(\tfrac{1}{6} - \tfrac{4}{3}r\right)$ of $t_1$ being negative for all $r > \tfrac{1}{8}$.
Hence, by Equation~(\ref{exam-c1-2}), the following properties hold: 
\begin{itemize}
	\item player~1 strictly prefers action~2 when $t_1 < \frac{1}{5}$;
	\item player~1 strictly prefers action~1 when $t_1 > \frac{1}{5}$.
\end{itemize}
This preference structure implies that $g_1^*$ remains player~1's best response to $g_2^*$. Consequently, the strategy profile $(g_1^*, g_2^*)$ constitutes a Bayesian Nash equilibrium.

Next, we are going to show that $g^*$ is perfect.
Consider a completely mixed strategy for player $2$, denoted as $g_2^k$, 
\[
g_2^k(t_2) = 
\begin{cases}
\frac{1}{k}\delta_{1} + 	\left(1 - \frac{1}{k}\right)\delta_{2}  &  \text{if } t_2 \in [0, r), \\
	\left(1 - \frac{1}{k}\right)\delta_{1} + \frac{1}{k}\delta_{2} &  \text{if } t_2 \in [r, 1],
\end{cases}
\]
where the positive integer $k \ge 2$ parameterizes the degree of mixing and $\delta_a$ denotes a point mass at action $a$.
Similarly, a perturbation of player $1$'s strategy $g_1^*$ is defined as 
$$g_1^k(t_1)=
\begin{cases}
 \frac{4}{k}\delta_{1}  +	\left(1 - \frac{4}{k}\right) \delta_{2} &    \text{if }t_1 \in [0,\frac{1}{5}), \\
		\left(1 - \frac{1}{k}\right) \delta_{1} + \frac{1}{k}\delta_{2}  &  \text{if } t_1 \in [\frac{1}{5}, 1],
\end{cases}
$$
 for $k \ge 2$.
Clearly, $\lim\limits_{k \to \infty} \rho ( g_i^{k}(t_i), \delta_{g_i^*(t_i)}) = 0$.
A straightforward computation leads to the following identities: 
\begin{align*}
	& V_1(a_1,t_1;g_2^k)  \\
	&= \int_{0}^{r} \Big[ u_1(a_1,2,t_1,t_2)\Big(1 - \tfrac{1}{k}\Big) + u_1(a_1,1,t_1,t_2)\tfrac{1}{k} \Big] \mathrm{d}t_2 \\
	&\quad + \int_{r}^{1} \Big[ u_1(a_1,2,t_1,t_2)\tfrac{1}{k} + u_1(a_1,1,t_1,t_2)\Big(1 - \tfrac{1}{k}\Big) \Big] \mathrm{d}t_2 \\
	&= \int_{0}^{r} \Big( u_1(a_1,2,t_1,t_2) + \Big[u_1(a_1,1,t_1,t_2) - u_1(a_1,2,t_1,t_2)\Big]\tfrac{1}{k} \Big) \mathrm{d}t_2 \\
	&\quad + \int_{r}^{1} \Big( u_1(a_1,1,t_1,t_2) + \Big[u_1(a_1,2,t_1,t_2) - u_1(a_1,1,t_1,t_2)\Big]\tfrac{1}{k} \Big) \mathrm{d}t_2 \\
	&= V_1(a_1,t_1; g_2^*) + \tfrac{1}{k} \int_{0}^{r} \Big[u_1(a_1,1,t_1,t_2) - u_1(a_1,2,t_1,t_2)\Big] \mathrm{d}t_2 \\
	&\quad + \tfrac{1}{k} \int_{r}^{1} \Big[u_1(a_1,2,t_1,t_2) - u_1(a_1,1,t_1,t_2)\Big] \mathrm{d}t_2.
\end{align*}
Since $V_1(2,t_1;g_2^*) - V_1(1,t_1;g_2^*)  =   \tfrac{4}{3}\left(t_1 - \tfrac{1}{5}\right)\left(\tfrac{1}{8} - r\right) $ (see Equation~(\ref{exam-c1-2})),  we can obtain that 
\begin{align*}
	& V_1(2,t_1;g_2^k) - V_1(1,t_1;g_2^k) \\ 
	& = \ \tfrac{1}{k} \int_{0}^{r} \Big[u_1(2,1,t_1,t_2) - u_1(2,2,t_1,t_2)\Big] \mathrm{d}t_2  + \tfrac{1}{k} \int_{r}^{1} \Big[u_1(2,2,t_1,t_2) - u_1(2,1,t_1,t_2)\Big] \mathrm{d}t_2 \\ 
	& \quad - \tfrac{1}{k} \int_{0}^{r} \Big[u_1(1,1,t_1,t_2) - u_1(1,2,t_1,t_2)\Big] \mathrm{d}t_2  - \tfrac{1}{k} \int_{r}^{1} \Big[u_1(1,2,t_1,t_2) - u_1(1,1,t_1,t_2)\Big] \mathrm{d}t_2 \\
	& \quad + V_1(2,t_1;g_2^*) - V_1(1,t_1;g_2^*) \\
	& =   \tfrac{1}{k}M + \tfrac{4}{3}\left(t_1 - \tfrac{1}{5}\right)\left(\tfrac{1}{8} - r\right),
\end{align*}
where 
\begin{align*}
	M & = \  \int_{0}^{r} \Big[u_1(2,1,t_1,t_2) - u_1(2,2,t_1,t_2)\Big] \mathrm{d}t_2  +  \int_{r}^{1} \Big[u_1(2,2,t_1,t_2) - u_1(2,1,t_1,t_2)\Big] \mathrm{d}t_2 \\ 
	& \quad -  \int_{0}^{r} \Big[u_1(1,1,t_1,t_2) - u_1(1,2,t_1,t_2)\Big] \mathrm{d}t_2  -  \int_{r}^{1} \Big[u_1(1,2,t_1,t_2) - u_1(1,1,t_1,t_2)\Big] \mathrm{d}t_2 
\end{align*}
is a finite constant (independent of $k$).
Hence, for sufficiently large $k$, the difference $V_1(2,t_1;g_2^k) - V_1(1,t_1;g_2^k)$ satisfies the following properties: 
	\begin{itemize}
		\item $V_1(2,t_1;g_2^k) - V_1(1,t_1;g_2^k) > 0$ when $t_1 < \tfrac{1}{5}$;
		\item $V_1(2,t_1;g_2^k) - V_1(1,t_1;g_2^k) < 0$ when $t_1 > \tfrac{1}{5}$.
	\end{itemize}
	Consequently, for any $t_1 \in [0,1] \setminus \{\tfrac{1}{5}\}$, 
$g_1^*(t_1) \in \mathrm{BR}_1(t_1,g_{2}^k)$ for sufficiently large $k$. 
We can also obtain that for any $k\ge 2$, 
\begin{align*}
	 V_2(2,t_2;g_1^k) - V_2(1,t_2;g_1^k)  
	= \left[ (-1)\times \tfrac{4}{5} + 7 \times \tfrac{1}{5}  \right] -     \left[ 1 \times \tfrac{4}{5} + (-1) \times \tfrac{1}{5}  \right]  
	 =  0.
\end{align*}
The first equality holds because player~$2$'s payoff is type-independent and player~$1$ follows strategy $g_1^k$, randomizing between actions with $\int_{T_1} g_1^k(t_1,\{1\}) \rmd t_1 =\frac{4}{5}$ and $\int_{T_1} g_1^k(t_1,\{2\}) \rmd t_1 =\frac{1}{5}$.  Hence, we know that for any $k\ge 2$, $g_2^*(t_2) \in \text{BR}_2 (t_2,g_{1}^k)$ for all $t_2 \in [0, 1]$. Thus, $(g_1^*, g_2^*)$ is a perfect equilibrium.

\item[\textbf{4.}] To show that this game does not possess any perfect monotone equilibrium, we only need to prove that the unique monotone equilibrium $(s_1^*, s_2^*)$ is not  perfect. 
%
%
Let $\{g_2^m\}_{m=1}^{\infty}$ be a sequence of completely mixed strategies for player~2 that approximates $s_2^*$ as in Definition~\ref{def-perfect}~a (1), which can be expressed as follows:\footnote{We slightly abuse notation by letting $\{g_2^m\}_{m=1}^{\infty}$ denote the sequence of completely mixed strategies, which is different from the sequence $\{g_2^k\}_{k=1}^{\infty}$ introduced above.}
			\[
			g_2^m(t_2) = 
			\begin{cases}
				(1 - \phi^m(t_2))\delta_{1} + \phi^m(t_2)\delta_{2} & \text{if } t_2 \in [0,\tfrac{7}{8}), \\
			 \phi^m(t_2)\delta_{1} + 	(1 - \phi^m(t_2))\delta_{2} &  \text{if }t_2 \in [\tfrac{7}{8},1],
			\end{cases}
			\]
			where the mixing function $\phi^m: [0,1] \to (0,1)$ satisfies
			$\phi^m(t_2) \to 0$ as $m \to \infty$ for almost all $t_2 \in [0, 1]$.
%
		  A direct computation yields 
\begin{align*}
&	V_1(a_1,t_1;g_2^m)  \\
	&= \int_{0}^{\frac{7}{8}} \Big[ u_1(a_1,1,t_1,t_2)(1 - \phi^m(t_2)) + u_1(a_1,2,t_1,t_2)\phi^m(t_2) \Big] \mathrm{d}t_2 \\
	&\quad + \int_{\frac{7}{8}}^1 \Big[ u_1(a_1,1,t_1,t_2)\phi^m(t_2) + u_1(a_1,2,t_1,t_2)(1 - \phi^m(t_2)) \Big] \mathrm{d}t_2 \\
	&= \int_{0}^{\frac{7}{8}} \Big( u_1(a_1,1,t_1,t_2) + \big[u_1(a_1,2,t_1,t_2) - u_1(a_1,1,t_1,t_2)\big]\phi^m(t_2) \Big) \mathrm{d}t_2 \\
	&\quad + \int_{\frac{7}{8}}^1 \Big( u_1(a_1,2,t_1,t_2) + \big[u_1(a_1,1,t_1,t_2) - u_1(a_1,2,t_1,t_2)\big]\phi^m(t_2) \Big) \mathrm{d}t_2 \\
	&= V_1(a_1,t_1;s_2^*) + \int_{0}^{\frac{7}{8}} \big[u_1(a_1,2,t_1,t_2) - u_1(a_1,1,t_1,t_2)\big]\phi^m(t_2) \mathrm{d}t_2 \\
	&\quad + \int_{\frac{7}{8}}^1 \big[u_1(a_1,1,t_1,t_2) - u_1(a_1,2,t_1,t_2)\big]\phi^m(t_2) \mathrm{d}t_2.
\end{align*}
By applying Equation~(\ref{exam-c1-1}) to the case $x_2 = \tfrac{7}{8}$, we have 
$V_1(2,t_1;s_2^*) - V_1(1,t_1;s_2^*) = 0$, which which leads to the following expression:  
			\begin{align*}
				& V_1(2,t_1;g_2^m) - V_1(1,t_1;g_2^m) \\ 
				&= \int_{0}^{\frac{7}{8}} \Big[u_1(2,2,t_1,t_2) - u_1(2,1,t_1,t_2)\Big]\phi^m(t_2) \mathrm{d}t_2 
			 + \int_{\frac{7}{8}}^{1} \Big[u_1(2,1,t_1,t_2) 
					  - u_1(2,2,t_1,t_2)\Big]\phi^m(t_2) \mathrm{d}t_2 \\
		& \quad	 - \int_{0}^{\frac{7}{8}} \Big[u_1(1,2,t_1,t_2) 
			- u_1(1,1,t_1,t_2)\Big]\phi^m(t_2) \mathrm{d}t_2 
			- \int_{\frac{7}{8}}^{1} \Big[u_1(1,1,t_1,t_2) - u_1(1,2,t_1,t_2)\Big]\phi^m(t_2) \mathrm{d}t_2 \\
				& =  \ \left(\tfrac{41}{24} - \tfrac{4}{3}t_1\right) \left( \int_{0}^{\frac{7}{8}} \phi^m(t_2) \mathrm{d}t_2 - \int_{\frac{7}{8}}^{1} \phi^m(t_2) \mathrm{d}t_2 \right) + \tfrac{2}{3} \left( \int_{0}^{\frac{7}{8}} t_2\phi^m(t_2) \mathrm{d}t_2 - \int_{\frac{7}{8}}^{1} t_2\phi^m(t_2) \mathrm{d}t_2 \right).
			\end{align*}
Define
$
X^m = \int_{0}^{\frac{7}{8}} \phi^m(t_2)\, dt_2 - \int_{\frac{7}{8}}^{1} \phi^m(t_2)\, dt_2$ and 
$Y^m = \int_{0}^{\frac{7}{8}} t_2\, \phi^m(t_2)\, dt_2 - \int_{\frac{7}{8}}^{1} t_2\, \phi^m(t_2)\, dt_2.
$
With these definitions, we have
$$
V_1(2, t_1; g_2^m) - V_1(1, t_1; g_2^m) = \left(\tfrac{41}{24} - \tfrac{4}{3} t_1\right) X^m + \tfrac{2}{3} Y^m.
$$
Since $ t_2 \le \frac{7}{8} $ for $ t_2 \in [0, \frac{7}{8}] $ and $ t_2 > \frac{7}{8} $ for $ t_2 \in (\frac{7}{8}, 1] $, it is clear that
$$
Y^m < \int_{0}^{\frac{7}{8}}\tfrac{7}{8} \phi^m(t_2)\, dt_2 - \int_{\frac{7}{8}}^{1} \tfrac{7}{8} \phi^m(t_2)\, dt_2 = \tfrac{7}{8} X^m. 
$$
\begin{itemize}
\item When $X^m > 0$, $V_1(2,t_1;g_2^m) - V_1(1,t_1;g_2^m)$ is strictly decreasing in $t_1$, which implies the existence of a cutoff type 
$\overline{x}_1^m \in [0,1]$ such that (except for at most one type)
\begin{equation} \label{eq-BR}
 \text{BR}_1(t_1,g_2^m) =  
			\begin{cases}
				\{2\} &   \text{if }t_1 \in [0,\overline{x}_1^m], \\
				\{1\}   &   \text{if }t_1 \in (\overline{x}_1^m, 1].
			\end{cases}
\end{equation}
 
\item  When $X^m \le 0$, we have $Y^m < 0$, $V_1(2,t_1;g_2^m) - V_1(1,t_1;g_2^m) < 0$ since $\left(\tfrac{41}{24} - \tfrac{4}{3} t_1\right) > 0$. Hence, we obtain that $\text{BR}_1(t_1,g_2^m) = \{ 1\}$ for any $t_1 \in [0, 1]$, which is the case $\overline{x}_1^m =0$ in Equation (\ref{eq-BR}). 

\end{itemize}
The sequence of cutoff types $\{\overline{x}_1^m\}_{ m  = 1}^{\infty}$ must have a convergent sequence $\{\overline{x}_1^{m_r}\}_{ r  = 1}^{\infty}$ with the limit $\overline{x}_1^*$. Let
$$ \hat g_1^*(t_1) =  
			\begin{cases}
				2 &  \text{if } t_1 \in [0,\overline{x}_1^*], \\
				1   &  \text{if } t_1 \in (\overline{x}_1^*, 1].
			\end{cases}
		$$ 
Hence, we have for almost all $t_1 \in [0, 1]$,
\begin{equation} \label{eq-BRlimit}
\lim\limits_{r \to \infty} d \left(\hat g_1^*(t_1), \text{BR}_1 (t_1,g_2^{m_r})\right) = 0. 
\end{equation}

Suppose that $(s_1^*, s_2^*)$ is a perfect equilibrium. Then there exists a sequence of completely mixed strategies $\{ g_2^m\}_{ m  = 1}^{\infty}$ that converges to  $s_2^*$ as in Definition~\ref{def-perfect}~a (1), and 
$\lim\limits_{m \to \infty} d \left(s_1^*(t_1), \text{BR}_1 (t_1,g_2^m)\right) = 0$ for almost all $t_1 \in [0, 1]$. 
By the triangle inequality, Equation (\ref{eq-BRlimit}) implies that $\hat g_1^*(t_1) = s_1^*(t_1)$ for almost all $t_1 \in [0, 1]$. 
However, $\hat{g}_1^*$ is a decreasing function, while $s_1^*$ is increasing and takes both values $1$ and $2$ on sets of positive measure in $[0,1]$. This contradiction establishes the result.
\end{enumerate}

The proof for Claim \ref{claim- SCC but not super} is now complete.
\end{proof}

\newpage

\setcounter{page}{1}


\section*{}
\addcontentsline{toc}{section}{Supplemental Appendix}  

\vspace{-1.3 cm}

\begin{center}
\large
\textbf{Supplement to ``Monotone Perfection''   }

\vspace{0.3 cm}

    Wei He,  \, \and Yeneng Sun, \,  \and Hanping Xu 
\end{center}

\setcounter{section}{2}  
\label{appendixB}  

The proof of Theorem~\ref{thm-discontinuous} is provided in Appendix~\ref{proof: discontinuous}.
 In Appendix \ref{proof: Facts}, we present the verifications of Facts~\ref{claim-super}-\ref{fact-no PME type}.
\setcounter{subsection}{0}

\subsection{Proof of  Theorem~\ref{thm-discontinuous}}\label{proof: discontinuous}

We divide the proof into two parts. In Part 1, we show the existence of a perfect monotone equilibrium. In Part 2, we prove that a perfect monotone equilibrium is limit admissible.

\begin{proof}[Proof of Part 1.]

	Let $G$ be a generalized  auction as we stated in Section~\ref{sec-discontinous applications}. We prove this part in a few steps. 
	\begin{itemize}
		\item Step~1.  For each positive integer $m$ with $m \ge 2$, we construct a game $G^m$ that differs from the original game $G$ only in its payoff functions. Specifically, given an action profile $b = (b_1, \ldots, b_n)$, player~$i$'s payoff in $G^m$ is that she receives when each player~$j$ plays $\tilde b_j^m = (1- \frac{1}{m})\delta_{b_j} + \frac{1}{2m}U[0 ,\overline{b}_j] + \frac{1}{2m}\delta_{Q}$ in game $G$, where $U[0,\overline{b}_j]$ is the uniform distribution on $[0,\overline{b}_j]$. That is, 
		$\tilde b_j^m$ is a completely mixed probability measure on $A_j$ putting probability $1- \frac{1}{m}$ on action $b_j$.
		A pure strategy  profile $(\alpha_1,\cdots, \alpha_n)$ in $G^m$ can be transformed as a completely mixed strategy profile $(\tilde \alpha_1^m,\cdots, \tilde \alpha_n^m)$ in $G$.

		Since $G^m$ may have discontinuous payoffs, we carefully further discretize the action spaces in $G^m$: for each $k  \ge 1$, we choose an increasing sequence of finite sets $\{A_i^k\}$ such that $\cup_{k = 1}^{\infty} A_i^k$ is dense and $(\cup_{k = 1}^{\infty} A_i^k) \cap (\cup_{k = 1}^{\infty} A_j^k) = \{Q\}$ for $i \ne j$. Then we obtain a sequence of Bayesian games $\{G^{mk}\}_{k = 1}^{\infty}$. Each game $G^{mk}$ satisfies Assumption~\ref{assum-1}, and hence possesses a monotone equilibrium $g^{mk}$.

		\item Step~2. There exists a subsequence of $\{ g^{mk}\}_{ k = 1}^{\infty}$ that converges to an increasing strategy $g^m$. The aim is to show that $g^m$ is a monotone equilibrium in $G^m$ for each $m \ge 2$. The argument would be trivial in the absence of ties; however, ties introduce discontinuities in the payoffs, making the results non-trivial.

		To handle the discontinuities, the common technique is to show that the probability under ${g}^m$, that two or more players simultaneously play the same highest action above $Q$, is zero. If $V^m$ were the original payoff in our auction game, then this approach would usually consist of the following involved argument.
		
		\begin{enumerate}
			\item If player~$i$'s variable payoff with an action $b_i$ at type $v_i$ is positive, then it is bounded from above by its right limit. It implies that
			$$V_i^m(b_i, v_i; g_{-i}^m) \le \lim\limits_{b_i' \to b_i^{+}} V_i^m(b_i', v_i; g_{-i}^m).
			$$ 
			
			\item For sufficiently large \( k \), the limit \( \lim_{b_i' \to b_i^{+}} V_i^m(b_i', v_i; g_{-i}^m) \) is approximately bounded by \( V_i^m(g_i^{mn_k}(v_i), v_i; g_{-i}^{mn_k}) \).  Importantly, as a by-product, $P_i(g^{mn_k}(v))$, which denotes the winning probability for player $i$ with players adopting action $g^{mn_k}(v)$, converges almost everywhere to some $\eta_i(v)$ taking values from $\{0, 1\}$.

			\item The limit of $V_i^m(g_{i}^{mn_k}(v_i), v_i; g_{-i}^{mn_k})$ is bounded above by $\sup_{b_i \in [0,\overline{b}_i] \cup \{Q\}} V_i^m(b_i, v_i; {g}_{-i}^m)$. Combining the two points above, we know that they are the same. 
			
			\item For almost all type $v$ such that player~$i$ is one of the winners, $\eta_i(v) = 1$ and player~$i$ gets a strictly positive payoff. That is, ties occur with probability zero.
			
			\item Therefore, the monotone equilibrium in $G^m$ is the limit of $\{ g^{mk}\}_{ m = 1}^{\infty}$, which is $g^{m}$.
		\end{enumerate}
		However, this argument cannot be directly applied to our setting. Recall the discussion in Remark~\ref{rmk-dis}. A strategy $\alpha_i$ in $G^m$ corresponds to a completely mixed strategy in $G$, and player~$i$'s payoff is written as the summation of finite many payoff terms. The difficulty is that we need to derive the above approximation for each term in the summation, not only the convergence of (interim) expected payoff in $G^m$.

		Recall that the completely mixed strategy in $G$ corresponding to $\alpha_i$ in $G^m$ involves a mixture of three possible strategies in the original game $G$: $U[0 ,\overline{b}_i]$, $\delta_{Q}$, and the strategy $\alpha_i$ itself. For each realization among the players $-i$, we divide these players other than $i$ into three subgroups: $I_1 = \{ j \ne i \colon \alpha_j = U[0,\overline{b}_j] \}$, consisting of players employing a uniform distribution over their action sets; $I_2 =   \{ j \neq i \colon \alpha_j = \delta_{ Q }\}$,  which includes players with a degenerate strategy placing all probability mass on $Q$; and $I_3 =  I\backslash (I_1 \cup I_2 \cup \{i\}) $, which contains the remaining players. Here, $I_1$, $I_2$, and $I_3$ form a partition of those players. One observation is that instead of studying the above approximation for each term in the summation, it suffices to study the approximation for each partition element.  We demonstrate this can be done in a coherent manner.

		\item Step~3. We obtain a monotone strategy $g$ as a limit of a subsequence of $\{g^m\}_{m=1}^{\infty}$. By a similar argument, we prove that $g$ is a monotone equilibrium in the game $G$.  
	\end{itemize}

	\

	\noindent \textbf{Step 1. }
	For each positive integer $m \geq 2$, let $\tilde{b}_i^m$ denote a completely mixed probability measure on $A_i = [0, \overline{b}_i] \cup \{Q\}$, which assigns probability $1 - \frac{1}{m}$ to the pure action $b_i$:
	\begin{equation}\label{eq:tilde_b_i^m}
		\tilde{b}_i^m = \left(1 - \tfrac{1}{m}\right)\delta_{b_i} + \tfrac{1}{2m} U[0, \overline{b}_i] + \tfrac{1}{2m} \delta_Q,
	\end{equation}
	where $U[0, \overline{b}_i]$ denotes the uniform distribution on the interval $[0, \overline{b}_i]$.

	Let \( B_1 \subseteq I \) denote the set of players who each adopt a mixed action given by the uniform distribution \( U[0, \overline{b}_j] \); that is, for each \( j \in B_1 \), player \( j \) selects \( U[0, \overline{b}_j] \). Let \( B_2 \subseteq I \setminus B_1 \) be the set of players who choose the pure action \( Q \), so that each \( j \in B_2 \) plays \( Q \). The remaining players comprise the set \( B_3 = I \setminus (B_1 \cup B_2) \), where each player \( j \in B_3 \) plays the pure action \( b_j \).
	Let $|K|$ denote the cardinality (i.e., the number of elements) of a set $K$.
	
	In the perturbed game $G^m$, the payoff function for player $i$ is defined as
	\begin{equation*}
		\begin{aligned}
			u_i^m(b, v) &= u_i(\tilde{b}^m, v) \\
			&= \sum_{B_1 \subseteq I} \sum_{B_2 \subseteq I \setminus B_1}
			\left( \prod_{j \in B_1} \tfrac{1}{\overline{b}_j} \right)
			\left(1 - \tfrac{1}{m} \right)^{|B_3|}
			\left( \tfrac{1}{2m} \right)^{|B_1| + |B_2|}  \\
			&\quad \times \int_{\mathop{\prod}_{j \in B_1} [0, \overline{b}_j]}
			u_i\left(\hat b_{B_1}, Q_{B_2}, b_{B_3}, v \right)
			\mathop{\otimes}_{j \in B_1} \mathrm{d}\hat b_j,
		\end{aligned}
	\end{equation*}
	where $\hat b_{B_1} = (\hat b_j)_{j \in B_1}$, $Q_{B_2} = (Q)_{j \in B_2}$, and $ b_{B_3} = ( b_j)_{j \in B_3}$. Note that $B_1$, $B_2$, and $B_3$ form a partition of $I$, and the summation is taken over all such partitions.
	
	It is easy to verify that $u_i^m$ is bounded and jointly measurable. Moreover, the Prohorov metric between $\tilde{b}_i^m$ and the Dirac measure $\delta_{b_i}$ satisfies
	$
	\rho(\tilde{b}_i^m, \delta_{b_i}) \to 0
	$
	as $m \to \infty$.
	Clearly, $u_i^m$ converges pointwise to $u_i$ as $m \to \infty$.

	Fix an increasing strategy profile $\alpha$. For each $m \geq 2$ and each player $i \in I$, consider a completely mixed strategy profile $\tilde{\alpha}_i^m$, where $\tilde{\alpha}_i^m(v_i)$ assigns the completely mixed probability measure $\widetilde{\alpha_i(v_i)}^m$ over the action set $A_i$. 
	By Equation~\eqref{eq:tilde_b_i^m}, the strategy $\tilde{\alpha}_i^m(v_i)$ places probability $1 - \frac{1}{m}$ on the pure action $\alpha_i(v_i)$, probability $\frac{1}{2m}$ on the action $Q$, and probability $\frac{1}{2m}$ on the uniform distribution over $[0, \overline{b}_i]$.
	
	Using a computation similar to that in Equation~\eqref{equa-v_im}, we obtain the following identity for player $i$'s interim payoff in game $G^m$:
	\begin{align}\label{eq:Vim}
		V_i^m(b_i, v_i; \alpha_{-i}) = V_i(\tilde{b}_i^m, v_i; \tilde{\alpha}_{-i}^m).    
	\end{align}

	Let \( I_1 \subseteq I \setminus \{i\} \) denote the set of players who follow the mixed strategy given by the uniform distribution \( U[0, \overline{b}_j] \) for all types. Let \( I_2 \subseteq I \setminus (I_1 \cup \{i\}) \) be the set of players who adopt the pure strategy \( Q \), i.e., each player \( j \in I_2 \) chooses \( Q \) for all types. The remaining players form the set \( I_3 = I \setminus (I_1 \cup I_2 \cup \{i\}) \), where each player \( j \in I_3 \) follows the monotone strategy \( \alpha_j \).
	It is easy to see that \( I_1 \), \( I_2 \), and \( I_3\) form a partition of the set \( I \setminus \{i\} \).	
	By Equation~\eqref{eq:Vim}, we can derive the following expression for player $i$'s interim payoff function in game $G^{m}$:
	\begin{align}\label{equa:general_payoff}
		V_i^m(b_i, v_i; \alpha_{-i}) 
		& = \left(1 - \tfrac{1}{m}\right) V_i(b_i, v_i; \tilde{\alpha}_{-i}^m) +  \tfrac{1}{2m} \cdot  \tfrac{1}{\overline{b}_i } \int_{[0,\overline{b}_i]}V_i( \hat b_i, v_i, \tilde \alpha^m_{-i}) \rmd \hat b_i \notag \\ 
		&= \left(1 - \tfrac{1}{m}\right) V_i(b_i, v_i; \tilde{\alpha}_{-i}^m) + R_i^m(v_i, \alpha_{-i}) \notag \\
		&= \sum_{I_1 \subseteq I \backslash \{i\}} \sum_{I_2 \subseteq I \setminus \left( I_1 \cup \{i\} \right) }
		\left( \prod_{j \in I_1} \tfrac{1}{\overline{b}_j} \right)
		\left(1 - \tfrac{1}{m} \right)^{|I_3| + 1 }
		\left( \tfrac{1}{2m} \right)^{|I_1| + |I_2|}  \notag \\
		&\quad \times  \int_{ \prod_{j \in I_1} [0, \overline{b}_j]}  V_i(b_i,v_i; \hat b_{I_1},  Q_{I_2}, \alpha_{I_3}) 
		\mathop{\otimes}_{j \in I_1} \mathrm{d}\hat b_j   +     R_i^m(v_i, \alpha_{-i}),
	\end{align}
	where  $\hat b_{I_I} = (\hat b_j)_{j \in I_I}$, $Q_{I_2} = (Q)_{j \in I_2}$,  $\alpha_{I_3} = (\alpha_j)_{j \in I_3}$ and $R_i^m(v_i, \alpha_{-i}) =  \frac{1}{2m} \int_{[0,\overline{b}_i]} \frac{1}{\overline{b}_i }V_i( \hat b_i, v_i, \tilde \alpha^m_{-i}) \rmd \hat b_i.
	$  
	We slightly abuse notation by interpreting any action \( \hat b_j \in [0,\overline{b}_j] \cup \{Q\} \) as a strategy in which player \( j \) selects the same action \( \hat b_j \) for every type \( t_j \in [0,1] \).
	
	Next, we show that for each \( i \in I \), the function \( V_i^m(b_i, v_i; \alpha_{-i}) \) satisfies the increasing differences condition in  \( (b_i, v_i) \), for any monotone strategy profile \( \alpha_{-i} \in \mathcal{F}_{-i} \).

	For any  $ b_i^H, b_i^L \in  A_i$ with $b_i^H > b_i^L$,  by Equation~\eqref{equa:general_payoff}, we have
	\begin{align}\label{eq-v^H-v^L}
		& \quad V_i^m(b_i^H,v_i;\alpha_{-i}) - V_i^m(b_i^L,v_i;\alpha_{-i})  \notag \\
		& =  	 \sum_{I_1 \subseteq I \backslash \{i\}} \sum_{I_2 \subseteq I \setminus \left( I_1 \cup \{i\} \right) }
		\left( \prod_{j \in I_1} \tfrac{1}{\overline{b}_j} \right)
		\left(1 - \tfrac{1}{m} \right)^{|I_3| + 1 }
		\left( \tfrac{1}{2m} \right)^{|I_1| + |I_2|}  \notag \\
		&\quad \times  \int_{ \prod_{j \in I_1} [0, \overline{b}_j]}  \left[V_i(b_i^H,v_i; \hat b_{I_1}, Q_{I_2},  \alpha_{I_3}) - V_i(b_i^L,v_i; \hat b_{I_1}, Q_{I_2}, \alpha_{I_3})  \right]
		\mathop{\otimes}_{j \in I_1} \mathrm{d}\hat b_j   \notag  \\
		&	=  \sum_{I_1 \subseteq I \backslash \{i\}} \sum_{I_2 \subseteq I \setminus \left( I_1 \cup \{i\} \right) }
		\left( \prod_{j \in I_1} \tfrac{1}{\overline{b}_j} \right)
		\left(1 - \tfrac{1}{m} \right)^{|I_3| + 1 }
		\left( \tfrac{1}{2m} \right)^{|I_1| + |I_2|}  \notag \\
		&	\int_{\prod\limits_{j \in I_1}[0, \overline{b}_j]} 
		\int_{[0,1]^{n-1}}  \left[u_i(b_i^H,  b_{I_1}, Q_{I_2}, \alpha_{I_3}(v_{I_3}), v) - u_i(b_i^L, b_{I_1}, Q_{I_2}, \alpha_{I_3}(v_{I_3}), v)\right]   f(v_{-i}|v_i) \mathrm{d} v_{-i}  \mathop{\otimes}_{j \in I_1} \mathrm{d} b_{j}.
	\end{align}

	Recall that the payoff function is given by:
	\[
	u_i(b, v) = \bigl( W_i(b_i, v) - C_i(b_i) \bigr) D(b) P_i(b) - \Phi_i(b_i).
	\]
	For notational simplicity, let \( \hat{P}_i(b) = D(b) P_i(b) \). Then, for any \( b_{-i} \in A_{-i} \) and \( v_{-i} \in [0,1]^{n-1} \), we have:
	\begin{align*}
		& u_i(b_i^H, b_{-i}, v) - u_i(b_i^L, b_{-i}, v) \\
		&= \underbrace{C_i(b_i^L) \hat{P}_i(b_i^L, b_{-i}) - C_i(b_i^H) \hat{P}_i(b_i^H, b_{-i}) + \Phi_i(b_i^L) - \Phi_i(b_i^H)}_{\text{(P1)}} \\
		&\quad + \underbrace{W_i(b_i^H, v) \bigl(\hat{P}_i(b_i^H, b_{-i}) - \hat{P}_i(b_i^L, b_{-i}) \bigr)}_{\text{(P2)}} +\underbrace{ \hat{P}_i(b_i^L, b_{-i}) \bigl( W_i(b_i^H, v) - W_i(b_i^L, v) \bigr)}_{\text{(P3)}}.
	\end{align*}
	
	\begin{itemize}
		\item The term \textnormal{(P1)} does not depend on \( v_i \).
		
		\item Since both \( D(b) \) and \( P_i(b_i, b_{-i}) \) are nonnegative and increasing in \( b_i \), it follows that \( \hat{P}_i(b_i, b_{-i}) \) is increasing in \( b_i \). Hence,
		$
		\hat{P}_i(b_i^H, b_{-i}) - \hat{P}_i(b_i^L, b_{-i}) \geq 0.
		$
		By Assumption~\ref{assum-d1} (1), the term (P2)		is increasing in \( v_i \).
		
		\item By Assumption~\ref{assum-d1} (4), the term 	(P3)		is also increasing in \( v_i \).		
	\end{itemize}

	Taking expectations, we conclude that 
	$
	V_i^m(b_i^H, v_i; \alpha_{-i}) - V_i^m(b_i^L, v_i; \alpha_{-i})
	$
	is increasing in \( v_i \). Therefore, \( V_i^m(b_i, v_i; \alpha_{-i}) \) satisfies the increasing differences condition  in \( (b_i, v_i) \) for any \( \alpha_{-i} \in \mathcal{F}_{-i} \).

	For each \( k \in \mathbb{Z}_{+} \) and player \( i \in I \), let \( A_i^k \subseteq [0, \overline{b}_i] \cup \{Q\} \) be a finite set such that \( \bigcup_{k = 1}^{\infty} A_i^k \) is dense in \( [0, \overline{b}_i] \cup \{Q\} \), with \( A_i^k \subseteq A_i^{k+1} \) for all \( k \). Assume also that for any \( i \neq j \), the intersection
	$
	( \cup_{k = 1}^{\infty} A_i^k ) \cap ( \cup_{k = 1}^{\infty} A_j^k ) = \{Q\}.
	$
	In game \( G^{mk} \), let \( u_i^m \) denote player \( i \)'s payoff, and let \( A_i^k \) be the corresponding (finite) action set for player \( i \). It is clear that \( V_i^m \) is player \( i \)'s interim payoff function in game \( G^{mk} \), and that \( V_i^m \) satisfies the increasing differences condition. By \citet[Theorem 1]{A2001}, each game \( G^{mk} \) admits a monotone equilibrium \( g^{mk} \).

	\

	\noindent \textbf{Step 2. }   
	By Helly's selection theorem, there exists a subsequence \( \{g^{mn_k}\}_{k=1}^{\infty} \) of \( \{g^{mk}\}_{k=1}^{\infty} \) that converges pointwise to a monotone strategy profile \( g^{m} \) for almost every \( v \in [0,1]^n \). That is,
	$
	\lim_{k \to \infty} d\big(g_i^{mn_k}(v_i), g_i^m(v_i)\big) = 0
	$
	for each player \( i \in I \) and almost every \( v_i \in [0,1] \). 
	We now show that \( g^m \) constitutes a monotone equilibrium in the game \( G^m \) for each \( m \).

	\bigskip
	
	In what follows, we restrict attention to such \( v_i \) for which the above pointwise convergence holds.
	We begin by characterizing player $i$'s maximum interim payoff at type $v_i \in [0,1]$ in the modified game $G^m$, when opponents employ monotone strategies $g_{-i}^m$. As preliminary analysis, we first examine the corresponding maximum interim payoff in the original game $G$, maintaining the assumption that opponents play according to $g_{-i}^m$.

	If $W_i(b_i,v) - C_i(b_i) \leq 0$ for all $v_{-i}$, then by Assumption~\ref{assum-d1} (2), we have $\Phi_i(b_i) \geq 0$. In this case, player~$i$'s interim payoff from playing $b_i$ against $g_{-i}^m$ at type $v_i$ is at most $0$, which is no better than the payoff from choosing $Q$ (i.e., quitting the game and receiving $0$). 
	We therefore focus on actions $b_i$ at type $v_i$ for which $W_i(b_i,v) - C_i(b_i) > 0$ holds for some $v_{-i}$. By Assumption~\ref{assum-d1} (5), this implies $W_i(b_i,v) - C_i(b_i) > 0$ for all $v_{-i} \in [0,1]^{n-1}$.

	Let 
	$
	\Omega_i = \{ v_{-i} \in [0,1]^{n-1} \,|\, \max_{j \ne i} g_j^m(v_j) \le b_i \}
	$
	denote the set of opponents' type profiles for which player~$i$ wins the auction by playing action $b_i$ against strategies $g_{-i}^m$.  
	We establish the following chain of inequalities for the interim payoff:
	\begin{align}\label{equa-general right limit}
		& \quad	V_i(b_i,v_i; g^m_{-i})  \notag \\
		&= \mathbb{P}(\Omega_i \mid v_i)\, \mathbb{E}\left[(W_i(b_i,v) - C_i(b_i))\, D(b_i, g^m_{-i}(v_{-i}))\, P_i(b_i, g^m_{-i}(v_{-i})) \,\big|\, v_i, \Omega_i\right] - \Phi_i(b_i) \notag \\
		&\leq \mathbb{P}(\Omega_i \mid v_i)\, \mathbb{E}\left[(W_i(b_i,v) - C_i(b_i))\, D(b_i, g^m_{-i}(v_{-i})) \,\big|\, v_i, \Omega_i\right]
		\mathbb{E}\left[P_i(b_i, g^m_{-i}(v_{-i})) \,\big|\, v_i, \Omega_i\right] - \Phi_i(b_i) \notag \\
		&\leq \mathbb{P}(\Omega_i \mid v_i)\, \mathbb{E}\left[(W_i(b_i,v) - C_i(b_i))\, D(b_i, g^m_{-i}(v_{-i})) \,\big|\, v_i, \Omega_i\right] - \Phi_i(b_i) \notag \\
		&= \lim_{b_i' \to b_i^+} V_i(b_i', v_i; g^m_{-i}).
	\end{align}
	Combining the condition \( W_i(b_i, v) - C_i(b_i) > 0 \) for all \( v_{-i} \in [0,1]^{n-1} \) with Assumption~\ref{assum-d1}(3), we conclude that both terms
	\[
	(W_i(b_i, v) - C_i(b_i))\, D(b_i, g_{-i}^m(v_{-i})) \quad \text{and} \quad 1 - P_i(b_i, g_{-i}^m(v_{-i}))
	\]
	are increasing and non-negative in \( v_{-i} \). Therefore, the first inequality in~\eqref{equa-general right limit} follows from \citet[Theorem 23]{MW1982}.
	The second inequality holds since the conditional expectation \( \mathbb{E}[P_i(b_i, g_{-i}^m(v_{-i})) \mid v_i, \Omega_i] \in [0,1] \), while the other term
	$
	\mathbb{E}[(W_i(b_i, v) - C_i(b_i))\, D(b_i, g_{-i}^m(v_{-i})) \mid v_i, \Omega_i]
	$ is non-negative.
	The final equality follows from two observations:
	\begin{itemize}
		\item[(i)] For all \( v_{-i} \in \Omega_i \), we have \( P_i(b_i', g_{-i}^m(v_{-i})) = 1 \) for any \( b_i' > b_i \).
		\item[(ii)] For all \( v_{-i} \notin \Omega_i \), \( P_i(b_i', g_{-i}^m(v_{-i})) \to 0 \) as \( b_i' \to b_i^+ \).
	\end{itemize}
	By the same argument, Inequality~\eqref{equa-general right limit} also holds for any monotone strategy profile \( \alpha_{-i} \in \mathcal{F}_{-i} \), namely,
	\begin{align}\label{equa-general right limit 2}
		V_i(b_i, v_i; \alpha_{-i}) \le \lim_{b_i' \to b_i^{+}} V_i(b_i', v_i; \alpha_{-i}).
	\end{align}
	
	

	We slightly abuse notation by using \( I_1, I_2 \), and \( I_3 \) to denote a partition of \( I \setminus \{i\} \), defined as follows. Let \( I_1 \subseteq I \setminus \{i\} \) denote the set of players who follow the mixed strategy given by the uniform distribution \( U[0, \overline{b}_j] \) for all types. Let \( I_2 \subseteq I \setminus (I_1 \cup \{i\}) \) be the set of players who adopt the pure strategy \( Q \), i.e., each player \( j \in I_2 \) chooses \( Q \) for all types. The remaining players form the set \( I_3 = I \setminus (I_1 \cup I_2 \cup \{i\}) \), where each player \( j \in I_3 \) follows the monotone strategy \( g^m_j \).
	By Equation~\eqref{equa:general_payoff}, we obtain
	\begin{align*}
		V_i(b_i, v_i; \tilde{g}_{-i}^m) 
		&= \sum_{I_1 \subseteq I \setminus \{i\}} \sum_{I_2 \subseteq I \setminus \left( I_1 \cup \{i\} \right)}
		\left( \prod_{j \in I_1} \tfrac{1}{\overline{b}_j} \right)
		\left(1 - \tfrac{1}{m} \right)^{|I_3| + 1}
		\left( \tfrac{1}{2m} \right)^{|I_1| + |I_2|} \\
		&\quad \times \int_{\prod_{j \in I_1} [0, \overline{b}_j]} V_i(b_i, v_i;\hat{b}_{I_1}, Q_{I_2}, g_{I_3} ) 
		\mathop{\otimes}_{j \in I_1} \mathrm{d} \hat{b}_j,
	\end{align*}
	where \( \hat{b}_{I_1} = (\hat{b}_j)_{j \in I_1} \),
	\( Q_{I_2} = (Q)_{j \in I_2} \), and   \( g^m_{I_3} = (g^m_j)_{j \in I_3} \).
	
	Since \( V_i \) is bounded, it follows from the dominated convergence theorem and Inequality~\eqref{equa-general right limit 2} that
	\begin{align}\label{equa-general monotone right limit}
		V_i(b_i, v_i; \tilde{g}^m_{-i}) \le \lim_{b_i' \to b_i^{+}} V_i(b_i', v_i; \tilde{g}^m_{-i}).
	\end{align}
	It then follows from Equation~\eqref{equa:general_payoff} that
	\begin{align}\label{equa-general payoff right limit}
		V_i^m(b_i, v_i; g_{-i}^m) \le \lim_{b_i' \to b_i^{+}} V_i^m(b_i', v_i; g_{-i}^m).
	\end{align}
	Since $Q$ is an isolated point in the action space, Inequality~\eqref{equa-general payoff right limit} trivially holds when $b_i = Q$ by continuity of the payoff function at this point.

	\bigskip

	We proceed to show that \( \lim_{b_i' \to b_i^+} V_i^m(b_i', v_i; g_{-i}^m) \) can be approximated from above by  \(V_i^m(g_i^{mn_k}(v_i), v_i; g_{-i}^{mn_k}) \) for sufficiently large \( k \). 
	
	To that end, fix any \( b_i \in [0, \overline{b}_i] \).
	By Inequality~\eqref{equa-general monotone right limit}, for any \( \epsilon > 0 \), there exists \( \delta^\epsilon > 0 \) such that for all \( \hat{b}_i \in (b_i, b_i + \delta^\epsilon) \),
	\[
	\lim_{b_i' \to b_i^+} V_i^m(b_i', v_i; g_{-i}^m) \le V_i^m(\hat{b}_i, v_i; g_{-i}^m) + \epsilon.
	\]
	Since each action set \( A_j^k \) is finite and becomes dense in \( [0, \overline{b}_j] \cup \{Q\} \) as \( k \to \infty \), and each player $j$'s strategy \( g_j^m \) has at most countably many mass points, we can choose a point \( \hat{b}_i \in (b_i, b_i + \delta^\epsilon) \) such that \( \hat{b}_i \notin \bigcup_{k=1}^\infty A_i^k \) and \( \hat{b}_i \) is not a mass point of any \( g_j^m \). Moreover, there exists a sequence \( \hat{b}_i^{n_k} \in A_i^{n_k} \) such that \( \lim_{k \to \infty} |\hat{b}_i^{n_k} - \hat{b}_i| = 0 \). 
	Then, for sufficiently large \( K \in \mathbb{Z}_+ \) and all \( k \ge K \), we have
	\begin{equation}\label{equa-general r1}
		\lim_{b_i' \to b_i^+} V_i^m(b_i', v_i; g_{-i}^m) 
		\le V_i^m(\hat{b}_i, v_i; g_{-i}^m) + \epsilon 
		\le V_i^m(\hat{b}_i^{n_k}, v_i; g_{-i}^{mn_k}) + 2\epsilon 
		\le V_i^m(g_i^{mn_k}(v_i), v_i; g_{-i}^{mn_k}) + 2\epsilon,
	\end{equation}
	where the last inequality holds because \( \hat{b}_i^{n_k} \in A_i^{n_k} \) is a feasible action in \( G^{mn_k} \), and \( g^{mn_k} \) is a Bayesian Nash equilibrium in that game. The Inequality~\eqref{equa-general r1} holds trivially when \( b_i = Q \).
	
	\

	We now show that the sequence \( \{V_i^m(g_i^{mn_k}(v_i), v_i; g_{-i}^{mn_k}) \}_{k = 1}^{\infty} \) admits a convergent subsequence, which we denote by the same notation for simplicity.
	
	For any partition \( I_1 \), \( I_2 \), and \( I_3 \) of \( I \setminus \{i\} \), suppose \( g_i^{mn_k}(v_i) \in [0, \overline{b}_i] \). A straightforward computation yields:
	\begin{align}\label{equa-general r3}
		&  \prod_{j \in I_1} \tfrac{1}{\overline{b}_j} \,  \int_{\prod_{j \in I_1}[0, \overline{b}_j]} V_i(g_i^{mn_k}(v_i), v_i; \hat b_{I_1}, Q_{I_2}, g^{mn_k}_{I_3}) 
		\mathop{\otimes}_{j \in I_1} \mathrm{d}\hat b_j \notag \\
		= & \prod_{j \in I_1} \tfrac{1}{\overline{b}_j} \,  \int_{\prod_{j \in I_1} [0, \min\{g_i^{mn_k}(v_i), \overline{b}_j\}]} 
		\mathbb{E}\big[ (W_i(g_i^{mn_k}(v), v) - C_i(g_i^{mn_k}(v_i))) \,   D(g_i^{mn_k}(v_i), \hat b_{I_1}, Q_{I_2}, g^{mn_k}_{I_3}(v_{I_3})) \notag \\
		& \hspace{4em} \times 
		\, P_i(g_i^{mn_k}(v_i), \hat b_{I_1}, Q_{I_2}, g^{mn_k}_{I_3}(v_{I_3})) \,\big|\, v_i \big] 
		\mathop{\otimes}_{j \in I_1} \mathrm{d}\hat b_j - \Phi_i(g_i^{mn_k}(v_i)) \notag \\
		= & \prod_{j \in I_1} \tfrac{1}{\overline{b}_j} \, \int_{\prod_{j \in I_1} [0, \min\{g_i^{mn_k}(v_i), \overline{b}_j\}]} 
		\mathbb{E}\big[ (W_i(g_i^{mn_k}(v), v) - C_i(g_i^{mn_k}(v_i)))  \,  D(g_i^{mn_k}(v_i), \hat b_{I_1}, Q_{I_2}, g^{mn_k}_{I_3}(v_{I_3})) \notag \\
		& \hspace{4em} \times 
		\, P_i(g_i^{mn_k}(v_i), Q_{I_1}, Q_{I_2}, g^{mn_k}_{I_3}(v_{I_3})) \,\big|\, v_i \big] 
		\mathop{\otimes}_{j \in I_1} \mathrm{d}\hat b_j - \Phi_i(g_i^{mn_k}(v_i)).
	\end{align}
	The first equality holds because if any player \( j \in I_1 \) chooses \(\hat b_j > g_i^{mn_k}(v_i) \), then player \( i \) cannot win, i.e., 
	$
	P_i(g_i^{mn_k}(v_i), \hat b_{I_1}, Q_{I_2}, g^{mn_k}_{I_3}(v_{I_3})) = 0.
	$
	The second equality holds since whenever \( \max_{j \in I_1} \hat b_j < g_i^{mn_k}(v_i) \), the winning probability simplifies to
	\[
	P_i(g_i^{mn_k}(v_i), \hat b_{I_1}, Q_{I_2}, g^{mn_k}_{I_3}(v_{I_3})) = P_i(g_i^{mn_k}(v_i), Q_{I_1}, Q_{I_2}, g^{mn_k}_{I_3}(v_{I_3})).
	\]
	Thus, for \( \max_{j \in I_1} \hat b_j < g_i^{mn_k}(v_i) \), we have
	\begin{align*}
		&  \quad V_i(g_i^{mn_k}(v_i), v_i; \hat b_{I_1}, Q_{I_2}, g^{mn_k}_{I_3}(v_{I_3})) \\
		&= \mathbb{E}\big[ \big( W_i(g_i^{mn_k}(v_i), v) - C_i(g_i^{mn_k}(v_i)) \big)\, 
		D(g_i^{mn_k}(v_i), \hat b_{I_1}, Q_{I_2}, g^{mn_k}_{I_3}(v_{I_3})) \\
		&\qquad \times P_i(g_i^{mn_k}(v_i), Q_{I_1}, Q_{I_2}, g^{mn_k}_{I_3}(v_{I_3})) \,\big|\, v_i \big]
		- \Phi_i(g_i^{mn_k}(v_i)).
	\end{align*}
	Since the probability that player \( i \) wins, \( P_i(b_i, b_{-i}) \), is increasing in \( b_i \) and decreasing in \( b_{-i} \), each function 
	$
	P_i(g_i^{mn_k}(v_i), Q_{I_1}, Q_{I_2}, g^{mn_k}_{I_3}(v_{I_3}))
	$
	is monotone in each component of \( v = (v_1, \ldots, v_n) \): increasing in \( v_i \) and decreasing in \( v_{-i} \). By Helly's selection theorem (extracting a subsequence if necessary), there exists a function \( \eta_i^{I_3} \colon [0,1]^n \to [0,1] \) such that
	\[
	P_i(g_i^{mn_k}(v_i), Q_{I_1}, Q_{I_2}, g^{mn_k}_{I_3}(v_{I_3})) \to \eta_i^{I_3}(v)
	\]
	almost everywhere, and $\eta^{I_3}$ is increasing in \( v_i \) and decreasing in \( v_{-i} \). By Assumption~\ref{assum-d1} (1)--(3), the functions \( W_i, C_i, D \) are continuous. Hence, by the dominated convergence theorem, for almost all \( v \in [0,1]^n \),
	\begin{align}\label{equa - general converges}
		&\mathbb{E}\big[ (W_i(g_i^{mn_k}(v_i), v) - C_i(g_i^{mn_k}(v_i)))\,
		D(g_i^{mn_k}(v_i), \hat  b_{I_1}, Q_{I_2}, g^{mn_k}_{I_3}(v_{I_3})) \notag \\
		&\qquad \times P_i(g_i^{mn_k}(v_i), Q_{I_1}, Q_{I_2}, g^{mn_k}_{I_3}(v_{I_3})) \,\big|\, v_i \big] \notag \\
		\to\;& \mathbb{E}\big[ (W_i(g_i^m(v_i), v) - C_i(g_i^m(v_i)))\, 
		D(g_i^m(v_i), \hat b_{I_1}, Q_{I_2}, g^m_{I_3}(v_{I_3}))\, \eta_i^{I_3}(v) \,\big|\, v_i \big].
	\end{align}
	Since the sets \( \{A_i^k\} \) are disjoint, we have \( P_i(g_i^{mn_k}(v_i), Q_{I_1}, Q_{I_2}, g^{mn_k}_{I_3}(v_{I_3})) \in \{0,1\} \) for all \( n_k \) and \( v \). It follows that \( \eta_i^{I_3}(v) \in \{0,1\} \) for almost all \( v \in [0,1]^n \). One may interpret \( \eta_i^{I_3} \) as a tie-breaking rule, satisfying \( \sum_{i=1}^n \eta_i^{I_3}(v) \le 1 \) almost everywhere.

	Moreover, since \( I_3 \subseteq I \setminus \{i\} \) ranges over finitely many subsets, there exists a common subsequence (denoted the same for simplicity) such that the convergence in Equation~\eqref{equa - general converges} holds for all such \( I_3 \).
	
	Applying \( b_i = g_i^{mn_k}(v_i) \) in Equation~\eqref{equa:general_payoff} and taking the limit as \( k \to \infty \), Equations~\eqref{equa-general r3} and \eqref{equa - general converges} together imply that, for almost all \( t_i \), we have
	\begin{align}\label{eq-limVim}
		\lim_{k \to \infty} V_i^m\big( g_i^{mn_k}(v_i), v_i; g_{-i}^{mn_k} \big)
		&= \sum_{I_1 \subseteq I \setminus \{i\}} \sum_{I_2 \subseteq I \setminus (I_1 \cup \{i\})}
		\left( \prod_{j \in I_1} \tfrac{1}{\overline{b}_j} \right)
		\left(1 - \tfrac{1}{m} \right)^{|I_3| + 1}
		\left( \tfrac{1}{2m} \right)^{|I_1| + |I_2|} \notag \\
		&\quad \times \int_{\prod_{j \in I_1} [0, \min\{g_i^m(v_i), \overline{b}_j\}]} \mathbb{E} \Big[ 
		(W_i(g_i^m(v), v) - C_i(g_i^m(v_i))) \notag \\
		&\quad \times D\big(g_i^m(v_i), \hat{b}_{I_1}, Q_{I_2}, g^m_{I_3}(v_{I_3})\big)
		\, \eta^{I_3}(v) \,\big|\, v_i \Big] \notag \\
		&\quad - (1 - \tfrac{1}{m}) \Phi_i(g_i^m(v_i)) + R_i^m(v_i, g_{-i}^m),
	\end{align}
	where
	\begin{align}\label{eq-Rim}
		R_i^m(v_i, g_{-i}^m)       
		&= \lim_{k \to \infty} R_i^m(v_i, g_{-i}^{mn_k})  \notag \\
		&= \lim_{k \to \infty} \tfrac{1}{2m} \cdot \tfrac{1}{\overline{b}_i} \int_{[0, \overline{b}_i]} V_i(\hat{b}_i, v_i, \tilde{g}_{-i}^{mn_k}) \, \mathrm{d} \hat{b}_i \notag \\
		&= \tfrac{1}{2m} \cdot \tfrac{1}{\overline{b}_i} \int_{[0, \overline{b}_i]} V_i(\hat{b}_i, v_i, \tilde{g}_{-i}^{m}) \, \mathrm{d} \hat{b}_i.
	\end{align}
	The existence of the limit \( \{ R_i^m(v_i, g_{-i}^{mn_k}) \}_{k=1}^{\infty} \) follows from the fact that each function in the set \( \{ g_j^{mn_k} \}_{k \in \mathbb{Z}_+, j \ne i} \) has at most countably many mass points. Therefore, these mass points (excluding \( Q \)) form a measure-zero subset of \( [0, \overline{b}_j] \). As a result, we have \( \lim_{k \to \infty} V_i(\hat{b}_i, v_i, \tilde{g}_{-i}^{mn_k}) = V_i(\hat{b}_i, v_i, \tilde{g}_{-i}^m) \) for almost all \( \hat{b}_i \in [0, \overline{b}_i] \). Hence, the limit exists and equals \( R_i^m(v_i, g_{-i}^m) \). It is straightforward to verify that Equation~\eqref{eq-limVim} also holds in the case where \( \lim\limits_{k \to \infty} g_i^{mn_k}(v_i) = g_i^m(v_i) = Q \).

	
	\bigskip
	
	Combining Inequality~\eqref{equa-general r1} with Equation~\eqref{eq-limVim} and taking \( \epsilon \to 0 \), we obtain
	\begin{align}\label{inequa-<=}
		\sup_{b_i \in [0, \overline{b}_i] \cup \{Q\}} V_i^m(b_i, v_i; g_{-i}^m) \le \lim_{k \to \infty} V_i^m\big( g_i^{mn_k}(v_i), v_i; g_{-i}^{mn_k} \big).
	\end{align}
	We will now show that this inequality is in fact binding. Since the inequality holds with equality trivially when \( \lim_{k \to \infty} g_i^{mn_k}(v_i) = g_i^m(v_i) = Q \), we restrict attention to the case where \( g_i^{mn_k}(v_i) \in [0, \overline{b}_i] \) for all sufficiently large \( k \).
	
	Since \( g^{mn_k} \) is a Bayesian Nash equilibrium in the game \( G^{mn_k} \), it follows that
	\[
	V_i^m(g_i^{mn_k}(v_i), v_i; g_{-i}^{mn_k}) - V_i^m(Q, v_i; g_{-i}^{mn_k}) \ge 0.
	\]
	By applying \( b_i = g_i^{mn_k}(v_i) \) and \( b_i = Q \) in Equation~\eqref{equa:general_payoff}, and setting \( \alpha_{-i} = g_{-i}^{mn_k} \), a straightforward computation yields
	\begin{align*}
		& \sum_{I_1 \subseteq I \setminus \{i\}} \sum_{I_2 \subseteq I \setminus (I_1 \cup \{i\})}
		\left( \prod_{j \in I_1} \tfrac{1}{\overline{b}_j} \right)
		\left(1 - \tfrac{1}{m} \right)^{|I_3| + 1}
		\left( \tfrac{1}{2m} \right)^{|I_1| + |I_2|} \\
		&\quad \times \int_{ \prod_{j \in I_1} [0, \overline{b}_j] }
		V_i(g_i^{mn_k}(v_i), v_i; \hat{b}_{I_1}, Q_{I_2}, g^{mn_k}_{I_3})
		\mathop{\otimes}_{j \in I_1} \mathrm{d} \hat{b}_j \ge 0.
	\end{align*}
	Hence, there must exist a partition \( I_1^*, I_2^*, I_3^* \) of \( I \setminus \{i\} \) such that
	\[
	\int_{ \prod_{j \in I_1^*} [0, \overline{b}_j] }
	V_i(g_i^{mn_k}(v_i), v_i; \hat{b}_{I_1^*}, Q_{I_2^*}, g^{mn_k}_{I_3^*})
	\mathop{\otimes}_{j \in I_1^*} \mathrm{d} \hat{b}_j \ge 0.
	\]
	Then, by Equation~\eqref{equa-general r3} and Assumption~\ref{assum-d1} (1) -- (3), it follows that there exist $\hat b_{I_1^*} \le g_i^{mn_k} $ such that 
	\begin{align}\label{equa-E>=0}
		& \mathbb{E} \Big[
		\big(W_i(g_i^{mn_k}(v_i), v) - C_i(g_i^{mn_k}(v_i))\big)
		\, D\big(g_i^{mn_k}(v_i), \hat b_{I_1^*}, Q_{I_2^*}, g^{mn_k}_{I_3^*}(v_{I_3^*})\big) \notag \\
		&\qquad \times P_i\big(g_i^{mn_k}(v_i), \hat b_{I_1^*}, Q_{I_2^*}, g^{mn_k}_{I_3^*}(v_{I_3^*})\big) \,\big|\, v_i \Big] \ge 0.
	\end{align}
	Finally, by Assumption~\ref{assum-d1} (5), we must have
	\[
	W_i(g_i^{mn_k}(v_i), v_i, v_{-i}) - C_i(g_i^{mn_k}(v_i)) \ge 0 \quad \text{for all } v_{-i}.
	\]
	As a result, Inequality~\eqref{equa-E>=0} holds for any partition \( I_1, I_2, I_3 \) of \( I \setminus \{i\} \). Then, by the convergence result in Equation~\eqref{equa - general converges}, we obtain
	\begin{align}\label{inequa-em>=0}
		\mathbb{E}\Big[ \big(W_i(g_i^m(v_i), v) - C_i(g_i^m(v_i))\big) \, D\big(g_i^m(v_i), \hat{b}_{I_1}, Q_{I_2}, g^m_{I_3}(v_{I_3})\big) \, \eta_i^{I_3}(v) \,\big|\, v_i \Big] \ge 0,
	\end{align}
	for any \( g_i^m(v_i) \in [0, \overline{b}_i] \cup \{Q\} \), any \( \hat{b}_{I_1} \) satisfying \( \max_{j \in I_1} \hat{b}_j \le g_i^m(v_i) \), and any partition \( I_1, I_2, I_3 \subseteq I \setminus \{i\} \).

	\

	Let $\hat{\Omega}_i^{I \backslash \{i\}} = \{ v_{-i}  \in [0,1]^{n-1}\, | \, \max_{j \ne i}  {g}_{j}^m(v_{j})  \le  {g}^m_i(v_i) \}$ be the set of types such that player \( i \)'s action \( g_i^m(v_i) \) is the highest action against $g_{-i}^m$. Then we have
	\begin{align}\label{equa-general l1}
		& \mathbb{E}\big[(W_i(g_i^m(v_i), v) - C_i(g_i^m(v_i))) \, D(g_i^m(v_i), g_{-i}^m(v_{-i})) \, \eta_i^{I \setminus \{i\}}(v) \,\big|\, v_i \big] - \Phi_i(g_i^m(v_i)) \notag \\
		= \; & \mathbb{P}(\hat{\Omega}_i^{I \setminus \{i\}} \mid v_i) \, 
		\mathbb{E}\big[(W_i(g_i^m(v_i), v) - C_i(g_i^m(v_i))) \, D(g_i^m(v_i), g_{-i}^m(v_{-i})) \, \eta_i^{I \setminus \{i\}}(v) \,\big|\, v_i, \hat{\Omega}_i^{I \setminus \{i\}} \big] \notag \\
		& \quad - \Phi_i(g_i^m(v_i)) \notag \\
		\le \; & \mathbb{P}(\hat{\Omega}_i^{I \setminus \{i\}} \mid v_i) \, 
		\mathbb{E}\big[(W_i(g_i^m(v_i), v) - C_i(g_i^m(v_i))) \, D(g_i^m(v_i), g_{-i}^m(v_{-i})) \,\big|\, v_i, \hat{\Omega}_i^{I \setminus \{i\}} \big] \notag \\
		& \quad \times \mathbb{E}\big[\eta_i^{I \setminus \{i\}}(v) \,\big|\, v_i, \hat{\Omega}_i^{I \setminus \{i\}} \big] - \Phi_i(g_i^m(v_i)) \notag \\
		\le \; & \mathbb{P}(\hat{\Omega}_i^{I \setminus \{i\}} \mid v_i) \, 
		\mathbb{E}\big[(W_i(g_i^m(v_i), v) - C_i(g_i^m(v_i))) \, D(g_i^m(v_i), g_{-i}^m(v_{-i})) \,\big|\, v_i, \hat{\Omega}_i^{I \setminus \{i\}} \big] 
		- \Phi_i(g_i^m(v_i)) \notag \\
		= \; & \lim_{\epsilon \to 0^+} V_i\big(g_i^m(v_i) + \epsilon, v_i; g_{-i}^m\big).
	\end{align}
	where the first inequality follows from \citet[Theorem 23]{MW1982}, and the second inequality follows from the bounds \( 0 \le \mathbb{E}[\eta_i^{I \setminus \{i\}}(v) \mid v_i, \hat{\Omega}_i^{I\backslash\{i\} }] \le 1 \), together with the fact that
	\[
	\mathbb{P}(\hat{\Omega}_i^{I\backslash\{i\} } \,  \mid v_i)  \,  \mathbb{E}\big[ (W_i(g_i^m(v_i), v) - C_i(g_i^m(v_i))) \, D(g_i^m(v_i), g_{-i}^m(v_{-i})) \,\big|\, v_i, \,  \hat{\Omega}_i^{I\backslash\{i\} } \big] \ge 0.
	\]	
	Given any partition \( I_1, I_2, I_3 \) of \( I \setminus \{i\} \), define the event $\hat \Omega_i^{I_3} = \{ v_{-i} \in [0,1]^{n-1} \mid \max_{j \in I_3} g_{j}^m(v_{j}) \le g_i^m(v_i) \}$ similarly.  By repeating the same argument used to derive Equation~\eqref{equa-general l1}, we can deduce that for any collection of actions \( \{ \hat{b}_j \}_{j \in I_1} \) satisfying \( \max_{j \in I_1} \hat{b}_j \le g_i^m(v_i) \), the following inequality holds:
	\begin{align*}
		& \mathbb{E}\Big[
		\big(W_i(g_i^m(v_i), v) - C_i(g_i^m(v_i))\big)  \,
		D\big(g_i^m(v_i), \hat b_{I_1}, Q_{I_2}, g_{I_3}^m(v_{-i})\big) \,
		\eta_i^{I_3}(v) 
		\,\Big|\, v_i \Big] 
		\\
		\le\ & 
		\mathbb{P}(\hat{\Omega}_i^{I_3} \mid v_i) \, 
		\mathbb{E}\Big[
		\big(W_i(g_i^m(v_i), v) - C_i(g_i^m(v_i))\big) \,
		D\big(g_i^m(v_i), \hat b_{I_1}, Q_{I_2}, g_{I_3}^m(v_{-i})\big)
		\,\Big|\, v_i, \, \hat{\Omega}_i^{I_3} \Big] 
	\end{align*}
	Then, the following inequality holds:
	\begin{align}\label{equa-general l2}
		& \prod_{j \in I_1} \frac{1}{\overline{b}_j }    
		\int_{\prod_{j \in I_1}[0,  \min\{ g_i^{m}(v_i), \overline{b}_j\} ]}   
		\mathbb{E}\Big[ 
		\big(W_i(g_i^{m}(v_i),v) - C_i(g_i^{m}(v_i))\big) \notag \\
		& \qquad  \qquad \times D\big(g_i^{m}(v_i), \hat b_{I_1}, Q_{I_2}, g^{m}_{I_3}(v_{I_3})\big) \, \eta_i^{I_3}(v) 
		\,\Big|\, v_i \Big] \, 
		\mathop{\otimes}_{j \in I_1} \mathrm{d} \hat b_j 
		\notag \\
		\le\ & 
		\prod_{j \in I_1} \frac{1}{\overline{b}_j }    
		\int_{\prod_{j \in I_1}[0,  \min\{ g_i^{m}(v_i), \overline{b}_j\} ]}   
		\mathbb{P}(\hat{\Omega}_i^{I_3}  \mid v_i) \, \mathbb{E}\Big[ 
		\big(W_i(g_i^m(v_i), v) - C_i(g_i^m(v_i))\big)  
		\notag \\
		& \quad  \qquad \times 
		D\big(g_i^{m}(v_i), \hat b_{I_1}, Q_{I_2}, g_{I_3}^{m}(v_{-i})\big) 
		\,\Big|\, v_i, \,  \hat{\Omega}_i^{I_3}  \Big] \,
		\mathop{\otimes}_{j \in I_1} \mathrm{d} \hat b_j   
	\end{align}
	By Equation~\eqref{eq-limVim} and Inequalities~\eqref{equa-general l1} and~\eqref{equa-general l2}, we obtain the following chain of inequalities:
	\begin{align}\label{inequality -general 2}
		\lim\limits_{k \to \infty} V_i^m( {g}^{mn_k}_i(v_i),v_i;  {g}^{mn_k}_{-i}) 	
		& \le \lim\limits_{\epsilon \to 0^+} V_i^m( {g}^m_i(v_i) + \epsilon, v_i; {g}_{-i}^m) 
		& \le 	\sup\limits_{b_i \in [0, \overline{b}_i] \cup \{Q\}} V_i^m(b_i, v_i;  {g}_{-i}^m).
	\end{align}	
	Hence, combine with Inequalities \eqref{inequa-<=} and \eqref{inequality -general 2}, we must have 
	\begin{align}\label{equa-equal}
		\lim\limits_{k \to \infty} V_i^m( {g}^{mn_k}_i(v_i),v_i;  {g}^{mn_k}_{-i}) 	 \, \, \,
		=  	\sup\limits_{b_i \in [0, \overline{b}_i] \cup \{Q\}} V_i^m(b_i, v_i;  {g}_{-i}^m).
	\end{align}
	Therefore, all inequalities appearing in Inequality~\eqref{equa-general l1} must be  binding.

	\bigskip

	Finally, we shall show that under the strategy profile \( g^m \), the probability that two or more players simultaneously play an action strictly above \( Q \) is zero. 
	Since all inequalities appearing in Inequality~\eqref{equa-general l1} must hold with equality, if \( \mathbb{P}(\hat{\Omega}_i^{I \setminus \{i\}} \mid v_i) > 0 \), then we have
	\begin{align}\label{equa - general f}
		0 
		& \le \mathbb{E}\big[
		(W_i(g_i^m(v_i), v) - C_i(g_i^m(v_i))) \, D(g_i^m(v_i), g_{-i}^m(v_{-i})) 
		\,\big|\, v_i, \hat{\Omega}_i^{I \setminus \{i\}}
		\big] \,
		\mathbb{E}\big[
		\eta_i^{I \setminus \{i\}}(v) \,\big|\, v_i, \hat{\Omega}_i^{I \setminus \{i\}} 
		\big] \notag \\
		& = \mathbb{E}\big[
		(W_i(g_i^m(v_i), v) - C_i(g_i^m(v_i))) \, D(g_i^m(v_i), g_{-i}^m(v_{-i})) 
		\,\big|\, v_i, \hat{\Omega}_i^{I \setminus \{i\}}
		\big].
	\end{align}
	We proceed to show that the inequality in Inequality~\eqref{equa - general f} is in fact strict, by analyzing the following two cases.

	\medskip
	
	\noindent	\textbf{Case 1:} Consider $v_i \in [\tilde{v}_i, \hat{v}_i]$ with $\tilde{v}_i < \hat{v}_i$, where the strategy $g_i^m$ takes a constant value on $[\tilde{v}_i, \hat{v}_i]$.
	
	For any \( v_i' \in (\tilde{v}_i, \hat{v}_i] \),  it is clear that the event
	$\hat  \Omega_i^{I \backslash \{i\}} =  \{ v_{-i} \in [0,1]^{n-1} \,  | \,  \max_{j \ne i} g_j^m(v_j) \\  \le g_i^m(v_i')      \}  $. 
	Then \( \mathbb{P}(\hat{\Omega}_i^{I \setminus \{i\}} \mid v_i') > 0 \), and substituting \( v_i \) by \( v_i' \) in Inequality~\eqref{equa - general f}, we obtain
	\[
	0 \le 
	\mathbb{E} \left[
	\left( W_i(g_i^m(v_i'), v) - C_i(g_i^m(v_i')) \right) \,
	D\big(g_i^m(v_i'), g_{-i}^m(v_{-i})\big)
	\,\big|\, v_i', \, \hat{\Omega}_i^{I \setminus \{i\}}
	\right].
	\]
	It then follows from Assumption~\ref{assum-d1} (3) and (5) that  $W_i( {g}_i^m(v_i'),v_i, v_{-i}) - C_i({g}_i^m(v_i')) \ge 0$ for any $v_{-i} \in [0,1]^{n-1}$. Moreover, by Assumption~\ref{assum-d1} (1) and (3), the function  $(W_i( {g}_i^m(v_i'),v) - C_i({g}_i^m(v_i'))) \, D(g_i^{m}(v_i'), g_{-i}^{m}(v_{-i})  )$ is strictly increasing in $v_i$ for $v_i \ge v_i'$  and increasing in $v_{-i}$. Then, by \citet[Theorem 5]{MW1982}, we obtain:
	\begin{align*}
		0 
		& \le \mathbb{E}\left[
		(W_i(g_i^m(v_i'), v) - C_i(g_i^m(v_i')))  \,
		D(g_i^m(v_i'), g_{-i}^m(v_{-i})) \,\middle|\, v_i', \hat{\Omega}_i^{I \backslash \{i\}}
		\right] \\
		& < \mathbb{E}\left[
		(W_i(g_i^m(v_i'), v) - C_i(g_i^m(v_i')))  \,
		D(g_i^m(v_i'), g_{-i}^m(v_{-i})) \,\middle|\, v_i, \hat{\Omega}_i^{I \backslash \{i\}}
		\right] \\
		& = \mathbb{E}\left[
		(W_i(g_i^m(v_i), v) - C_i(g_i^m(v_i))) \,
		D(g_i^m(v_i), g_{-i}^m(v_{-i})) \,\middle|\, v_i, \hat{\Omega}_i^{I \backslash \{i\}}
		\right].
	\end{align*}
	As a result, the inequality in Inequality~\eqref{equa - general f} holds strictly on the subinterval almost everywhere.\footnote{More precisely, the inequality is strict for all such subintervals except possibly at their lower endpoints. Since there are at most countably many such subintervals, the set of all exceptional points has  measure zero.}

	\medskip
	
	\noindent	\textbf{Case 2:}  Consider \( v_i \) such that \( g_i^m \) is continuous and strictly increasing.\footnote{Since \( g_i^m \) is monotone, it has at most countably many jumps. We ignore the set of those \( v_i \) as it is of measure zero.}

	As \( \{ g_j^m \}_{j \in I \setminus \{i\}} \) has at most countably many mass points, the set of \( v_i \) such that \( g_i^m(v_i) \) is a mass point of \( \{ g_j^m \}_{j \in I \setminus \{i\}} \) but not a mass point of \( g_i^m \) is of measure zero. 
	We focus on those \( v_i \) such that \( g_i^m(v_i) \) is not a mass point of \( \{ g_j^m \}_{j \in I } \).
	
	Since  $g_i^m$ is continuous at $v_i$ and $\{g_j^m\}_{j \in I}$ has at most countably many mass points, there exists some $v_i' < v_i$ such that $g_i^m(v_i') $ is not a mass point of $\{g_j^m\}_{j \in I}$ and $\mathbb{P} (\hat  \Omega_i'  \, | \,  v'_i) > 0$, where $\hat  \Omega_i' = \{ v_{-i} \in [0,1]^{n-1} \, | \,  \max_{ j \ne i}  g_j^m(v_j) \le g_i^m(v_i')   \} $. Then, we  obtain 
	\begin{align*}
		0 
		& \le \mathbb{E}\left[\left(W_i(g_i^m(v_i'), v_i', v_{-i}) - C_i(g_i^m(v_i'))\right) D\left(g_i^m(v_i'), g_{-i}^m(v_{-i})\right) \,\middle|\, v_i', \, \hat \Omega_i'\right] \\
		& < \mathbb{E}\left[\left(W_i(g_i^m(v_i'), v_i, v_{-i}) - C_i(g_i^m(v_i'))\right) D\left(g_i^m(v_i'), g_{-i}^m(v_{-i})\right) \,\middle|\, v_i,  \, \hat \Omega_i'\right].
	\end{align*}
	The first inequality follows from Equation~(\ref{equa - general f}) as $\mathbb{P} (\hat  \Omega_i'  \, | \,  v'_i) > 0$. The second inequality holds as $(W_i( {g}_i^m(v_i'),v) - C_i({g}_i^m(v_i'))) \,  D(g_i^{m}(v_i'), g_{-i}^{m}(v_{-i})  )$  is increasing in $v_{-i}$ and strictly increasing in $v_i$ for $v_i \ge v_i'$, see \citet[Theorem 5]{MW1982}.  Moreover, by Assumption~\ref{assum-d1}~(5), it follows that  $W_i( {g}_i^m(v_i'),v_i, v_{-i})) - C_i({g}_i^m(v_i')) > 0$.

	In addition, we have
	\begin{align*}
		V_i^m(  g_i^m(v_i), v_i; g_{-i}^m)  
		= &\ \lim_{k \to \infty} V_i^m\big( g_i^{mn_k}(v_i), v_i; g_{-i}^{mn_k} \big) 	\\
		= &\ \sup_{b_i \in [0, \overline{b}_i] \cup \{Q\}} V_i^m(b_i, v_i; g_{-i}^m) \\
		\ge &\ V_i^m( g_i^m(v_i'), v_i; g_{-i}^m).
	\end{align*}
	The first equality holds since \( g_i^m(v_i) \) is not a mass point of \( \{g_j^m\}_{j \in I} \). The second equality follows from Equation~\eqref{equa-equal}, and the last inequality follows from the definition of the supremum.

	Recall the calculation of $V_i^m$ in  Equation~\eqref{equa:general_payoff}. Then $V_i^m(  g_i^m(v_i'), v_i; g_{-i}^m )$ can be written as the summation of two terms, where the second term does not depend on $g_i^m(v_i')$, and the first term is to subtract $\Phi_i(g_i^m(v_i'))$ from an integral over $W_i( g_i^m(v_i'),v_i, v_{-i})) - C_i({g}_i^m(v_i'))$.  Similarly, $V_i^m(  g_i^m(v_i), v_i; g_{-i}^m )$ can be written as the summation of two terms, where the second term is the same as that in the previous case, and the first term is to subtract $\Phi_i(g_i^m(v_i))$ from the integral over $W_i( g_i^m(v_i),v_i, v_{-i})) - C_i({g}_i^m(v_i))$. Since
	\begin{itemize}
		\item $\Phi_i(g_i^m(v_i)) \ge \Phi_i(g_i^m(v'_i))$,
		
		\item $W_i( {g}_i^m(v_i'),v_i, v_{-i})) - C_i({g}_i^m(v_i')) > 0$ implies that its integral is also positive, and
		
		\item $V_i^m(  g_i^m(v_i), v_i; g_{-i}^m)   \ge  V_i^m(  g_i^m(v_i'), v_i; g_{-i}^m )$ from the inequality above,
	\end{itemize}
	the integral over $W_i( g_i^m(v_i),v_i, v_{-i})) - C_i({g}_i^m(v_i))$ is positive. By Assumption~\ref{assum-d1}~(5) again, $W_i( g_i^m(v_i),v_i, v_{-i})) - C_i({g}_i^m(v_i)) > 0$ for all $v_{-i}$, which implies that
	$$\bE[(W_i( {g}_i^m(v_i),v) - C_i({g}_i^m(v_i))) \, D(g_i^{m}(v_i), g_{-i}^{m}(v_{-i})  ) |v_i , \hat  \Omega_i^{I \backslash \{i\}}] > 0.
	$$

	From Inequality~(\ref{equa - general f}), we can conclude that $\bE[\eta_i^{I \backslash \{i\}}(v) \, | \,  v_i, \hat  \Omega_i^{I \backslash \{i\}}] = 1$ for almost all $v_i$ such that $\mathbb{P}(\hat \Omega_i^{I \backslash \{i\}}  \, |  \, v_i) > 0$. Given an arbitrary nonempty subset$ H \subseteq \{1,2,\ldots, n\}$, and define the set
	$$T_H = \{ v \in [0,1]^n : g_i^m(v_i) = \max_{j \in I} g_j^m(v_j) > Q, \, \forall i \in H \}.
	$$
	We analyze the probability $\mathbb{P}(T_H)$.  Suppose $\mathbb{P}(T_H) > 0$. Then for every $i \in H$, , by the above result, we have 
	$\eta_i^{I \backslash \{ i\}}(v) = 1$ for almost all $v \in T_H$. Since $\sum_{i = 1}^{n} \eta_i^{I \backslash \{ i\}}(v) \le 1$ for almost all $v \in T$, it must be that $|H| = 1$. It implies that the probability that two or more players simultaneously play the highest action above $Q$ under ${g}^m$ is $0$. Then for every $i$ and almost all $v_i$, the function $V_i^m(\cdot,v_i;\cdot)$ is continuous at $(g_i^m(v_i), g_{-i}^m(v_{-i}))$, and we have
	$$ V_i^m(g_i^m(v_i), v_i; g_{-i}^m) \, =  \, \lim_{k \to \infty} V_i^m(g_i^{mn_k}(v_i), v_i; g_{-i}^{mn_k}) \,  = \sup_{b_i \in [0, \overline{b}_i] \cup \{Q\}} V_i^m(b_i, v_i;  {g}_{-i}^m).
	$$
	Therefore, ${g}^m$ is a monotone equilibrium.
	
	\

	{\noindent\textbf{Step 3. }}
	By Helly's selection theorem, there exists a subsequence \( \{ g^{m_k} \}_{k=1}^{\infty} \) of the sequence \( \{ g^m \}_{m=1}^{\infty} \) that converges to a monotone strategy profile \( g \) for almost all type profiles \( v \in [0,1]^n\). We will show that \( g \) constitutes a perfect monotone equilibrium.
	
	By construction of \( g \), we have
	\[
	\lim_{k \to \infty} d(g_i^{m_k}(v_i), g_i(v_i)) = 0 \quad \text{for all } i \text{ and almost all } v_i \in [0,1].
	\]	
	Let \( \tilde{g}_j^{m_k}(t_j) = \left(1 - \frac{1}{m_k}\right) \delta_{g_j^m(t_j)} + \frac{1}{2m_k} U[0, \overline{b}_j] + \frac{1}{2m_k} \delta_{\{Q\}} \) for each \( t_j \in [0,1] \) and \( k \in \mathbb{Z}_{+} \), be the corresponding completely mixed strategy profile associated with \( g^{m_k} \). It is clear that
	\[
	\lim\limits_{k \to \infty} \rho(\tilde{g}_i^{m_k}(t_i), \delta_{g_i(t_i)}) = 0.
	\]
	Since \( g^{m_k} \) is a Bayesian Nash equilibrium in \( G^{m_k} \), we have
	\[
	V_i^{m_k}(b_i, v_i; g_{-i}^{m_k}) \le V_i^{m_k}(g_i^{m_k}(v_i), v_i; g_{-i}^{m_k}) \quad \text{for all } b_i \in [0, \overline{b}_j] \cup \{ Q \}.
	\]
	By substituting \( m \) with \( m_k \), setting \( \alpha_{-i} = g_{-i}^{m_k} \), and applying \( b_i = g_i^{m_k}(v_i) \) in Equation~\eqref{equa:general_payoff}, the expression becomes equivalent to:
	\[
	V_i(b_i, v_i; \tilde{g}_{-i}^{m_k}) \le V_i(g_i^{m_k}(v_i), v_i; \tilde{g}_{-i}^{m_k}) \quad \text{for all } b_i \in [0, \overline{b}_i] \cup \{ Q \}.
	\]
	Thus, \( g_i^{m_k}(v_i) \in \mathrm{BR}_i(v_i, \tilde{g}^{m_k}_{-i}(v_i)) \) for almost all \( v_i \in [0,1] \).
	Then, by the triangle inequality, we have
	$$
	0 \le \lim_{k \to \infty} d \big( {g_i(v_i)}, \mathrm{BR}_i(t_i, \tilde{g}_{-i}^{m_k}) \big) 
	\le
	\lim_{k \to \infty} d \big( {g_i(v_i)}, {g}_i^{m_k}(v_i) \big)  + 
	\lim_{k \to \infty} d \big( {g}_i^{m_k}(t_i), \mathrm{BR}_i(v_i, \tilde{g}_{-i}^{m_k}) \big) 
	= 0.
	$$
	Hence, 
	$g$ is perfect.
	
	\
	
	Next, we will show that \( g \) is a monotone equilibrium. Note that each \( g^{m_k} \) is a monotone equilibrium in the game \( G^{m_k} \), and that the sequence of games \( G^{m_k} \) converges to \( G \), while the strategies \( g^{m_k} \) converge pointwise to \( g \). What remains is to verify that the probability of two or more players simultaneously playing the same highest action above \( Q \) is zero.
	The argument proceeds analogously to that in Step~2 above. For simplicity, in what follows, we outline only the necessary modifications to that argument.
	
	The key step is to examine the maximum payoff attainable by each player \( i \) at her type \( v_i \in [0,1] \) in the game \( G \). It suffices to consider those actions \( b_i \) for which \( W_i(b_i, v) - C_i(b_i) \ge 0 \) holds. By applying the same reasoning as in Inequality~\eqref{equa-general right limit}, we obtain the following inequality:
	\begin{align}\label{equa:step3-<=}
		V_i(b_i, v_i; g_{-i}) \le \lim_{b_i' \to b_i^{+}} V_i(b_i', v_i; g_{-i}).
	\end{align}
	Again, by applying similar arguments as those used in Inequality~\eqref{equa-general r1}, we obtain the following sequence of inequalities:
	\begin{align}\label{equa:step3lim}
		\lim_{b_i' \to b_i^{+}} V_i(b_i', v_i; g_{-i}) &\le V_i(\hat{b}_i, v_i; g_{-i}) + \epsilon  \notag \\
		&\le V_i(\hat{b}_i, v_i; g_{-i}^{m_k}) + 2\epsilon \quad \text{for } k \ge K \notag \\
		&\le V_i(g_i^{m_k}(v_i), v_i; g_{-i}^{m_k}) + 3\epsilon \quad \text{for } k \ge K,
	\end{align}
	where the first and second inequalities hold for some \( \hat{b}_i \) sufficiently close to \( b_i \), with \( \hat{b}_i \) not being a mass point of \( g \). The final inequality follows from the fact that \( g^{m_k} \) is a Bayesian Nash equilibrium in \( G^{m_k} \). Hence, when \( k \) is sufficiently large, it must be that
	\[
	V_i(\hat{b}_i, v_i; g_{-i}^{m_k}) \le V_i(g_i^{m_k}(v_i), v_i; g_{-i}^{m_k}) + \epsilon.
	\]
	Notice that
	\[
	V_i(g_i^{m_k}(v_i), v_i; g_{-i}^{m_k}) = \mathbb{E}[ (W_i(g_i^{m_k}(v_i), v) - C_i(g_i^{m_k}(v_i))) D(g^{m_k}(v)) P_i(g^{m_k}(v)) \mid v_i] - \Phi_i(g_i^{m_k}(v_i)).
	\]
	By the monotonicity of \( g \) and Helly's selection theorem (extracting a subsequence if necessary), we obtain
	\begin{align*}
		& \mathbb{E}[(W_i(g_i^{m_k}(v_i), v) - C_i(g_i^{m_k}(v_i))) D(g^{m_k}(v)) P_i(g^{m_k}(v)) \mid v_i] \\ 
		& \quad \to \mathbb{E}[(W_i(g_i(v_i), v) - C_i(g_i(v_i))) D(g(v)) \eta_i^* \mid v_i],
	\end{align*}
	where \( \eta_i^* \colon [0,1]^n \to [0,1] \), and \( \sum\limits_{j = 1}^n \eta_j^*(v_i) \le 1 \). Hence, the limit of the sequence \( \{ V_i(g_i^{m_k}(v_i), v_i; g_{-i}^{m_k}) \}_{k = 1}^{\infty} \) exists (after extracting a subsequence if necessary).
	%
	Then, it follows from Inequalities~\eqref{equa:step3-<=} and~\eqref{equa:step3lim} that
	\begin{align}\label{equa:step3sup<=}
		\sup_{b_i \in [0, \overline{b}_i] \cup \{ Q \}} V_i(b_i, v_i; g_{-i}) \, \, \le \, \, \lim_{k \to \infty} V_i(g_i^{m_k}(v_i), v_i; g_{-i}^{m_k}).
	\end{align}
	By the same arguments as those used in  Inequalities \eqref{equa-E>=0} -- \eqref{equa-general l1}, we obtain
	\begin{align}\label{equa:step3sup>=}
		\lim_{k \to \infty} V_i(g_i^{m_k}(v_i), v_i; g_{-i}^{m_k})\, \, \,  \le \sup_{b_i \in [0, \overline{b}_i] \cup \{ Q \}} V_i(b_i, v_i; g).
	\end{align}
	Hence, Inequalities~\eqref{equa:step3sup<=} and \eqref{equa:step3sup>=} imply that
	\[
	\lim_{k \to \infty} V_i(g_i^{m_k}(v_i), v_i; g_{-i}^{m_k}) \, \, \,  = \sup_{b_i \in [0, \overline{b}_i] \cup \{ Q \}} V_i(b_i, v_i; g).
	\]

	Finally, repeating the proof in Step 2, we can show that the probability of two or more players simultaneously playing the highest action above $ Q $ under $g$ is zero. Thus, 
	$$\lim_{k \to \infty} V_i(g_i^{m_k}(v_i), v_i; g_{-i}^{m_k}) = V_i(g_i(v_i), v_i; g_{-i}).
	$$
	This  implies that $g$ is a monotone equilibrium and completes the proof.
\end{proof}

\begin{proof}[Proof of Part 2.]

	Let $\alpha = (\alpha_1, \cdots, \alpha_n),$ be a perfect monotone equilibrium. There exists a sequence of completely mixed strategies $\alpha^k = (\alpha_1^k, \cdots, \alpha_n^k)$ and zero-measure sets $\{S_i\}$ such that for each player $i$ and $v_i \in T_i \backslash S_i$ with $\lambda_i(T_i \backslash S_i) = 1$,
	\begin{itemize}
		\item[(i)]  $\lim\limits_{k \to \infty} \rho \left( \alpha_i^k(v_i), \delta_{\alpha_i(v_i)} \right) = 0$;
		\item[(ii)]  $\lim\limits_{k \to \infty} d \left( \alpha_i(v_i), \text{BR}_i(v_i , \alpha_{-i}^k) \right) = 0$.
	\end{itemize}
	Hereafter, we consider $v_i \in T_i \setminus (S_i \cup \{0\})$. 
	Since $\alpha_i$ is a pure strategy profile and 
	$
	\lim_{k \to \infty} d\big( \alpha_i(v_i), \text{BR}_i(v_i , \alpha_{-i}^k) \big) = 0,
	$
	we can find a sequence of actions $r_i^k(v_i) \in \text{BR}_i(v_i, \alpha_{-i}^k)$ such that 
	$
	\lim_{k \to \infty} d\big( \alpha_i(v_i), r_i^k(v_i) \big) = 0.
	$
	To show that $\alpha_i(v_i)$ is limit undominated, it suffices to show that each $r_i^k(v_i)$ is undominated at type $v_i$. Moreover, for each $k$, it suffices to show that every $b_i' \in \text{BR}_i(v_i, \alpha_{-i}^k)$ is undominated at type $v_i$ for player $i$.


	For any $b_i' \in \text{BR}_i(v_i, \alpha_{-i}^k)$, suppose $b_i'$ is dominated at type $v_i$ for player $i$. Then there exists $\sigma_i \in \mathcal{M}(A_i)$ such that
	\[
	\int_{A_i} V_i(b_i, v_i; \alpha_{-i}^k) \, \mathrm{d} \sigma_i(b_i) \ge V_i(b_i', v_i; \alpha_{-i}^k).
	\]
	In this case, we must have $\sigma_i \in \mathcal{M}(\text{BR}_i(v_i, \alpha_{-i}^k))$. Therefore, to show that any $b_i' \in \text{BR}_i(v_i, \alpha_{-i}^k)$ is undominated at type $v_i$, it suffices to prove that $b_i'$ cannot be dominated by any $\sigma_i \in \mathcal{M}(\text{BR}_i(v_i, \alpha_{-i}^k) \setminus \{b_i'\})$.
	
	We now demonstrate that $Q \notin \text{BR}_i(v_i, \alpha_{-i}^k)$. 
	By Assumption~\ref{assum-d1} (1) and (2), we have $\Phi_i(0) = C_i(0) = 0$ and $W_i(0, v_i, v_{-i}) > 0$ for any $v_i > 0$. 
	Since $\alpha_{-i}^k$ is a completely mixed strategy profile, playing $0$ at type $v_i$ against $\alpha_{-i}^k$ gives player~$i$ a strictly positive winning probability. 
	Therefore, for any $v_i > 0$, we have
	\[
	V_i(0, v_i; \alpha_{-i}^k) > 0 = V_i(Q, v_i; \alpha_{-i}^k),
	\]
	which implies that $Q \notin \text{BR}_i(v_i, \alpha_{-i}^k)$ for any $v_i > 0$.

	Suppose that $b_i' \in \text{BR}_i(v_i, \alpha_{-i}^k)$ and $b_i' > \max_{j \ne i} \overline{b}_j$ (recall that $\overline{b}_j$ is the upper bound of player~$j$'s action space); that is, player~$i$ is the unique winner. 	
	If $b_i' > 0$, then there exists $\epsilon > 0$ such that 
	$
	b_i' - \epsilon > \max\{0, \max_{j \ne i} \overline{b}_j\}.
	$
	In this case, player~$i$ can strictly benefit by playing $b_i' - \epsilon$, contradicting the assumption that $b_i' \in \text{BR}_i(v_i, \alpha_{-i}^k)$. 
	Hence, it must be that $b_i' = 0 > \max_{j \ne i} \overline{b}_j$, which is impossible.

	Therefore, for any $b_i' \in \text{BR}_i(v_i, \alpha_{-i}^k)$, the analysis above implies that $b_i' \ne Q$ and $b_i' \le \max_{j \ne i} \overline{b}_j$. 
	We now consider the following three cases.

	
	\noindent \textbf{Case~(I)}:		 $b_i' =  \inf \{  b_i |  b_i \in \text{BR}_i(v_i, \alpha_{-i}^k) \}  = \max_{j \ne i} \overline{b}_j$. 
	
	Since any $b_i \in \text{BR}_i(v_i, \alpha_{-i}^k)$ is  less than or equal to $\max_{j \ne i} \overline{b}_j$, it follows that $\text{BR}_i(v_i, \alpha_{-i}^k) = \{b_i'\}$, i.e., it is a singleton set.  It implies that 
	$
	V_i\big(b_i', v_i; \alpha_{-i}^k\big) > V_i\big(b_i, v_i; \alpha_{-i}^k\big)$ for all $b_i \ne b_i',
	$
	it is clear that $b_i'$ is undominated at type $v_i$ for player $i$.


	\noindent \textbf{Case~(II)}:	  $b_i' = \inf \{  b_i |  b_i \in \text{BR}_i(v_i, \alpha_{-i}^k)  \} <  \max_{ j \ne i} \overline{b}_j$.  
	
	By Case (I), we know that if $\text{BR}_i(v_i, \alpha_{-i}^k) = \{b_i'\}$ is a singleton set, then $b_i'$ is undominated.  
	We now focus on the case where there are multiple optimal actions; that is, $\text{BR}_i(v_i, \alpha_{-i}^k)$ is not a singleton set.

	Since 
	$
	V_i(b_i', v_i; \alpha_{-i}^k) \ge V_i(0, v_i; \alpha_{-i}^k) > 0,
	$
	by Assumption~\ref{assum-d1} (2) and (5), we have $W_i(b_i', v) - C_i(b_i') > 0$ for any $v_{-i} \in [0,1]^{n-1}$. 
	Hence, by Assumption~\ref{assum-d1} (3), the expression 
	$
	(W_i(b_i', v) - C_i(b_i')) D(b_i', b_{-i}) - \Phi_i(b_i')
	$
	is increasing in $b_{-i}$. Therefore, there exists $\hat{b}_{-i}$ such that $\max_{j \ne i} \hat{b}_j < b_i'$ and
	$
	V_i(b_i', v_i; \hat{b}_{-i}) > 0.
	$
	
	Now pick any $b_i^o \in \text{BR}_i(v_i, \alpha_{-i}^k) \setminus \{b_i'\}$. By the condition of this case, we have $b_i^o > b_i'$, which implies 
	$
	P_i(b_i^o, \hat{b}_{-i}) = 1.
	$
	Then, by Assumption~\ref{assum-d1} (6), it follows that	
	\begin{align*}
		V_i(b_i', v_i; \hat{b}_{-i}) 
		&= \mathbb{E}\left[ (W_i(b_i', v) - C_i(b_i')) D(b_i', \hat{b}_{-i}) - \Phi_i(b_i') \,\middle|\, v_i \right] \\
		&> \mathbb{E}\left[ (W_i(b_i^o, v) - C_i(b_i^o)) D(b_i', \hat{b}_{-i}) - \Phi_i(b_i^o) \,\middle|\, v_i \right] \\
		&= V_i(b_i^o, v_i; \hat{b}_{-i}).
	\end{align*}		
	Therefore, for any $\sigma_i \in \mathcal{M}(\text{BR}_i(v_i, \alpha_{-i}^k) \setminus \{b_i'\})$, we have
	\[
	V_i(b_i', v_i; \hat{b}_{-i}) > \int_{A_i} V_i(b_i, v_i; \hat{b}_{-i}) \, \mathrm{d} \sigma_i(b_i),
	\]
	which implies that $b_i'$ is undominated at type $v_i$ for player~$i$.

	\noindent \textbf{Case~(III)}:			  $b_i' > \inf \{  b_i |  b_i \in \text{BR}_i(v_i, \alpha_{-i}^k) \} \ge 0 $. 
	
	We first show that for any $ b_i^o \in  \text{BR}_i(v_i, \alpha_{-i}^k)$ such that $ b_i^o  < b_i'$, 	there exists  $ b^o_{-i}$ such that $ b^o_i   < \max_{j \ne i}  b^o_j  < b_i'$  and 
	$	V_i(b_i',v_i; {b}^o_{-i}) > 0.
	$
	
	Since $ b_i^o \in  \text{BR}_i(v_i, \alpha_{-i}^k)$,  we have
	$	V_i(b_i',v_i; \alpha^k_{-i})
	=  V_i( b_i^o, v_i; \alpha^k_{-i}) > 0.$
	Following the argument in Case~(II), there exists $\hat{b}_{-i}$ such that
	$$\max_{j \ne i} \hat{b}_j < b_i' \qquad \mbox{and} \qquad 
	V_i( b_i', v_i; \hat{b}_{-i}) > 0.
	$$
	Let $\overline{b}_{j^*} =  \max_{ j \ne i} \overline{b}_j$. Then  $[{b}_i^o ,b_i'] \subseteq [0, \overline{b}_{j^*}]$. Choose ${b}^o_{j^*} \in (\max\{ {b}^o_i , \max_{j \ne i} \hat{b}_j\}, b_i')$ and set ${b}_j^o = \hat b_j$ for $j \ne i$ and $j^*$. Clearly, 
	$$  {b}^o_i <  {b}^o_{j^*}   =  \max_{j \ne i}   b^o_j < b_i'.
	$$ 
	By Assumption~\ref{assum-d1} (3) and (5),  the function $(W_i(b_i',v) - C_i(b_i'))D(b_i',b_{-i}) - \Phi_i(b_i')$ is  increasing $b_{-i}$, which implies
	$$V_i( b_i', v_i; b^o_{-i}) \ge V_i( b_i', v_i; \hat b_{-i}) > 0.$$

	\medskip

	Next, fix $\sigma_i \in \mathcal{M}(\text{BR}_i(v_i, \alpha_{-i}^k) \setminus \{b_i'\})$, and let
	$
	\check{b}_i = \sup \left\{ b_i \,\middle|\, b_i \in \text{supp}(\sigma_i),\ b_i < b_i' \right\},
	$
	where $\text{supp}(\sigma_i)$ denotes the support of $\sigma_i$. We consider the following two cases.
	
	\begin{enumerate}
		\item [\textbf{(I')}]  $ \check b_i  < b_i'$.

		There exists some $\check{b}_{-i}$ such that $\check{b}_i < \max_{j \ne i} \check{b}_j < b_i'$ and $V_i(b_i', v_i; \check{b}_{-i}) > 0$. 
		Since $\Phi_i(b_i)$ is nonnegative and increasing, we have $V_i(b_i', v_i; \check{b}_{-i}) > 0 \ge V_i(\hat{b}_i, v_i; \check{b}_{-i})$ for any $\hat{b}_i \le \check{b}_i$. 
		In addition, $V_i(b_i', v_i; \check{b}_{-i}) > V_i(\hat{b}_i, v_i; \check{b}_{-i})$ for any $\hat{b}_i > b_i'$, since $(W_i(b_i, v_i) - C_i(b_i)) D(b_i, b_{-i}) - \Phi_i(b_i)$ is decreasing in $b_i$. 
		Because $\text{supp}(\sigma_i) \subseteq [0, \check{b}_i] \cup [b_i', \overline{b}_i]$, we have
		\[
		V_i(b_i', v_i; \check{b}_{-i}) > \int_{A_i} V_i(b_i, v_i; \check{b}_{-i}) \, \mathrm{d}\sigma_i(b_i).
		\]
		Therefore, $\sigma_i$ does not dominate $b_i'$ at type $v_i$ for player~$i$.

		\item[\textbf{(II')}] $\check{b}_i =  b_i'$. 
		
		If $\sigma_i\big((b_i', \overline{b}_i]\big) = 1$, then by the same argument as in Case~(II), we can conclude that $\sigma_i$ does not dominate $b_i'$.

		Next, we focus on the case where $\sigma_i\big((b_i', \overline{b}_i]\big) < 1$. If $b_i' \in (0, \overline{b}_j]$, let $b_j^{\epsilon} = b_i' - \epsilon > 0$ for sufficiently small $\epsilon$; if $b_i' > \overline{b}_j$, let $b_j^{\epsilon} = \overline{b}_j$. Then $b_i' > \max_{j \ne i} b_j^{\epsilon} \ge b_i' - \epsilon > 0$.\footnote{Since $b_i' \le \max_{j \ne i} \overline{b}_j$, there exists $j^*$ such that $b_i' \in (0,\overline{b}_{j^*}].$} 		
		Since $\check{b}_i = b_i'$, $\sigma_i\big((b_i', \overline{b}_i]\big) < 1$, the statement right after Case~(III), and the fact that $(W_i(b_i', v_i) - C_i(b_i')) D(b_i', b_{-i}) - \Phi_i(b_i')$ is increasing in $b_{-i}$, we obtain that $V_i(b_i', v_i; b_{-i}^{\epsilon}) > 0$ for sufficiently small $\epsilon$. 			
		This further implies that $V_i(b_i', v_i; b_{-i}^{\epsilon})$ is decreasing in $\epsilon$.

		
		For $\epsilon > 0$ such that $\sigma_i\big(\{ b_i \mid b_i < b_i' - \epsilon \}\big) > 0$\footnote{Such $\epsilon$ exists because $\sigma_i\big([0, b_i']\big) > 0$.}, define $\sigma_i^{\epsilon} \in \mathcal{M}(\text{BR}_i(v_i, \alpha_{-i}^k) \setminus \{b_i'\})$ by
		\[
		\sigma_i^{\epsilon}(K) = \frac{ \sigma_i\big( \{ b_i \mid b_i < b_i' - \epsilon \} \cap K \big) }{ \sigma_i\big( \{ b_i \mid b_i < b_i' - \epsilon \} \big) }
		\]
		for any measurable subset $K \subseteq \{Q\} \cup [0, \overline{b}_i]$. That is, $\sigma_i^{\epsilon}$ is the probability measure induced by restricting $\sigma_i$ to the set $[0, b_i' - \epsilon)$ and renormalizing.
		
		Denote
		\begin{align*}
			D(\epsilon) 
			&= \left( V_i(b_i', v_i; b_{-i}^{\epsilon}) - \int_{A_i} V_i(b_i, v_i; b_{-i}^{\epsilon}) \, \mathrm{d}\sigma_i^{\epsilon}(b_i) \right) \cdot \sigma_i\left( \{ b_i \mid b_i < b_i' - \epsilon \} \right) \\
			&= V_i(b_i', v_i; b_{-i}^{\epsilon}) \cdot \sigma_i\left( \{ b_i \mid b_i < b_i' - \epsilon \} \right) - \int_{[0, b_i' - \epsilon)} \left( -\Phi_i(b_i) \right) \, \mathrm{d}\sigma_i(b_i),
		\end{align*}
		where the second equality holds because player~$i$ always loses when playing according to $\sigma_i^{\epsilon}$ against $b_{-i}^{\epsilon}$ and thus receives a payoff of $-\Phi_i(b_i)$. 		
		It is clear that $D(\epsilon) > 0$ for sufficiently small $\epsilon$, and that $D(\epsilon)$ is decreasing in $\epsilon$. Let $M$ be a bound on $u_i$, i.e., $|u_i| \leq M$.

		Hence, for sufficiently small $\epsilon$,  we can obtain:
		\begin{align*}
			&  V_i(b_i',v_i; b_{-i}^{\epsilon}) -   \int_{A_i}V_i(b_i,v_i; b_{-i}^{\epsilon})   \rmd \sigma_i( b_i)  \\
			&	= \quad \left(V_i(b_i',v_i; b_{-i}^{\epsilon}) -   \int_{A_i}V_i(b_i,v_i; b_{-i}^{\epsilon})   \rmd \sigma_i^{\epsilon}(b_i) \right) \sigma_i\left([0, b_i' - \epsilon)\right)  \\
			& \quad + \left(V_i(b_i',v_i; b_{-i}^{\epsilon}) -   \int_{A_i}V_i(b_i,v_i; b_{-i}^{\epsilon})  \rmd \sigma_i^{1}( b_i) \right) \sigma_i\left( [b_i' - \epsilon, b_i')\right)  \\
			&  \quad +\left(V_i(b_i',v_i; b_{-i}^{\epsilon}) -   \int_{A_i}V_i(b_i,v_i; b_{-i}^{\epsilon}) \rmd  \sigma_i^2( b_i) \right) \sigma_i \left([ b_i',\overline{b}_i]\right)   \\
			&	>  D(\epsilon)  -   2M \sigma_i \left( [b_i' - \epsilon, b_i')\right) + 0 > 0
		\end{align*}   
		The first equality follows by decomposing $\sigma_i$ into three parts supported on $[0, b_i' - \epsilon)$, $[b_i' - \epsilon, b_i')$, and $[b_i', \overline{b}_i]$, respectively, and renormalizing these parts into probability measures $\sigma_i^{\epsilon}$, $\sigma_i^{1}$, and $\sigma_i^{2}$.
		\begin{itemize}
			\item Since $u_i$ is bounded by $M$, we have
			\[
			V_i(b_i', v_i; b_{-i}^{\epsilon}) - \int_{A_i} V_i(b_i, v_i; b_{-i}^{\epsilon}) \, \mathrm{d}\sigma_i^{1}(b_i) \geq -2M.
			\]
			\item Since $(W_i(b_i, v) - C_i(b_i))D(b_i,b_{-i}) - \Phi_i(b_i)$ is strictly decreasing in $b_i$, it follows that
			\[
			V_i(b_i', v_i; b_{-i}^{\epsilon}) - \int_{A_i} V_i(b_i, v_i; b_{-i}^{\epsilon}) \, \mathrm{d}\sigma_i^{2}(b_i) > 0.
			\]
		\end{itemize}
		The first inequality arises from the above observations, while the second inequality follows because $D(\epsilon)$ is decreasing in $\epsilon$ with $\lim_{\epsilon \to 0^+} D(\epsilon) > 0$, and $\lim_{\epsilon \to 0^+} \sigma_i\big([b_i' - \epsilon, b_i')\big) = 0$. 
		Therefore, $\sigma_i$ does not dominate $b_i'$.
	\end{enumerate}
	Combining the above results, we conclude the proof.
\end{proof}	



\subsection{Verifications of Facts~\ref{claim-super}-\ref{fact-no PME type}}\label{proof: Facts}

\medskip

\begin{proof}[Verification of Fact~\ref{claim-super}]
%
%
%
%
%
%
		
		
		Let $t_i$ denote the type of player $i$ for $i \in \{1,2\}$. Since each player's payoff function is independent of types, the condition that $u_i$ is supermodular in $(a_i,t_j)$ reduces to $u_i$ being supermodular in $a_i$ alone. This condition holds trivially.
		
		Furthermore, for each player $i \in \{1,2\}$, we observe that:
		\[
		u_i(0,0,t_1,t_2) + u_i(1,1,t_1,t_2) = 1 > u_i(0,1,t_1,t_2) + u_i(1,0,t_1,t_2) = 0.
		\]
		This inequality establishes that $u_i$ is supermodular in $(a_1,a_2)$. Thus, this game is a supermodular game.
		
			%
		
		\
		
		Next, we show that this game admits only two Bayesian Nash equilibria. Suppose that $(s_1,s_2)$ is a Bayesian Nash equilibrium of this game. There are two cases.
		\begin{itemize}
			\item[Case 1.] The action distribution induced by $s_2$ (denoted by $\hat \lambda_2$) assigns positive probability to action $0$ (i.e., $\hat \lambda_2(\{0\}) > 0$). Since action $1$ is weakly dominated by action $0$, it follows immediately that
			\[
			\text{BR}_1(t_1, s_2) \equiv \{0\} \quad \Longrightarrow \quad s_1 \equiv 0.
			\]
			It further implies that $s_2 \equiv 0$.
			
			\item[Case 2.] The action distribution induced by $s_2$ assigns zero probability to action $0$ (i.e., $s_2 \equiv 1$).\footnote{We omit measure-zero sets for simplicity.} As established in Case 1, we must have $s_1 \equiv 1$; otherwise, player $2$ would prefer action $0$ regardless of his type. Thus, $s_1 = s_2 \equiv 1$ in this scenario.
		\end{itemize}
		
		In the first Bayesian Nash equilibrium (denoted by $(g_1, g_2)$), both players always choose the maximum action \(1\) (\textit{i.e.}, \(g_1 = g_2 \equiv 1\)). It is clear that for both players,  action $1$ is weakly dominated by action $0$. As a result, the monotone equilibrium \((g_1 ,g_2 )\) is not perfect.
		
		In the second Bayesian Nash equilibrium (denoted by $(f_1, f_2)$), both players always choose the minimum action \(0\) (\textit{i.e.}, \(f_1 = f_2 \equiv 0\)). We are going to show that $(f_1, f_2)$ is a perfect monotone equilibrium.
		

		For each player $i \in \{1,2\}$, consider the sequence $\{f_i^m\}_{m=3}^{\infty}$ of completely mixed strategies defined by
		\[
		f_i^m(t_i)  =  \left(1 - \tfrac{1}{m}\right)\delta_0 + \tfrac{1}{m}\delta_1 \quad \text{for all } t_i \in [0,1].
		\]
		This sequence satisfies $\lim_{m \to \infty} \rho(f_i^m(t_i), \delta_{f_i(t_i)}) = 0$ for all $t_i \in [0,1]$. When player $i$ employs strategy $f_i^m$, the induced action distribution assigns positive probability to action $0$. Consequently, player $j$'s best response is action $0$ regardless of type. Formally, for all $m \geq 3$, the best response correspondence satisfies
		\[
		\text{BR}_j(t_j, f_i^m) = \{0\} \quad \text{for all } t_j \in [0,1].
		\]
		Hence, for each player \( j \) and almost all \( t_j \in [0, 1] \), we have \( f_j(t_j) \in \text{BR}_j(t_j, f_i^m) \) for any \( m \ge 3 \). This implies that the strategy profile \( (f_1, f_2) \) constitutes a perfect monotone equilibrium.		

The verification of Fact~\ref{claim-super} is now complete.
\end{proof}

Before verifying Facts~\ref{fact:non constant ME} and \ref{fact-no PME type}, we present the payoff table of the bidders in Example~\ref{exam-multi-unit fail PME}.

\begin{table}[h!]
	\centering
	\scalebox{0.85}{
		\begin{tabular}{c|c|c|c}
			\multicolumn{1}{c|}{\textbf{Bidder 1's action}} & 
			\multicolumn{3}{c}{\textbf{Bidder 2's action}} \\
			\hline
			& $0$ & $1$ & $2$ \\
			$(b_{11},0), b_{11} \ge 0$
			& $\tfrac{1}{2} [u_1(t_1)+u_1(2t_1)]$ & $u_1(t_1)$ & $u_1(t_1)$ \\
			$(b_{11},1), b_{11} \ge 1$ 
			& $u_1(2t_1)$ & $\tfrac{1}{2} [u_1(t_1 - 1)+u_1(2t_1 - 2)]$ & $u_1(t_1 - 1)$ \\
			$(b_{11},2), b_{11} \ge 2$ 
			& $u_1(2t_1)$ & $u_1(2t_1-2)$ &  $\tfrac{1}{2} [u_1(t_1 - 2)+u_1(2t_1 - 4)]$  \\
			$(3,3)$ / $(4,3)$ / $(5,3)$
			& $u_1(2t_1)$ & $u_1(2t_1-2)$ & $u_1(2t_1 - 4)$\\
			$(4,4)$ / $(5,4)$
			& $u_1(2t_1)$ & $u_1(2t_1-2)$ & $u_1(2t_1 - 4)$  \\
			$(5,5)$ 
			& $u_1(2t_1)$ & $u_1(2t_1-2)$ & $u_1(2t_1 - 4)$  \\
	\end{tabular}}
\end{table}

\begin{table}[h!]\label{table:bidder 1}
	\centering
	\scalebox{0.85}{
		\begin{tabular}{c|c|c|c}
			\multicolumn{1}{c|}{\textbf{Bidder 1's action}} & 
			\multicolumn{3}{c}{\textbf{Bidder 2's action}} \\
			\hline
			& $3$ & $4$ & $5$ \\
			$(b_{11},0), b_{11} \ge 0$
			& $u_1(t_1)$ & $u_1(t_1)$ & $u_1(t_1)$ \\
			$(b_{11},1), b_{11} \ge 1$ 
			& $u_1(t_1 - 1)$ & $u_1(t_1 - 1)$ 	& $u_1(t_1-1)$ \\
			$(b_{11},2), b_{11} \ge 2$ 
			& $u_1(t_1 - 2)$ & $u_1(t_1 - 2)$ & $u_1(t_1-2)$ \\
			$(3,3)$ / $(4,3)$ / $(5,3)$
			&  $\tfrac{1}{2} [u_1(t_1 - 3)+u_1(2t_1 - 6)]$  & $u_1(t_1 - 3)$& $u_1(t_1-3)$ \\
			$(4,4)$  / $(5,4)$
			& $u_1(2t_1 - 6)$ &  $\tfrac{1}{2} [u_1(t_1 - 4)+u_1(2t_1 - 8)]$ & $u_1(t_1-4)$   \\
			$(5,5)$
			& $u_1(2t_1 - 6)$ &  $u_1(2t_1 - 8)$ & $\tfrac{1}{2}[u_1(t_1-5) + u_1(2t_1 - 10)]$   \\
	\end{tabular}}
	\caption{Bidder 1's Payoff Table}
\end{table}

 \begin{table}[h!]	\label{table:bidder 2}
	\centering
	\scalebox{0.85}{
		\begin{tabular}{c|c|c|c|c|c|c}
			\multicolumn{1}{c|}{\textbf{Bidder 1's action}} & 
			\multicolumn{5}{c}{\textbf{Bidder 2's action}} \\
			\hline
			& $0$ & $1$ & $2$ & $3$ & $4$ & $5$ \\
			$(b_{11},0)$ with $b_{11} \ge 0$
			& $\tfrac{1}{2} t_2$ & $t_2$ & $t_2$ & $t_2$ & $t_2$ & $t_2$ \\
			$(b_{11},1)$ with $b_{11} \ge 1$
			& $0$ & $\tfrac{1}{2} (t_2 - 1)$ & $t_2 - 1$ & $t_2 - 1$ & $t_2 - 1$ & $t_2 - 1$ \\
			$(b_{11},2)$ with $b_{11} \ge 2$
			& $0$ & $0$ & $\tfrac{1}{2} (t_2 - 2)$ & $t_2 - 2$ & $t_2 - 2$& $t_2 - 2$\\
			$(3,3)$ / $(4,3)$ / $(5,3)$ 
			& $0$ & $0$ & $0$ & $\tfrac{1}{2} (t_2 - 3)$ & $t_2 - 3$&  $t_2 - 3$ \\
			$(4,4)$ / $(5,4)$
			& $0$ & $0$ & $0$ & $0$ & $\tfrac{1}{2} (t_2 - 4)$ & $t_2 - 4$ \\
			$(5,5)$ 
			& $0$ & $0$ & $0$ & $0$ & $0$ & $\tfrac{1}{2} (t_2 - 5)$ \\
	\end{tabular}}
	\caption{Bidder 2's Payoff Table}
\end{table}

\begin{proof}[Verification of Fact~\ref{fact:non constant ME}]
	\begin{enumerate}
		\item[\textbf{1.}] 	Given that bidder 2 plays the strategy $	f_2(t_2) = 
		\begin{cases}
			1 & \text{if } t_2 \in [0, \tfrac{2}{3})\\ 
			2 & \text{if } t_2 \in [\tfrac{2}{3}, 2]
		\end{cases}$, bidder 1's payoff for 
		$t_1 \in [3, \tfrac{13}{4}]$, $t_1 \in (\tfrac{13}{4}, \tfrac{17}{4})$, 
		$t_1 \in [\tfrac{17}{4}, \tfrac{9}{2}]$, and $t_1 \in (\tfrac{9}{2},5]$ 
		is given as a $4$-dimensional vector:  
		
		\[
		\begin{cases}
			(t_1, \; t_1, \; t_1, \; \tfrac{9}{2}), & \text{if } (b_{11}, b_{12}) = (b_{11},0) \text{ with } b_{11} \ge 1, \\[1mm]
			(2t_1-\tfrac{10}{3}, \; \tfrac{4}{3}t_1 -  \tfrac{7}{6}, \; \tfrac{9}{2}, \; \tfrac{9}{2}), & \text{if } (b_{11}, b_{12}) \text{ satisfies } b_{12} \ge 3.
		\end{cases}
		\]
		
		The payoff of bidder 1 when choosing $(b_{11}, b_{12}) = (b_{11},1)$ or $(b_{11}, b_{12}) = (b_{11},2)$  are  less than that of choosing $(5,5)$;  hence we omit these payoffs here.  
		
		For $t_1 \in (3, \tfrac{13}{4})$, we have $t_1 > 2t_1 - \tfrac{10}{3}$, and 
		$\tfrac{4}{3} t_1 - \tfrac{7}{6} > t_1$ for all $t_1 \in (\tfrac{7}{2},\tfrac{17}{4})$.
		It is then easy to see that $	f_1(t_1) = 
		\begin{cases}
			(1,0) & \text{if } t_1 \in [3, \tfrac{7}{2}) \\ 
			(3,3) & \text{if } t_1 \in [\tfrac{7}{2}, 5]
		\end{cases}$ is a best response against $f_2$.  
		
		Moreover, given that bidder 1 plays the strategy $f_1$, $f_2$ is bidder 2's best response by Table~7.
		
		\medskip
		
		\item[\textbf{2.}]  	Suppose there exists a perfect monotone equilibrium $(f_1^*, f_2^*)$.  
		Since $u_2(x) = x \leq 0$ whenever $x \leq 0$ and $t_2 \in [0,2]$, Table~7 shows that, for bidder~2, actions $3$, $4$, and $5$ are weakly dominated by action $2$ for all $t_2 \in [0,2]$. Hence, these actions cannot occur in any perfect monotone equilibrium. Moreover, when $t_2 > 1$, action $0$ is weakly dominated by action $1$; when $t_2 < 1$, action $2$ is weakly dominated by action $1$.  
		
		These observations imply that, in any perfect monotone equilibrium, bidder~2's strategy must take the form
		\begin{align}\label{equa:bidder 2 strategy}
			f_2^*(t_2) =
			\begin{cases}
				0, & \text{if } t_2 \in [0,2p_0), \\[6pt]
				1, & \text{if } t_2 \in [2p_0,2(p_0+p_1)), \\[6pt]
				2, & \text{if } t_2 \in [2(p_0+p_1),2],
			\end{cases}
		\end{align}
		where $0 \leq p_0 \leq \tfrac{1}{2}$, $p_1 \geq 0$, and $p_0+p_1 \geq \tfrac{1}{2}$.

		Since bidder 1 is satiated by  four  and half units of surplus, by Table~6, we know that for $t_1 \ge 4.5$, action $(b_{11},b_{12})$ with $b_{12} \ge 1$ are weakly dominated by action $(b_{11},0)$ for bidder 1. Hence, in any perfect monotone equilibrium,  bidder~1's strategy must take the form
		\begin{align}\label{equa:bidder 1 strategy}
			f_1^*(t_1) =
			\begin{cases}
				(0,0), & \text{if } t_1 \in [3, 3+2 \tilde p_0), \\[6pt]
				(1,0), & \text{if } t_1 \in [3+2 \tilde p_0, 3 + 2(\tilde p_0+ \tilde p_1)), \\[6pt]
				(2,0), & \text{if } t_1 \in [3 + 2(\tilde p_0+\tilde p_1),3 + 2(\tilde p_0+ \tilde p_1+ \tilde p_2)), \\[6pt]
				(3,0), & \text{if } t_1 \in [3 + 2(\tilde p_0+\tilde p_1+ \tilde p_2),3 + 2(\tilde p_0+ \tilde p_1+ \tilde p_2+ \tilde p_3)), \\[6pt]
				(4,0), & \text{if } t_1 \in [3 + 2(\tilde p_0+\tilde p_1+\tilde p_2+\tilde p_3),3 + 2(\tilde p_0+\tilde p_1+ \tilde p_2+\tilde p_3+ \tilde p_4)), \\[6pt]
				(5,0), & \text{if } t_1 \in [3 + 2(\tilde p_0+ \tilde p_1+ \tilde p_2+\tilde p_3+\tilde p_4), 5],
			\end{cases}
		\end{align}
		where $\tilde p_0, \tilde p_1,\tilde p_2,\tilde p_3,\tilde p_4 \ge 0$ and $\tilde p_0+ \tilde p_1+\tilde p_2+\tilde p_3+\tilde p_4 \le 1$.
		
		Given that bidder 1 plays the strategy $f_1^*$ as in Equation~\eqref{equa:bidder 1 strategy}, by bidder 2's payoff table (Table~7) we know that action $0$ is not bidder 2's best response against $f_1^*$ for any $t_2 > 0$. That is, $p_0 = 0$ in Equation~\eqref{equa:bidder 2 strategy},  which leads to
		
		\begin{align}\label{equa:2 more}
			f_2^*(t_2) =
			\begin{cases}
				1, & \text{if } t_2 \in [0,2p_1), \\[2mm]
				2, & \text{if } t_2 \in [2p_1,2],
			\end{cases}
		\end{align}
		where $\tfrac{1}{2} \le p_1 \le 1$.  
		
		Given that bidder 2 plays the strategy $f_2^*$ as in Equation~\eqref{equa:2 more}, by bidder 1's payoff table (Table~6) we know that $p_1 \cdot u_1(2t_1 - 2)  + (1-p_1) \cdot u_1(2t_1 - 4)  \ge \tfrac{1}{2} \times  [ (2t_1 - 2) + (2t_1 - 4)] = 2t_1 - 3 > t_1 = u_1(t_1)  $ for $t_1 \in (3,\tfrac{13}{4})$. That is, $(b_{11}, 0)$ with $b_{11} \ge 1$ is not a best response for bidder $1$ for all $t_1 \in (3,\tfrac{13}{4})$, which leads to a contradiction with Equation~\eqref{equa:bidder 1 strategy}. Hence, a perfect monotone equilibrium does not exist.
	\end{enumerate}

This completes the verification of Fact \ref{fact:non constant ME}.
\end{proof}

\medskip

\begin{proof}[Verification of Fact~\ref{fact-no PME type}]

Suppose there is a perfect monotone equilibrium $(f_1^*, f_2^*)$. For bidder~2, if his type $t_2 \in [0,1]$, it is clear that actions $2$--$5$ are weakly dominated by action $1$; if his type $t_2 \in [4,5]$, then actions $0$--$3$ are weakly dominated by action $4$. That is, $f_2^*$ must be 
\begin{align}\label{equa:2 type}
	f_2^*(t_2) =
	\begin{cases}
		0, & \text{if } t_2 \in [0,p_0), \\[2mm]
		1, & \text{if } t_2 \in [p_0,1], \\[2mm]
		4, & \text{if } t_2 \in [4,4+p_4), \\[2mm]
		5, & \text{if } t_2 \in [4+p_4,5],
	\end{cases}
\end{align}
where $0 \le p_0, p_4 \le 1$. 

Given that bidder~2 plays the strategy $f_2^*$, if bidder~1's type $t_1 \in [2,3]$, then bidder~2 will randomize over actions $0$ and $1$. Hence, bidder~1's best response will be to choose among actions $(b_{11}, b_{12})$ with $b_{12} \ge 2$. If bidder~1's type $t_1 \in [3,4]$, then bidder~2 will randomize over actions $4$ and $5$. Hence, bidder~1's best response will be to choose among actions $(b_{11},0)$. The above analysis shows that $f_1^*$ is not monotone. Therefore, there does not exist any perfect monotone equilibrium.
\end{proof}


\begin{thebibliography}{XX}
\addcontentsline{toc}{section}{References}
\setlength{\itemsep}{2.5pt}

\bibitem[Athey(2001)]{A2001} Susan~Athey, Single crossing properties and the existence of pure strategy equilibria in games of incomplete information, \textit{Econometrica} \textbf{69} (2001), 861--889.


\bibitem[Amir and Rietzke(2025)]{AR2025} Rabah~Amir and David Rietzke, A Comment on: ``Monotone comparative statics'', \textit{Econometrica} \textbf{93} (2025), 1481--1490.

\bibitem[Arrow(1951)]{Arrow1951} Kenneth~J.~Arrow, Alternative approaches to the theory of choice in risk-taking situations, \textit{Econometrica} \textbf{19} (1951), 404--437.

		
\bibitem[Bajoori, Flesch and Vermeulen(2016)]{Bajoori2016} Elnaz~Bajoori, J\v{a}nos~Flesch and Dries~Vermeulen, Behavioral perfect equilibrium in Bayesian games, \textit{Games and Economic Behavior} \textbf{98} (2016),  78--109.
		

		

\bibitem[Brandenburger, Friedenberg and Keisler(2008)]{BFK2008} Adam~Brandenburger, Amanda~Friedenberg and Jerome~H.~Keisler, Admissibility in games, \textit{Econometrica} \textbf{76} (2008), 307--352.



\bibitem[Carbonell-Nicolau and McLean(2014)]{CM2014} Oriol~Carbonell-Nicolau and Richard~P.~McLean, Refinements of Nash equilibrium in potential games, \textit{Theoretical Economics} \textbf{9} (2014), 555--582.

\bibitem[Carbonell-Nicolau and McLean(2015)]{CM2015} Oriol~Carbonell-Nicolau and Richard~P.~McLean, On equilibrium refinements in supermodular games, \textit{International Journal of Game Theory} \textbf{44} (2015), 869--890.

\bibitem[Carbonell-Nicolau and McLean(2018)]{CM2018} Oriol~Carbonell-Nicolau and Richard~P.~McLean, On the existence of Nash equilibrium in Bayesian games, \textit{Mathematics of Operations Research} \textbf{43} (2018), 100--129.

\bibitem[Carbonell-Nicolau(2021)]{Carbonell-Nicolau2021} Oriol~Carbonell-Nicolau, Perfect equilibria in games of incomplete information, \textit{Economic Theory} \textbf{71} (2021), 1591--1648.

\bibitem[Che, Kim and Kojima(2026)]{CKK2026} Yeon-Koo Che, Jinwoo Kim and Fuhito Kojima, Monotone comparative statics without lattices, working paper, 2026. Available at \href{https://arxiv.org/abs/1911.06442}{\texttt{https://arxiv.org/abs/1911.06442}}.


\bibitem[Dziewulski and Quah(2024)]{DQ2024} Pawel~Dziewulski and John K‐H. Quah, Comparative statics with linear objectives: normality, complementarity, and ranking multi‐prior beliefs, \textit{Econometrica} \textbf{92} (2024), 167--200.


\bibitem[Fu and Yu(2018)]{FY2018} Haifeng~Fu and Haomiao~Yu, Pareto refinements of pure-strategy equilibria in games with public and private information, \textit{Journal of Mathematical Economics} \textbf{79} (2018), 18--26.





\bibitem[He and Sun(2019)]{HS2019} Wei~He and Yeneng~Sun, Pure-strategy equilibria in Bayesian games, \textit{Journal of Economic Theory} \textbf{180} (2019), 11--49.

\bibitem[He, Sun and Xu(2025)]{HSX2025} Wei~He, Yeneng~Sun and Hanping~Xu, Monotone perfection,  working paper, 2025. Available at \href{https://arxiv.org/abs/2509.01358}{\texttt{https://arxiv.org/abs/2509.01358}}.


\bibitem[Jackson et al.(2002)]{JSSZ2002} Matthew~O.~Jackson, Leo~Simon, Jeroen~M.~Swinkels and William~R.~Zame, Communication and equilibrium in discontinuous games of incomplete information, \textit{Econometrica} \textbf{70} (2002), 1711–1740.

\bibitem[Jackson and Swinkels(2005)]{JS2005} Matthew~O.~Jackson and Jeroen~M.~Swinkels, Existence of equilibrium in single and double private value auctions, \textit{Econometrica} \textbf{73} (2005), 93--139.


\bibitem[Kohlberg and Mertens(1986)]{KM1986} Elon Kohlberg and Jean-Francois Mertens, On the strategic stability of equilibria, \textit{Econometrica} \textbf{54} (1986), 1003--1037.

\bibitem[Luce and Raiffa(1957)]{LR1957} Duncan~Luce and Howard~Raiffa, Games and Decisions: Introduction and Critical Survey, Wiley, New York, 1957.

\bibitem[McAdams(2003)]{M2003} David~McAdams, Isotone equilibrium in games of incomplete information, \textit{Econometrica} \textbf{71} (2003), 1191--1214.

\bibitem[McAdams(2006)]{M2006} David~McAdams, Monotone equilibrium in multi-unit auctions, \textit{The Review of Economic Studies} \textbf{73} (2006), 1039--1056.

\bibitem[McAdams(2007a)]{M2007a} David~McAdams, Uniqueness in symmetric first-price auctions with affiliation, \textit{Journal of Economic Theory} \textbf{136} (2007a), 144--166.

\bibitem[McAdams(2007b)]{M2007} David~McAdams, On the failure of monotonicity in uniform-price auctions, \textit{Journal of Economic Theory} \textbf{137} (2007b), 729--732.


\bibitem[Milgrom and Roberts(1990)]{MR1990}  Paul R.  Milgrom and John Roberts, Rationalizability, learning, and equilibrium in games with strategic complementarities, \textit{Econometrica} \textbf{58} (1990), 1255--1277.

\bibitem[Milgrom and Shannon(1994)]{MS1994}  Paul  R. Milgrom and Chris Shannon, Monotone comparative statics, \textit{Econometrica} \textbf{62} (1994), 157--180.

\bibitem[Milgrom and Weber(1982)]{MW1982}  Paul R.  Milgrom and Robert  J. Weber, A theory of auctions and competitive bidding, \textit{Econometrica} \textbf{50} (1982), 1089--1122.

\bibitem[Milgrom and Weber(1985)]{MW1985} Paul~R.~Milgrom and Robert~J.~Weber, Distributional strategies for games with incomplete information, \textit{Mathematics of Operations Research} \textbf{10} (1985), 619--632.

\bibitem[Myerson and Reny(2020)]{MR2020} Roger~B.~Myerson and Philip~J.~Reny, Perfect conditional $\epsilon$‐equilibria of multi‐stage games with infinite sets of signals and actions, \textit{Econometrica} \textbf{88} (2020), 495--531.

\bibitem[Prokopovych and Yannelis(2017)]{PY2017} Pavlo~Prokopovych and Nicholas~C.~Yannelis, On strategic complementarities in discontinuous games with totally ordered strategies, \textit{Journal of Mathematical Economics} \textbf{70} (2017), 147--153.
		
\bibitem[Prokopovych and Yannelis(2019)]{PY2019} Pavlo~Prokopovych and Nicholas~C.~Yannelis, On monotone approximate and exact equilibria of an asymmetric first-price auction with affiliated private information, \textit{Journal of Economic Theory} \textbf{184} (2019), 104925.

\bibitem[Quah(2007)]{Quah2007} John~K-H.~Quah, The comparative statics of constrained optimization problems, \textit{Econometrica} \textbf{75} (2007), 401--431.

\bibitem[Quah and Strulovici(2009)]{QS2009} John~K-H.~Quah and Bruno~Strulovici, Comparative statics, informativeness, and the interval dominance order, \textit{Econometrica} \textbf{77} (2009), 1949--1992.

\bibitem[Quah and Strulovici(2012)]{QS2012} John~K-H.~Quah and Bruno~Strulovici, Aggregating the single crossing property, \textit{Econometrica} \textbf{80} (2012), 2333--2348.

\bibitem[Radner and Rosenthal(1982)]{RR1982} Roy~Radner and Robert~W.~Rosenthal, Private information and pure-strategy equilibria, \textit{Mathematics of Operations Research} \textbf{7} (1982), 401--409.

\bibitem[Rath(1994)]{Rath1994} Kali~P.~Rath, Some refinements of Nash equilibria of large games, \textit{Games and Economic Behavior} \textbf{7} (1994), 92--103. 

\bibitem[Rath(1998)]{Rath1998} Kali~P.~Rath, Perfect and proper equilibria of large games, \textit{Games and Economic Behavior}  \textbf{22} (1998), 331--342.

\bibitem[Reny(1999)]{Reny1999} Philip~J.~Reny, On the existence of pure and mixed strategy Nash equilibria in discontinuous games, \textit{Econometrica} \textbf{67} (1999), 1029--1056.

\bibitem[Reny(2011)]{R2011} Philip~J.~Reny, On the existence of monotone pure-strategy equilibria in Bayesian games, \textit{Econometrica} \textbf{79} (2011), 499--553.

\bibitem[Reny and Zamir(2004)]{RZ2004} Philip~J.~Reny and Shmuel Zamir, On the existence of pure strategy monotone equilibria in asymmetric first-price auctions, \textit{Econometrica} \textbf{72} (2004), 1105--1125.

\bibitem[Selten(1975)]{S1975} Reinhard~Selten, Reexamination of the perfectness concept for equilibrium points in extensive games, \textit{International Journal of Game Theory} \textbf{4} (1975), 25--55.

\bibitem[Shannon(1995)]{Shannon1995} Chris~Shannon, Weak and strong monotone comparative statics, \textit{Economic Theory} \textbf{5} (1995), 209--227.

\bibitem[Simon and Stinchcombe(1995)]{SS1995}  Leo K. Simon and Maxwell B. Stinchcombe, Equilibrium refinement for infinite normal-form games, \textit{Econometrica} \textbf{63} (1995), 1421--1443.


\bibitem[Sun and Zeng(2020)]{SZ2020} Xiang~Sun and Yishu~Zeng, Perfect and proper equilibria in large games, \textit{Games and Economic Behavior} \textbf{119} (2020), 288--308.

\bibitem[Topkis(1978)]{T1978}  Donald M. Topkis, Minimizing a submodular function on a lattice, \textit{Operations Research} \textbf{26} (1978), 305--321.

		
				
\bibitem[Van Zandt and  Vives(2007)]{ZV2007} Timothy~Van~Zandt and  Xavier~Vives, Monotone equilibria in Bayesian games of strategic complementarities, \textit{Journal of Economic Theory} \textbf{134} (2007), 339--360.

\bibitem[Vives(1990)]{Vives1990} Xavier~Vives, Nash equilibrium with strategic complementarities, \textit{Journal of Mathematical Economics} \textbf{19} (1990), 305--321.
\end{thebibliography}
\end{document}